\documentclass[showpacs,aps,superscriptaddress,letterpaper,nofootinbib]{revtex4}

\usepackage{xspace}
\usepackage{graphicx}
\usepackage{epsfig}
\usepackage{float}
\usepackage[nointegrals]{wasysym}
\usepackage{hyperref}
\usepackage{url}
\usepackage{subfig}
\usepackage{amsmath}
\usepackage{xcolor}
\usepackage{amssymb,array}

\newcommand{\offshell}{off-shell}

\newcommand{\onshell}{on-shell}

\newcommand{\be}{\begin{equation}}
\newcommand{\ee}{\end{equation}}
\newcommand{\ba}{\begin{eqnarray}}
\newcommand{\ea}{\end{eqnarray}}

\newcommand{\Z}{Z}

\newcommand{\PW}{W}
\newcommand{\PH}{H}

\newcommand{\qqbar}{q\bar{q}}

\newcommand{\Hboson}{$H$~boson\xspace}
\newcommand{\Hff}{H\!f\!\bar{f}}

\begin{document}

\begin{flushright}
\vbox{
\begin{tabular}{l}
{HU-EP-21/38}
\end{tabular}
}
\end{flushright}

\vspace{0.6cm}

\title{Constraining anomalous Higgs boson couplings to virtual photons}
\author{Jeffrey Davis  \thanks{e-mail: jdavi231@jhu.edu}}
\affiliation{Department of Physics and Astronomy, Johns Hopkins University, Baltimore, MD 21218, USA}
\author{Andrei V. Gritsan \thanks{e-mail: gritsan@jhu.edu}}
\affiliation{Department of Physics and  Astronomy, Johns Hopkins University, Baltimore, MD 21218, USA}
\author{Lucas~S.~Mandacar\'{u}~Guerra  \thanks{e-mail: lmandac1@jhu.edu}}
\affiliation{Department of Physics and Astronomy, Johns Hopkins University, Baltimore, MD 21218, USA}
\author{Savvas Kyriacou \thanks{e-mail: skyriac2@jhu.edu}}
\affiliation{Department of Physics and Astronomy, Johns Hopkins University, Baltimore, MD 21218, USA}
\author{Jeffrey Roskes \thanks{e-mail: hroskes@jhu.edu}}
\affiliation{Department of Physics and Astronomy, Johns Hopkins University, Baltimore, MD 21218, USA}
\author{Markus Schulze \thanks{e-mail:  markus.schulze@physik.hu-berlin.de}}
\affiliation{Institut f\"ur Physik, Humboldt-Universit\"at zu Berlin, D-12489 Berlin, Germany}

\date{September 27, 2021}

\begin{abstract}
\vspace{2mm}
We present a study of Higgs boson production in vector boson fusion and in association with a vector boson
and its decay to two vector bosons, with a focus on the treatment of virtual loops and virtual photons. 
Our analysis is performed with the JHU generator framework.  Comparisons are made to several other frameworks, 
and the results are expressed in terms of an effective field theory. New features of this study include a proposal on how
to handle singularities involving Higgs boson decays to light fermions via photons, calculation of the partial 
Higgs boson width in the  presence of anomalous couplings to photons, a comparison of the next-to-leading-order
electroweak corrections to effects from effective couplings, and phenomenological observations regarding the special role of 
intermediate photons in analysis of LHC data in the effective field theory framework. Some of these features 
are illustrated with projections for experimental measurements with the full LHC and HL-LHC datasets. 
\end{abstract}


\maketitle

\thispagestyle{empty}


\section{Introduction}
\label{sect:eft_intro}

The large amount of data analyzed by the ATLAS and CMS experiments on the Large Hadron Collider (LHC) is consistent with the predictions 
of the standard model (SM) of particle physics. 
Among the measurements performed, the discovery and characterization of the Higgs ($H$) boson have been 
crucial in completing the SM~\cite{Aad:2012tfa,Chatrchyan:2012xdj,Chatrchyan:2012jja}.
Yet, open questions remain, such as the low value of the \Hboson's mass, its Yukawa coupling hierarchy, 
the source of $CP$ violation required for matter abundance, and the connection of the SM to other cosmological observations. 

Studies of electroweak production (VBF and $VH$) and decay ($H\to VV$) of the \Hboson probe $HVV$ interactions 
over a large range of momentum transfer, which can expose possible new particles that couple through loops. 
Such electroweak processes lead to rich information in kinematic distributions of the \Hboson decay products
and associated particles. Analysis of such distributions can shed light on the nature of the $HVV$ interactions
and has been discussed extensively\cite{Nelson:1986ki,Soni:1993jc,Chang:1993jy,Barger:1993wt,
Arens:1994wd,BarShalom:1995jb,Gunion:1996xu,Han:2000mi,Plehn:2001nj,Choi:2002jk,Buszello:2002uu,Hankele:2006ma,
Accomando:2006ga,Godbole:2007cn,Hagiwara:2009wt,Gao:2010qx,DeRujula:2010ys,Christensen:2010pf,
Gainer:2011xz,Bolognesi:2012mm,Ellis:2012xd,Chen:2012jy,Gainer:2013rxa,Artoisenet:2013puc,Anderson:2013afp,Chen:2013waa,
Chen:2013ejz,Gainer:2014hha,Gonzalez-Alonso:2014eva,Dolan:2014upa,Demartin:2014fia,Chen:2014gka,Buckley:2015vsa,
Greljo:2015sla,Gritsan:2016hjl,deFlorian:2016spz,Hartmann:2015aia,Dawson:2018pyl,Dedes:2018seb,Dawson:2018liq,
Brivio:2019myy,Gritsan:2020pib}.
Such studies can be naturally performed within the effective field theory (EFT) framework,
with examples of application to LHC data documented in 
Refs.~\cite{Chatrchyan:2012jja,Aad:2013xqa,Chatrchyan:2013mxa,Khachatryan:2014kca,
Khachatryan:2015mma,Aad:2015mxa,Aad:2016nal,Khachatryan:2016tnr,Sirunyan:2017tqd,Aaboud:2017oem,Aaboud:2017vzb,
Aaboud:2018xdt,Sirunyan:2019twz,Sirunyan:2019nbs,Aad:2020mnm,Sirunyan:2021fpv}. 

In our earlier studies using the JHU generator 
framework~\cite{Gao:2010qx,Bolognesi:2012mm,Anderson:2013afp,Gritsan:2016hjl,Gritsan:2020pib,Martini:2021uey}, 
we relied on dedicated Monte Carlo simulation, and demonstrated optimal discrimination, reweighting techniques, 
and analysis of a bosonic resonance with the most general anomalous couplings. 
We build upon this framework of the JHU generator and MELA analysis package
with the goal of demonstrating its application to the \Hboson's interactions in electroweak processes
with massless vector bosons, such as in $HZ\gamma$, $H\gamma\gamma$, and $H\mathrm{gg}$ vertices. 
Such couplings are generated in the SM through loops of SM particles. They also lead to divergence of fixed-order calculations 
for virtual $\gamma^*$ states when the four-momentum squared $q^2_{\gamma^*}$ approaches zero. In the perturbative expansion, 
such terms are poorly defined at low values of $q^2$. Some prior discussion of this effect can be found 
in Refs.~\cite{Hartmann:2015aia,Dawson:2018pyl,Dedes:2018seb,Dawson:2018liq,Brivio:2019myy,Denner:2019zfp}.

We review the parameterization of anomalous \Hboson couplings in Section~\ref{sect:eft_couplings} and
discuss applications of the JHU generator framework to EFT studies in Section~\ref{sect:eft-basis}.
In Section~\ref{sect:eft-xs}, the partial \Hboson width and production cross sections are calculated in the presence 
of anomalous couplings to massless vector bosons. 
In Section~\ref{sect:eft-sm}, the treatment of the next-to-leading-order (NLO) electroweak (EW) effects is discussed. 
In Section~\ref{sect:eft-lowq}, we make a proposal on how to handle singularities involving intermediate photons 
in the \Hboson decays.
In Section~\ref{sect:eft-analysis}, several phenomenological observations are made in application to LHC data.


\section{Parameterization of anomalous interactions}
\label{sect:eft_couplings}

We start with the $HVV$ scattering amplitude of a spin-zero boson $H$ and two vector bosons $VV$ with polarization vectors 
and momenta $\varepsilon_{1}^\mu$, $q_1^\mu$ and $\varepsilon_{2}^\mu$, $q_2^\mu$.  The amplitude is parameterized by
\begin{eqnarray}
  A({H V_1 V_2}) \!=\! \frac{1}{v} &\bigg\{  &
                          M_{V_1}^2 \bigg(g_{1}^{VV} 
+ \frac{\kappa_1^{VV}q_{1}^2 + \kappa_2^{VV} q_{2}^{2}}{\left(\Lambda_{1}^{VV} \right)^{2}} + \frac{\kappa_3^{VV} (q_1+q_2)^2}{\left(\Lambda_{Q}^{VV} \right)^{2}} 
+ \frac{2 q_1\cdot q_2}{M_{V_1}^2} g_2^{VV}\bigg) (\varepsilon_1 \cdot \varepsilon_2) 
\nonumber \\ &&
                        -2 g_2^{VV} \, {(\varepsilon_1 \cdot q_2)(\varepsilon_2 \cdot q_1)}
                        -2  g_4^{VV} \, {\varepsilon_{\varepsilon_1\,\varepsilon_2\,q_1\,q_2}}
                               \bigg\}
                               \,,
\label{eq:HVV}
\end{eqnarray}
where $v$ is the vacuum expectation value, under the conventions $\varepsilon_{0123}=+1$ and $(q^\mu)=(E,\vec{q})$.
This amplitude represents the three possible tensor structures of the \Hboson's interaction with two vector bosons,
with expansion of the terms up to $q_i^2$. 
By symmetry we have $\kappa_1^{ZZ}=\kappa_2^{ZZ}$, but we do not enforce 
$\kappa_1^{WW}=\kappa_2^{WW}$ for $W^\pm$ bosons.
Note that 
$\kappa_1^{\gamma\gamma}=\kappa_2^{\gamma\gamma}=\kappa_1^{\rm gg}=\kappa_2^{\rm gg}=\kappa_1^{Z\gamma}=0$, 
while $\kappa_2^{Z\gamma}/(\Lambda_1^{Z\gamma})^2$ may contribute.
The coupling $\kappa_3^{VV}/(\Lambda_Q^{VV})^2$ allows for scenarios which violate the gauge symmetries of the SM. 

An effective $HVV$ interaction may be generated by loops of fermions, in which case the couplings $\kappa_f$ 
and $\tilde\kappa_f$ describe the \Hboson interactions as 
\begin{eqnarray}
&& { A}(\Hff) = - \frac{m_f}{v}
\bar{\psi}_{f} \left ( \kappa_f  + \mathrm{i} \, \tilde\kappa_f  \gamma_5 \right ) {\psi}_{f}\,,
\label{eq:ampl-spin0-qq} \label{eq:Hffcoupl}
\end{eqnarray}
where $\bar{\psi}_{f}$ and ${\psi}_{f}$ are the Dirac spinors and $m_f$ is the fermion mass.
For the SM fermions, $(\kappa_f, \tilde\kappa_f)= (1,0)$.

The equivalent Lagrangian for \Hboson interactions with gauge bosons (in the mass eigenstate parameterization) reads
\begin{eqnarray}
\label{eq:EFT_hvv}
 {\cal L}_{\rm hvv} =& & {h \over v} \left [ 
  M_Z^2 \left (1 +  \delta c_z \right )   Z_\mu Z^\mu
+ \frac{M_Z^2}{v^2} c_{zz} Z_{\mu \nu} Z^{\mu\nu}  
+ \frac{e^2}{s_w^2} c_{z \Box} Z_\mu \partial_\nu Z^{\mu \nu}
+ \frac{M_Z^2}{v^2} \tilde c_{zz}  Z^{\mu \nu} \tilde Z_{\mu\nu}
\right . \nonumber \\ & & \left . 
 +2 M_W^2 \left (1 +  \delta c_w \right ) W_\mu^+ W^{-\mu} 
+ 2 \frac{M_W^2}{v^2} c_{ww}  W_{\mu \nu}^+  W^{-\mu\nu} 
+ \frac{e^2}{s_w^2} c_{w \Box} \left (W_\mu^- \partial_\nu W^{+\mu \nu} + {\mathrm h.c.} \right )  
 + \frac{e^2}{2 s_w^2} \tilde c_{ww}  W^{+\mu \nu}   \tilde W_{\mu\nu}^- 
 \right . \nonumber \\ & & \left . 
+ \frac{e^2}{2 s_w c_w} c_{z \gamma}  Z_{\mu \nu} A^{\mu\nu} 
+ \frac{e^2}{2 s_w c_w} \tilde c_{z \gamma} Z_{\mu \nu} \tilde A^{\mu\nu}
+ \frac{e^2}{s_w c_w} c_{\gamma \Box}  Z_\mu \partial_\nu A^{\mu \nu}
\right . \nonumber \\ & & \left . 
+ c_{\gamma \gamma} {e^2 \over 4} A_{\mu \nu} A^{\mu \nu} 
+ \tilde c_{\gamma \gamma} {e^2 \over 4} A^{\mu \nu} \tilde A_{\mu \nu} 
+  c_{gg} {g_s^2 \over 4 } G_{\mu \nu}^a G^{a \mu \nu}   
+  \tilde c_{gg} {g_s^2 \over 4} G^{a\mu \nu} \tilde G_{\mu \nu}^a  
\right ]\,,
\end{eqnarray} 
in accordance with Eq.~(II.2.20) in Ref.~\cite{deFlorian:2016spz},
where $e^2=4\pi \alpha$ and $g_s^2= 4\pi \alpha_s$ are the squared electromagnetic and strong coupling constants, 
respectively, and $s_w=\sqrt{1-c_w^2}$ is the sine of the weak mixing angle. 
The covariant derivative used to derive this expression is 
$D_\mu = \partial_\mu -\mathrm{i} \frac{e}{2 s_w} \sigma^i W_\mu^i - \mathrm{i} \frac{e}{2 c_w} B_\mu$~\cite{deFlorian:2016spz, Falkowski:2001958}.  
We note that the convention $\varepsilon_{0123}=+1$ defines the relative sign of the $CP$-odd $\tilde{c}_i$ and 
$CP$-even ${c}_i$ couplings~\cite{Gritsan:2020pib}, while the relative sign in front of the $W_\mu^i$ and $B_\mu$ terms 
in the covariant derivative defines the sign of the $Z\gamma$ couplings relative to the $ZZ$ and $\gamma\gamma$. 
The latter could be viewed as the sign of $s_w$, if a different convention is
adopted.\footnote{In the actual parameterization of the \textsc{JHUGen} framework discussed in Section~\ref{sect:eft-basis}
and Refs.~\cite{Gao:2010qx,Bolognesi:2012mm,Anderson:2013afp,Gritsan:2016hjl,Gritsan:2020pib,Martini:2021uey}, 
the $D_\mu = \partial_\mu -\mathrm{i} \frac{e}{2 s_w} \sigma^i W_\mu^i + \mathrm{i} \frac{e}{2 c_w} B_\mu$
convention was adopted for historical reasons. A transformation $g_i^{Z\gamma}\to -g_i^{Z\gamma}$ or $\kappa_i^{Z\gamma}\to -\kappa_i^{Z\gamma}$
of the input parameters in this framework would lead to the convention 
$D_\mu = \partial_\mu -\mathrm{i} \frac{e}{2 s_w} \sigma^i W_\mu^i - \mathrm{i} \frac{e}{2 c_w} B_\mu$,
which is needed for consistent application of the formalism discussed in this paper.
}

The generality of our amplitude parameterization allows us to uniquely represent each EFT 
coefficient in Eq.~(\ref{eq:EFT_hvv}) by an anomalous coupling in Eq.~(\ref{eq:HVV}). 
\begin{eqnarray}  &&
\label{eq:EFT_ci}
    \delta c_z  = \frac12 g_1^{ZZ} - 1\,,
    \quad\quad
    c_{zz} = -\frac{2 s_w^2 c_w^2}{e^2} g_2^{ZZ}\,,
    \quad\quad
    c_{z \Box} = \frac{M_Z^2 s_w^2}{e^2} \, \frac{\kappa_1^{ZZ}}{(\Lambda_1^{ZZ})^2}\,,
    \quad\quad
    \tilde c_{zz} = -\frac{2 s_w^2 c_w^2}{e^2} g_4^{ZZ}\,,
    \nonumber \\ &&
     \delta c_w = \frac12 g_1^{WW} - 1\,,
    \quad\quad
    c_{ww} = -\frac{2 s_w^2 }{e^2} g_2^{WW}\,,
    \quad\quad
    c_{w \Box} = \frac{M_W^2 s_w^2}{e^2} \, \frac{\kappa_1^{WW}}{(\Lambda_1^{WW})^2}\,,
    \quad\quad
    \tilde c_{ww} = -\frac{2 s_w^2}{e^2} g_4^{WW}\,,
    \nonumber\\ &&
     c_{z \gamma} = -\frac{2 s_w c_w}{e^2} g_2^{Z\gamma}\,,
    \quad\quad
    \tilde c_{z \gamma} = -\frac{2 s_w c_w}{e^2} g_4^{Z\gamma}\,,
    \quad\quad
    c_{\gamma \Box} = \frac{s_w c_w}{e^2} \, \frac{M_Z^2}{(\Lambda_1^{Z\gamma})^2} \kappa_2^{Z\gamma}\,,
   \nonumber\\ &&
     c_{\gamma \gamma} = -\frac{2}{e^2} g_2^{\gamma\gamma}\,,   
    \quad\quad
   \tilde c_{\gamma \gamma} = -\frac{2}{e^2} g_4^{\gamma\gamma}\,,
    \quad\quad
     c_{gg} = -\frac{2}{g_s^2} g_2^{gg}\,,
   \quad\quad
   \tilde c_{gg} = -\frac{2}{g_s^2} g_4^{gg}\,.
\end{eqnarray} 
Note that not every anomalous coupling in Eq.~(\ref{eq:HVV}) has a corresponding term in the EFT Lagrangian of Eq.~(\ref{eq:EFT_hvv}). 
For example, the term $\kappa_3^{VV}/(\Lambda_Q^{VV})^2$  is not gauge invariant and is not present in Eq.~(\ref{eq:EFT_hvv}). 
Similarly, $\kappa_1^{WW}=\kappa_2^{WW}$ due to charge symmetry.

So far we have discussed the \Hboson interactions without considering additional symmetries. 
The $\mathrm{SU(3)\times SU(2)\times U(1)}$ symmetry of the standard model effective field theory 
(SMEFT)~\cite{Weinberg:1979sa, Buchmuller:1985jz, Leung:1984ni, Dedes:2017zog} is a motivated
framework which allows relating EFT operators. 
Not all of the EFT coefficients are independent when limiting the discussion to dimension-six interactions
with this symmetry. 
The linear relations for the dependent coefficients can be found in Ref.~\cite{deFlorian:2016spz} and they translate into 
relations amongst our anomalous couplings as follows:
\begin{eqnarray}
  g_1^{WW} &=& g_1^{ZZ} + \frac{\Delta M_W}{M_W} \,,    
  \label{eq:deltaMW}
  \\
  g_2^{WW} &=& c_w^2 g_2^{ZZ} + s_w^2 g_2^{\gamma\gamma} + 2 s_w c_w g_2^{Z\gamma}\,,
  \label{eq:g2WW}
  \\
  g_4^{WW} &=& c_w^2 g_4^{ZZ} + s_w^2 g_4^{\gamma\gamma} + 2 s_w c_w g_4^{Z\gamma}\,,
  \label{eq:g4WW}
  \\
  \frac{\kappa_1^{WW}}{(\Lambda_1^{WW})^2} (c_w^2-s_w^2) &=& \frac{\kappa_1^{ZZ}}{(\Lambda_1^{ZZ})^2}
                                                           +2 s_w^2 \frac{g_2^{\gamma\gamma}-g_2^{ZZ}}{M_Z^2} 
                                                          +2 \frac{s_w}{c_w} (c_w^2-s_w^2) \frac{g_2^{Z\gamma}}{M_Z^2}\,,
 \label{eq:kappa1WW}
  \\
  \frac{\kappa_2^{Z\gamma}}{(\Lambda_1^{Z\gamma})^2} (c_w^2-s_w^2) &=&  2 s_w c_w \left( \frac{\kappa_1^{ZZ}}{(\Lambda_1^{ZZ})^2} 
                                                                              + \frac{ g_2^{\gamma\gamma} - g_2^{ZZ}}{M_Z^2}  \right)
                                                                               +2 (c_w^2-s_w^2) \frac{g_2^{Z\gamma}}{M_Z^2}\,.
 \label{eq:kappa2Zgamma}
 \end{eqnarray}

The Lagrangian for \Hboson interactions with gauge bosons can be written in the Warsaw basis~\cite{Grzadkowski:2010es}
which preserves the $\mathrm{SU(3)\times SU(2)\times U(1)}$ symmetry of SMEFT.
The relationship between operators in the Warsaw basis and the mass-eigenstate basis is discussed in Section~\ref{sect:eft-basis}.


\section{The JHU generator framework and the EFT bases}
\label{sect:eft-basis}

The JHU generator framework ({\textsc{JHUGen}) includes a Monte Carlo generator and matrix element techniques for 
optimal analysis of the data. It is built upon the earlier developed framework of the JHU generator and \textsc{MELA} analysis 
package~\cite{Gao:2010qx,Bolognesi:2012mm,Anderson:2013afp,Gritsan:2016hjl,Gritsan:2020pib,Martini:2021uey}
and extensively uses matrix elements provided by 
\textsc{MCFM}~\cite{Campbell:2010ff,Campbell:2011bn,Campbell:2013una,Campbell:2015vwa,Campbell:2015qma}.
The SM processes in \textsc{MCFM} are extended to add the most general scalar and 
gauge couplings and possible additional states. 
This framework includes many options for production and decay of the $H$ boson, which include the gluon fusion, 
vector boson fusion, and associated production with a vector boson ($VH$) in both \onshell\ $H$ and \offshell\ $H^*$ 
production~\cite{Gritsan:2020pib}. In the \offshell\ case, interference with background processes or a second resonance is included. 
The processes with direct sensitivity to fermion $\Hff$ couplings, such as $t\bar{t}H$, $b\bar{b}H$, $tqH$, 
$tWH$, or $H\to\tau^+\tau^-$, are discussed in Refs.~\cite{Gritsan:2016hjl,Martini:2021uey}.

The {\textsc{JHUGen} framework was adopted in Run-I analyses using LHC
data~\cite{Chatrchyan:2012xdj,Chatrchyan:2012jja,Chatrchyan:2013mxa,Chatrchyan:2013iaa,
Aad:2013xqa,Khachatryan:2014iha,Khachatryan:2014ira,Khachatryan:2014kca,Khachatryan:2015mma,Khachatryan:2015cwa,
Aad:2015mxa,Khachatryan:2016tnr}
and employed in recent Run-II measurements of the $HVV$ anomalous couplings from the first joint analysis 
of \onshell\ production and decay~\cite{Sirunyan:2017tqd,Sirunyan:2019nbs}, 
from the first joint analysis of \onshell\ and \offshell\ \Hboson\ production~\cite{Sirunyan:2019twz}, 
for the first measurement of the $CP$ structure of the Yukawa interaction between the \Hboson and top quark~\cite{Sirunyan:2020sum},
in the search for a second resonance in interference with the continuum background~\cite{Sirunyan:2018qlb,Sirunyan:2019pqw},
and in EFT approach to the $HVV$, $H$gg, and $\Hff$ interactions~\cite{Sirunyan:2021fpv}.

\subsection{EFT basis considerations}

The framework is based on the amplitude parameterization in Eqs.~(\ref{eq:HVV}) and~(\ref{eq:Hffcoupl}).
In order to simplify translation between different coupling conventions and operator bases, including the Higgs and Warsaw bases, 
within the JHU generator framework, we provide the \textsc{JHUGenLexicon} program, which includes an interface 
to the generator and matrix element library and can also be used for standalone or other applications~\cite{Gritsan:2020pib}. 
The relationship of the amplitude parameterization to the mass eigenstate basis of the EFT formulations 
in Eq.~(\ref{eq:EFT_hvv}) is performed through the simple linear relationship in Eq.~(\ref{eq:EFT_ci}). 
The functionality of this program is similar to  \textsc{Rosetta}~\cite{Falkowski:2015wza}, but it is limited in scope
to application to the \Hboson interactions and provides additional options to introduce certain symmetries or constraints,
as illustrated below. 

We count five $CP$-even and three $CP$-odd independent electroweak $HVV$ operators, as well as one $CP$-even and 
one $CP$-odd $H$gg operators in the mass-eigenstate basis in Section~\ref{sect:eft_couplings}. The same number
of independent \Hboson operators exists in the Warsaw basis. The relationship between the six
$CP$-even operators is quoted explicitly in Eq.~(14) of Ref.~\cite{Falkowski:2015wza}. This relationship is direct, 
with the exception of the $\delta{v}$ parameter defined in the Warsaw basis in Eq.~(15) of Ref.~\cite{Falkowski:2015wza}. 
One could remove an extra parameter from transformation with constraints from precision electroweak data. 
For example, we can set $\Delta M_W=0$ in Eq.~(\ref{eq:deltaMW}), because $M_W$ is measured precisely. 
This allows us to express $\delta{v}$ through the other $HVV$ operators in the Warsaw basis.
The \textsc{JHUGenLexicon} program provides such an option and the following studies in this paper 
will be presented with such a constraint. 

With the above symmetries and constraints, including $\Delta M_W=0$, the translation between the Warsaw basis 
and the independent amplitude coefficients is
\begin{eqnarray}
\label{eq:Warsaw-to-Higgs}
   \delta g_1^{ZZ} &=& \frac{v^2}{\Lambda^2} \left (2 C_{\PH \Box} + \frac{6e^2}{s_w^2} C_{\PH W B} + \left(\frac{3c_w^2}{2s_w^2} -\frac{1}{2}\right) C_{\PH D} \right),
   \nonumber \\
   \kappa_1^{ZZ}  &=& \frac{v^2}{\Lambda^2} \left (-\frac{2e^2}{s_w^2} C_{\PH W B} + \left(1-\frac{1}{2 s_w^2}\right) C_{\PH D} \right),
   \nonumber \\
   g_2^{ZZ} &=& -2 \frac{v^2}{\Lambda^2} \left( s_w^2 C_{\PH B} + c_w^2 C_{\PH W} + s_w c_w C_{\PH W B} \right),
   \nonumber \\
   g_2^{Z\gamma} &=& -2 \frac{v^2}{\Lambda^2} \left( s_w c_w \left( C_{\PH W}-C_{\PH B}\right)  + \frac12 \left(s_w^2-c_w^2\right) C_{\PH W B} \right),
   \nonumber \\
   g_2^{\gamma\gamma}  &=& -2 \frac{v^2}{\Lambda^2} \left( c_w^2 C_{\PH B} + s_w^2 C_{\PH W} - s_w c_w C_{\PH W B} \right),
   \nonumber \\
   g_2^{\rm gg} &=& -2 \frac{v^2}{\Lambda^2} C_{\PH G} ,
   \nonumber \\
   g_4^{ZZ} &=& -2 \frac{v^2}{\Lambda^2} \left( s_w^2 C_{\PH\widetilde{B}} + c_w^2 C_{\PH\widetilde{W}} + s_w c_w C_{\PH\widetilde{W}B}  \right),
   \nonumber \\
   g_4^{Z\gamma} &=& -2 \frac{v^2}{\Lambda^2} \left( s_w c_w \left( C_{\PH\widetilde{W}}-C_{\PH\widetilde{B}}\right)  + \frac12 \left(s_w^2-c_w^2\right) C_{\PH\widetilde{W}B}  \right),
   \nonumber \\
   g_4^{\gamma\gamma} &=& -2 \frac{v^2}{\Lambda^2} \left( c_w^2 C_{\PH\widetilde{B}} + s_w^2 C_{\PH\widetilde{W}} - s_w c_w C_{\PH\widetilde{W}B} \right),
   \nonumber \\
   g_4^{\rm gg} &=& -2 \frac{v^2}{\Lambda^2} C_{\PH\widetilde{G}},
\end{eqnarray}
where $\Lambda$ is the scale of new physics, which we set to $\Lambda=1$\,TeV as a convention, 
and $\delta g_1^{ZZ}$ is the correction to the SM value of $g_1^{ZZ}=2$.
According to Eq.~(\ref{eq:deltaMW}), $\delta g_1^{WW}=\delta g_1^{ZZ}$, and
the other dependent amplitude coefficients can be derived from Eqs.~(\ref{eq:g2WW}--\ref{eq:kappa2Zgamma}).
 
A numerical example of the relationship between the $C_{\PH X}=1$ contribution of a single operator in the Warsaw basis
and the couplings in the mass-eigenstate amplitude in Eq.~(\ref{eq:HVV}) is shown in Table~\ref{tab:relate-couplings}, which 
corresponds to the reverse of Eq.~(\ref{eq:Warsaw-to-Higgs}).

\begin{table}[!t]
\centering
\captionsetup{justification=centerlast}
\caption{The values of the couplings in the mass-eigenstate amplitude in Eq.~(\ref{eq:HVV}) corresponding to 
the $C_{\PH X}=1$ contribution of a single operator in the Warsaw basis with $\Lambda=1$\,TeV. 
The relationship corresponds to the reverse of Eq.~(\ref{eq:Warsaw-to-Higgs}).
When quoting the $\kappa_2^{Z\gamma}$ and $\kappa_1^{ZZ}=\kappa_2^{ZZ}$ values, 
we set $\Lambda_1^{Z\gamma}=\Lambda_1^{ZZ}=100$\,GeV in Eq.~(\ref{eq:HVV}).
}
\label{tab:relate-couplings}
\begin{tabular}{lccccccccccccc}
\vspace{-0.3cm} \\
   \hline
            &  $\delta{g}_1^{ZZ}=\delta{g}_1^{WW}$  & $\kappa_1^{ZZ}$ 
            &  $g_2^{ZZ}$  &  $g_2^{Z\gamma}$ &  $g_2^{\gamma\gamma}$  
            &  $g_4^{ZZ}$  &  $g_4^{Z\gamma}$ &  $g_4^{\gamma\gamma}$ 
            & $\kappa_2^{Z\gamma}$ & $\kappa_1^{WW}$  & $g_2^{WW}$  &  $g_4^{WW}$  
             \\
   \hline
     $C_{\PH\Box}$    & 0.1213 & 0 & 0 & 0 & 0 & 0 & 0 & 0 & 0 & 0 & 0 & 0  \\
     $C_{\PH{D}}$    & 0.2679 & $-0.0831$ & 0 & 0 & 0 & 0 & 0 & 0 & $-0.1320$ & $-0.1560$ & 0 & 0 \\
     $C_{\PH\PW}$    & 0 & 0 & $-0.0929$ & $-0.0513$ & $-0.0283$ & 0 & 0 & 0 & 0 & 0 & $-0.1212$ & 0 \\
     $C_{\PH\PW{B}}$    & 0.1529 & $-0.0613$ & $-0.0513$ & ~~0.0323 & ~~0.0513 & 0 & 0 & 0 & ~~0.1763 & ~~0.0360 & 0 & 0 \\
     $C_{\PH{B}}$    & 0 & 0 & $-0.0283$ &~~0.0513 & $-0.0929$ & 0 & 0 & 0 & 0 & 0 & 0 & 0 \\
     $C_{\PH\widetilde{\PW}}$    & 0 & 0 & 0 & 0 & 0 & $-0.0929$ & $-0.0513$ & $-0.0283$ & 0 & 0 & 0 & $-0.1212$ \\
     $C_{\PH\widetilde{W}B} $    & 0 & 0 & 0 & 0 & 0 & $-0.0513$ & ~~0.0323 & ~~0.0513 & 0 & 0 & 0 & 0 \\
     $C_{\PH\widetilde{B}}$    & 0 & 0 & 0 & 0 & 0 & $-0.0283$ & ~~0.0513 & $-0.0929$ & 0 & 0 & 0 & 0 \\
   \hline
\end{tabular}
\end{table}

\subsection{Application to the VBF, $VH$, and $H\to VV$ processes}

One of the new features in this paper, compared to the earlier work, is the study of the
$g_2^{Z\gamma}$,  $g_2^{\gamma\gamma}$, $g_4^{Z\gamma}$, and $g_4^{\gamma\gamma}$ anomalous couplings
in electroweak production of the \Hboson. 
Their effect in the $H\to 4\ell$ process was studied with LHC data~\cite{Khachatryan:2014kca} and with phenomenological 
tools~\cite{Chen:2012jy,Chen:2014gka}. In the following we re-examine the $H\to 4\ell$ decay and investigate the VBF 
and $VH$ processes. In the case of $VH$ production, we consider three final states $Z(\to f\bar{f})H$, $\gamma^*(\to f\bar{f})H$, and $\gamma H$,
and both $q\bar{q}$ or gg production channels, as all are affected by the $HVV$ couplings of our interest. 
While the gluon fusion process formally appears at higher order in QCD,
the large gluon parton luminosity at the LHC makes this channel interesting to examine. 

In this study, we only examine the operators affecting the \Hboson interactions in Table~\ref{tab:relate-couplings}  
and study their effect on the $HVV$ couplings. Other operators, such as $HZff$ contact terms for example, 
are included in the \textsc{JHUGen} framework, but they are not the primary interest in this study
because their existence would become evident in resonance searches and in electroweak measurements,
without the need for \Hboson production. Moreover, such contact terms are equivalent to the combination of
the $\kappa_1^{ZZ}$ and $\kappa_2^{Z\gamma}$ couplings if flavor universality is assumed~\cite{Gritsan:2020pib}.
Not only $HVV$ interactions may be affected by the above operators in the processes under study. 
For example, the  $C_{\PH\PW{B}}$ operator also affects the $Zff$ couplings. However, these $Zff$ couplings
should be well constrained in electroweak measurements. For this reason, should one of the considered
operators affect the $Zff$ interactions, we assume some other operators not affecting the direct \Hboson 
interactions must also contribute to bring the $Zff$ couplings to the SM values. 

Numerical results of the relative contributions of operators to the $H\to VV\to 4\ell$, VBF, $q\bar{q}$ 
or gg\,$\to V(\to\ell^+\ell^-)H$, and  $\gamma H$ processes are shown in Appendix~\ref{app:A}. 
The general observations from Tables~\ref{tab:warsaw-4l}--\ref{tab:warsaw-VH} is that the relative importance of
the $g_2^{Z\gamma}$,  $g_2^{\gamma\gamma}$, $g_4^{Z\gamma}$, and $g_4^{\gamma\gamma}$ couplings
changes between the processes. Taking the example of the $C_{\PH\PW{B}}$ operator, these couplings lead
to an overwhelming contribution in the $H\to4\ell$ process. However, their contribution in the VBF and $VH$ 
processes is not significant and is especially tiny in the case of the $VH$ process. These features will affect our ability 
to use different processes to constraint anomalous couplings with photons. 
We note that the $VH$ process with $V\to\ell^+\ell^-$ includes both $ZH$ and $\gamma^*H$ production
mechanisms, where $\gamma^*$ leads to low-$q^2$ contributions in the $m_{\ell\ell}$ invariant mass,
which can be observed in kinematic distributions. 

The gg\,$\to ZH$ process has been shown to have no contributions of the two anomalous $HVV$ tensor 
structures appearing in Eq.~(\ref{eq:HVV}) in the triangular loop diagram~\cite{Gritsan:2020pib}. 
Therefore, only the SM-like tensor structure with the $g_1$ and $\kappa_1^{ZZ}$ couplings contributes 
to this diagram, as shown in Table~\ref{tab:warsaw-ggVH}. The \offshell\ photon does not couple to the 
triangular fermion loop either~\cite{Gritsan:2020pib}, and, therefore,  the $\kappa_2^{Z\gamma}$ coupling 
does not contribute. The box diagram is sensitive to the fermion couplings of the \Hboson, which we do not 
vary in this study of anomalous $HVV$ interactions. As the result, the gg\,$\to ZH$ process 
features a rather limited set of EFT operators and we will not study this process in more detail in this paper,
leaving further details to Ref.~\cite{Gritsan:2020pib}. 

The $\gamma H$ production process has been largely neglected in analysis of LHC data. However, this process was used
in the search for the \Hboson with anomalous couplings in $e^+e^-$ production prior to the \Hboson discovery~\cite{L3:2004vpt} 
and proposed in application to $CP$-even EFT operator constraints at the LHC~\cite{Khanpour:2017inb,Shi:2018lqf}. 
From Table~\ref{tab:warsaw-gammaH}, it is evident that only the
$g_2^{Z\gamma}$,  $g_2^{\gamma\gamma}$, $g_4^{Z\gamma}$, and $g_4^{\gamma\gamma}$ 
couplings contribute, and this channel does not receive tree-level SM contributions. 
Because the photon is \onshell, it does not receive contribution from $\kappa_2^{Z\gamma}$ either.
This process is generated by the dimension-6 operators squared in the EFT expansion in
combination with the EW loops generated by the SM particles. As an approximation to the SM production 
cross section, we use the calculation with the $g_2^{Z\gamma,\rm SM}$ and $g_2^{\gamma\gamma,\rm SM}$
values calculated in Section~\ref{sect:eft-sm}. These point-like couplings reproduce the SM decay width of the 
processes $\PH\to Z\gamma$ and $\gamma\gamma$, respectively. Due to the \offshell\ $V=Z/\gamma^*$ in the 
process $\qqbar\to V\to \gamma H$, these point-like couplings are not expected to reproduce the full EW loop 
calculation in the SM, but they are expected to provide a good estimate, which we use as $\sigma_{\rm SM}^{\gamma\PH}$ 
in Table~\ref{tab:warsaw-gammaH}. 
The $\gamma H$ process may be of particular interest in isolating the $CP$-odd couplings $g_4^{Z\gamma}$ and 
$g_4^{\gamma\gamma}$ in combination with $CP$-even couplings $g_2^{Z\gamma}$ and $g_2^{\gamma\gamma}$,
which is complementary to the $\PH\to Z\gamma$ and $\gamma\gamma$ decays.

\subsection{Kinematic distributions with EFT effects} 

The kinematic effects in the $H\to VV$, VBF, and $VH$ processes can typically be described with 
five angular observables and two invariant masses, or $q^2_i$ of the two vector bosons,
as illustrated in Fig.~\ref{fig:kinematics}~\cite{Gao:2010qx,Anderson:2013afp,Gritsan:2020pib}. 
The distributions of two of these angles, $\theta^*$ and $\Phi_1$, are random for a spin-zero \Hboson,
but are less trivial for a higher-spin resonance or non-resonant production. 
In the following, we disentangle the relative contributions of the $ZZ$, $WW$, $Z\gamma$, and $\gamma\gamma$
intermediate vector-boson states to simulation with a given operator in the Warsaw basis. 
Such a decomposition reveals interesting kinematic effects and also allows us to validate the 
tools used for simulation of EFT effects and match their conventions. 

\begin{figure}[t]
\centerline{
\includegraphics[width=0.45\linewidth]{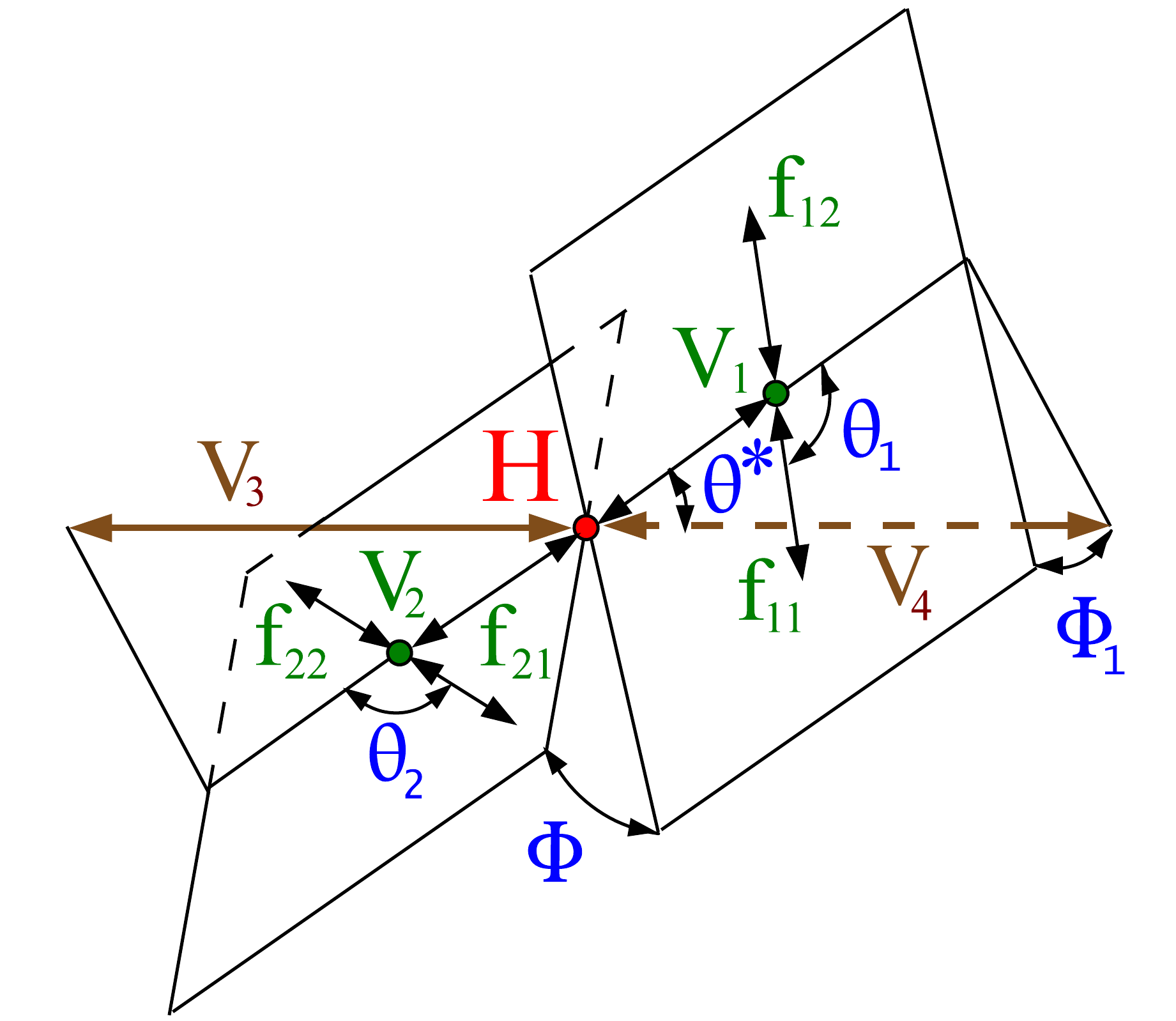}
}
\captionsetup{justification=centerlast}
\caption{
Three kinematic topologies of the \Hboson\ production and decay~\cite{Gritsan:2020pib}:
vector boson fusion $q_{12}q_{22}\to q_{11}q_{21}(V_1V_2\to H\to V_3V_4)$,
$VH$ production $q_{11}q_{12}\to V_1 \to V_2(H\to V_3V_4)$, and
four-fermion decay $V_3V_4\to H \to V_1V_2 \to 4f$.
}
\label{fig:kinematics}
\end{figure}

We use the \textsc{JHUGen} program to generate several models which allow us to visualize 
the relative contributions of the mass eigenstates of the vector bosons. 
We also model the inclusive kinematic distributions with the \textsc{SMEFTsim} program~\cite{Brivio:2017btx}
using \textsc{MadGraph5} simulation~\cite{Alwall:2014hca}.
Once the sign conventions are matched, as discussed in Section~\ref{sect:eft_couplings},
we find good agreement.
A similar comparison with SM couplings using the \textsc{Prophecy4f}~\cite{Denner:2019fcr} 
and \textsc{HAWK}~\cite{Denner:2014cla} generators is shown in Section~\ref{sect:eft-sm}.
Throughout this paper and unless otherwise noted, the calculations are performed at LO in QCD and EW, 
with the $\overline{\mathrm{MS}}$-mass for the top quark $m_{t} =162.7$\,GeV, the \onshell\ mass 
for the bottom quark $m_{b} =4.18$\,GeV, QCD scale $\mu = M_H/2$, $\alpha_s=0.1188$,
$\alpha=1/128$, $s^2_w=0.23119$, $G_F = 1.16639 \times 10^{-5}$ GeV$^{-2}$~\cite{ParticleDataGroup:2020ssz}, 
and the NNPDF 3.0 parton distribution functions~\cite{NNPDF:2014otw}.

In the $H\to4\ell$ and $VH$ processes, we require $m_{\ell\ell}>1$\,GeV.
In the VBF process, we apply the selection requirements 
$m_{jj}>300$\,GeV, $p_T^\mathrm{jet}  > 1$\,GeV, $|\eta^\mathrm{jet} | < 5$, 
$\Delta\eta_{jj} > 1$, $\Delta{R}_{jj} > 0.3$, $\sqrt{q_V^2} > 15$\,GeV.
In the $H\to4\ell$ decay, we model the $C_{HWB}=1$ contribution to the SM, as shown in Fig.~\ref{fig:Lexicon-H4l}
with cross section decomposition presented in Table~\ref{tab:warsaw-4l}. 
In VBF or $VH$, we model the $C_{\PH\widetilde{W}B}=10$ or $C_{HB}=100$ contribution to the SM, 
as shown in Fig.~\ref{fig:Lexicon-VBF} or Fig.~\ref{fig:Lexicon-VH},
with the cross section decomposition presented in Table~\ref{tab:warsaw-VBF} or Table~\ref{tab:warsaw-VH}.
The size of anomalous contributions is chosen to be large compared to SM for visibility of their contributions. 

In the $H\to4\ell$ process, the larger and the smaller invariant masses of the dilepton pairs $m_1$ and $m_2$
are the two observables representing $q^2_1$ and $q^2_2$. 
In Fig.~\ref{fig:Lexicon-H4l}, there are clear peaks towards $m_2\to 0$
in the case of couplings with photons, $HZ\gamma$ and $H\gamma\gamma$, a feature to which we will return in 
Sections~\ref{sect:eft-xs}, \ref{sect:eft-sm} and~\ref{sect:eft-lowq}. In the case of $H\gamma\gamma$, 
this extends to $m_1\to 0$ as well. Modeling such contributions becomes essential, and we will discuss
extensions of such modeling to $m_{\ell\ell}<1$\,GeV later. Moreover, in analysis of experimental data, 
detector effects change significantly for either $\gamma^*$ or $Z$ intermediate states, and dedicated 
simulation of such effects with the full detector modeling becomes important. 
In Fig.~\ref{fig:Lexicon-H4l}, the $m_1$ and $m_2$ distributions are shown separately 
in the $H\to4e / 4\mu$ and $H\to2e2\mu$ decays. The interference of two diagrams with 
permutation of identical leptons in the case of $H\to4e / 4\mu$ leads to suppression of the 
peaks at $m_1\to 0$ and $m_2\to 0$ in the case of $H\gamma\gamma$. 
This feature becomes important in analysis of the $H\gamma\gamma$ couplings. 

Since $Zff$ couplings have been constrained with precision EW data, we do not allow their change
in these studies and assume that modification of other operators, not contributing to the \Hboson couplings,
can compensate any possible shift of the $Zff$ couplings due to $C_{\PH{W}B}$. 
However, in Fig.~\ref{fig:Lexicon-H4l} we also show distributions with modification of
the $Zff$ couplings, indicated with $\delta g^{Zff}$. Corrections to the multidimensional angular distributions 
are expected due to non-zero values of the $R_i$ and $A_f$ parameters discussed 
in Refs.~\cite{Gao:2010qx,Bolognesi:2012mm,Anderson:2013afp}.
These corrections are visible in the projection on the $\Phi$ observable in Fig.~\ref{fig:Lexicon-H4l},
but are very small for any practical purpose with the typical values of $C_{\PH{W}B}$
in the present studies. These corrections become sizable with larger values of $C_{\PH{W}B}$.

In the VBF process, we can calculate the $q_{1,2}^{\rm VBF}=\sqrt{-q_{1,2}^2}$ values using the momenta of
the fully reconstructed \Hboson and two jets and using the direction of incoming partons along the proton beams.
In Fig.~\ref{fig:Lexicon-VBF}, there is a clear preference of lower $q_{1,2}^{\rm VBF}$ values in the case
of couplings with photons. There is a strong correlation between the $q_{1,2}^{\rm VBF}$ values and the
transverse momentum $p_T$ of the jets, which leads to different detector effects. 
We note the asymmetric distribution of the $\Phi^{\rm VBF}$ angle in Fig.~\ref{fig:Lexicon-VBF},
which is most visible in the $HZZ$ process but can also be seen in the combined distribution. This happens due 
to interference of the $CP$-even SM amplitude and $CP$-odd $C_{\PH\widetilde{W}B}=10$ contributions. 

In the $VH$ process, $q^2_1$ and $q^2_2$ represent the $VH$ and the $V\to\ell^+\ell^-$ invariant masses, respectively. 
There are particularly dramatic effects in the $m_{\ell\ell}$ distribution, shown in Fig.~\ref{fig:Lexicon-VH}, 
where the virtual photon $\gamma^*$ results in the low-mass enhancement, as opposed to the peak at $m_Z$. 
A dedicated analysis of the small invariant masses in the $\gamma^*H$ production may be needed
for effective EFT analysis of the process.

\begin{figure}[tb]
\centering
\centerline{
\includegraphics[width=0.32\textwidth]{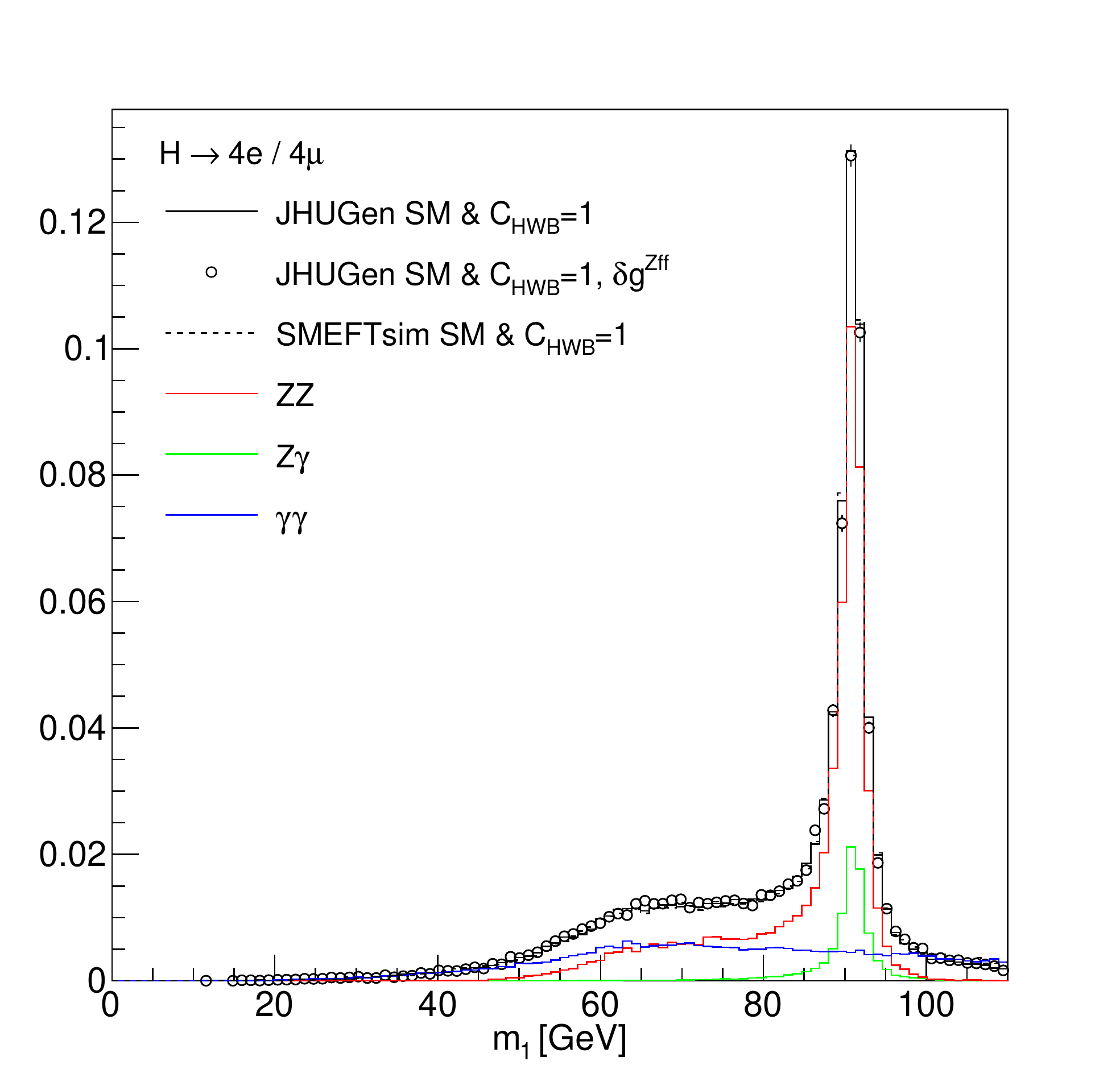}
\includegraphics[width=0.32\textwidth]{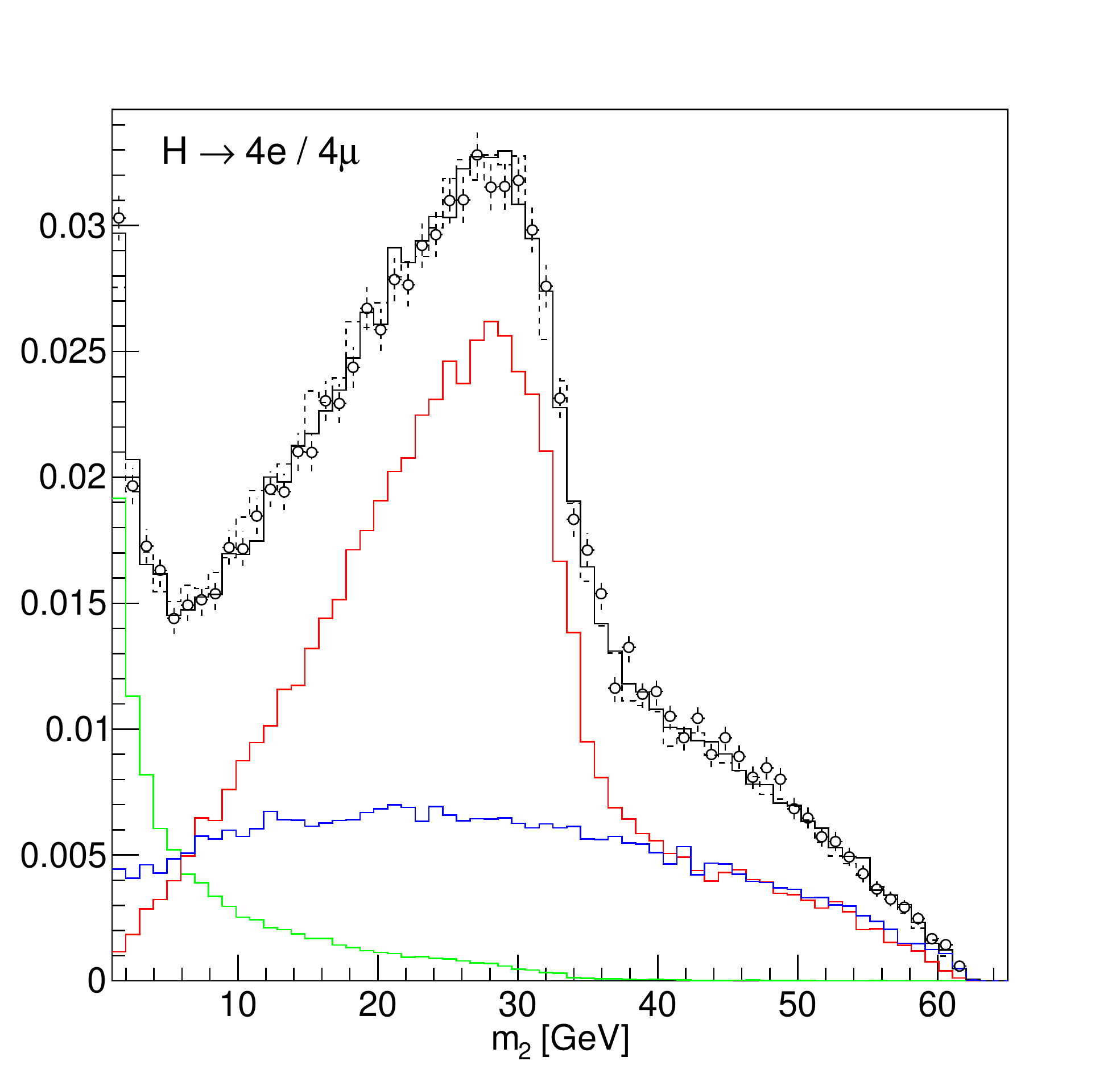}
\includegraphics[width=0.32\textwidth]{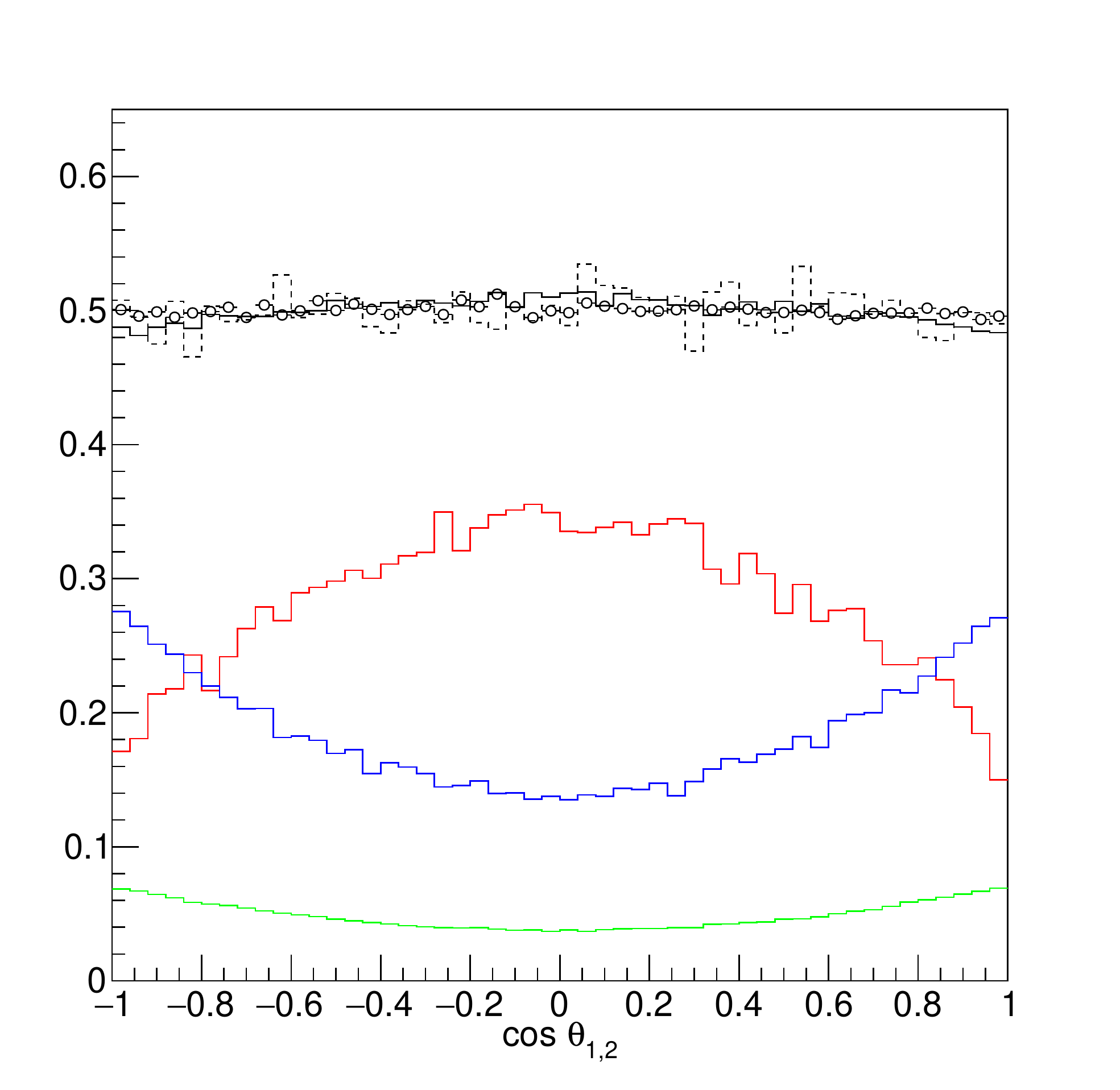}
}
\centerline{
\includegraphics[width=0.32\textwidth]{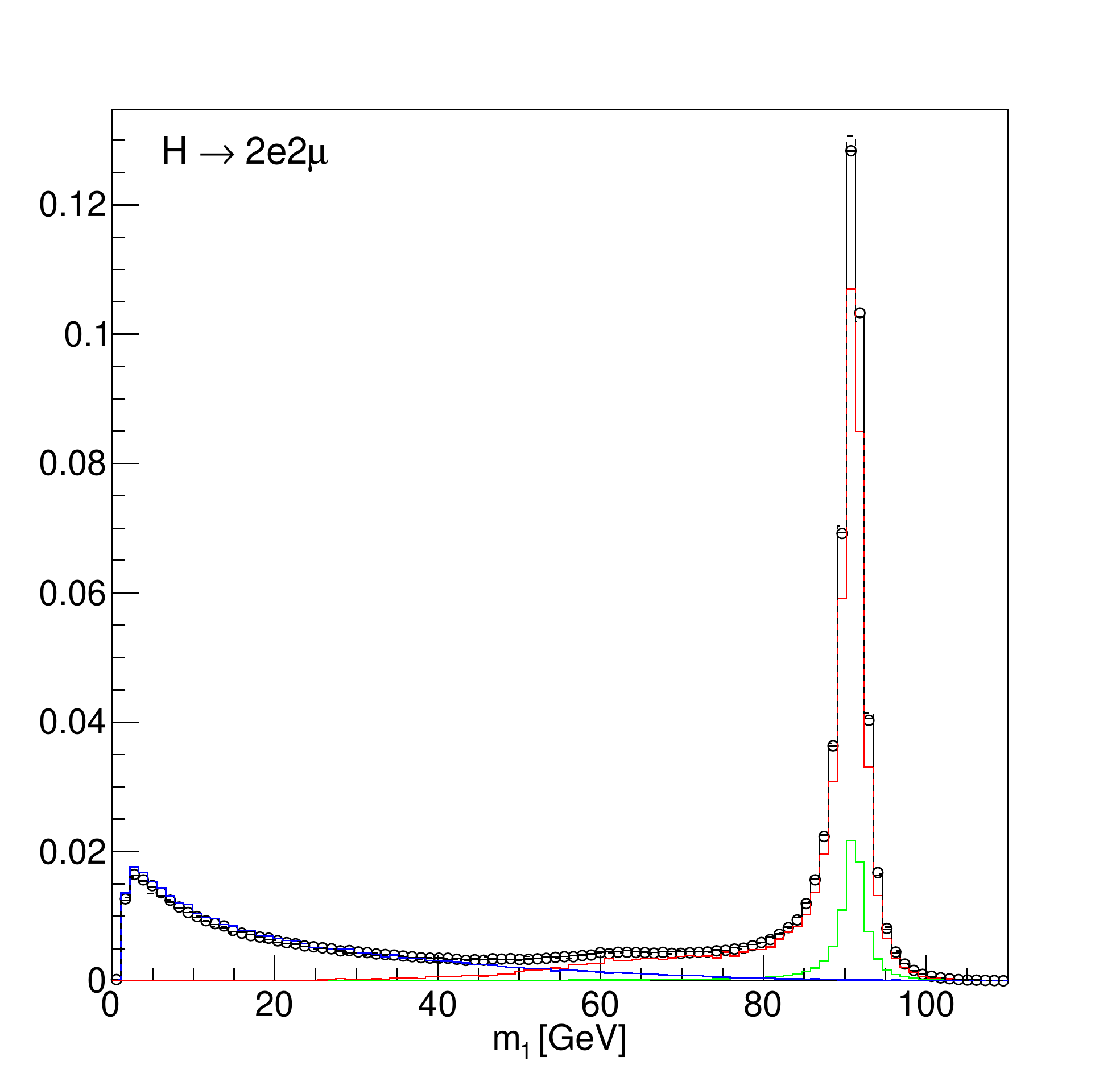}
\includegraphics[width=0.32\textwidth]{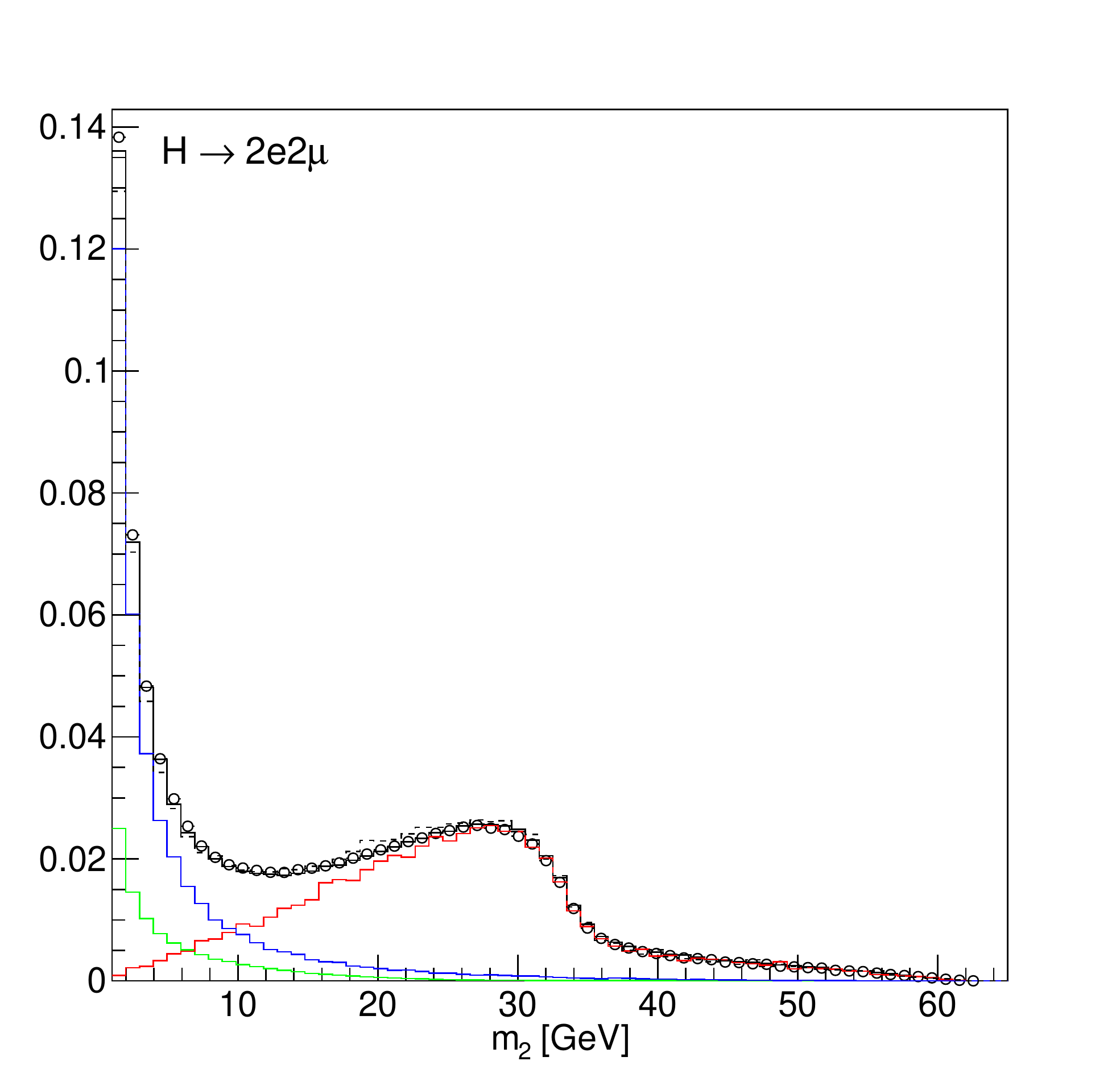}
\includegraphics[width=0.32\textwidth]{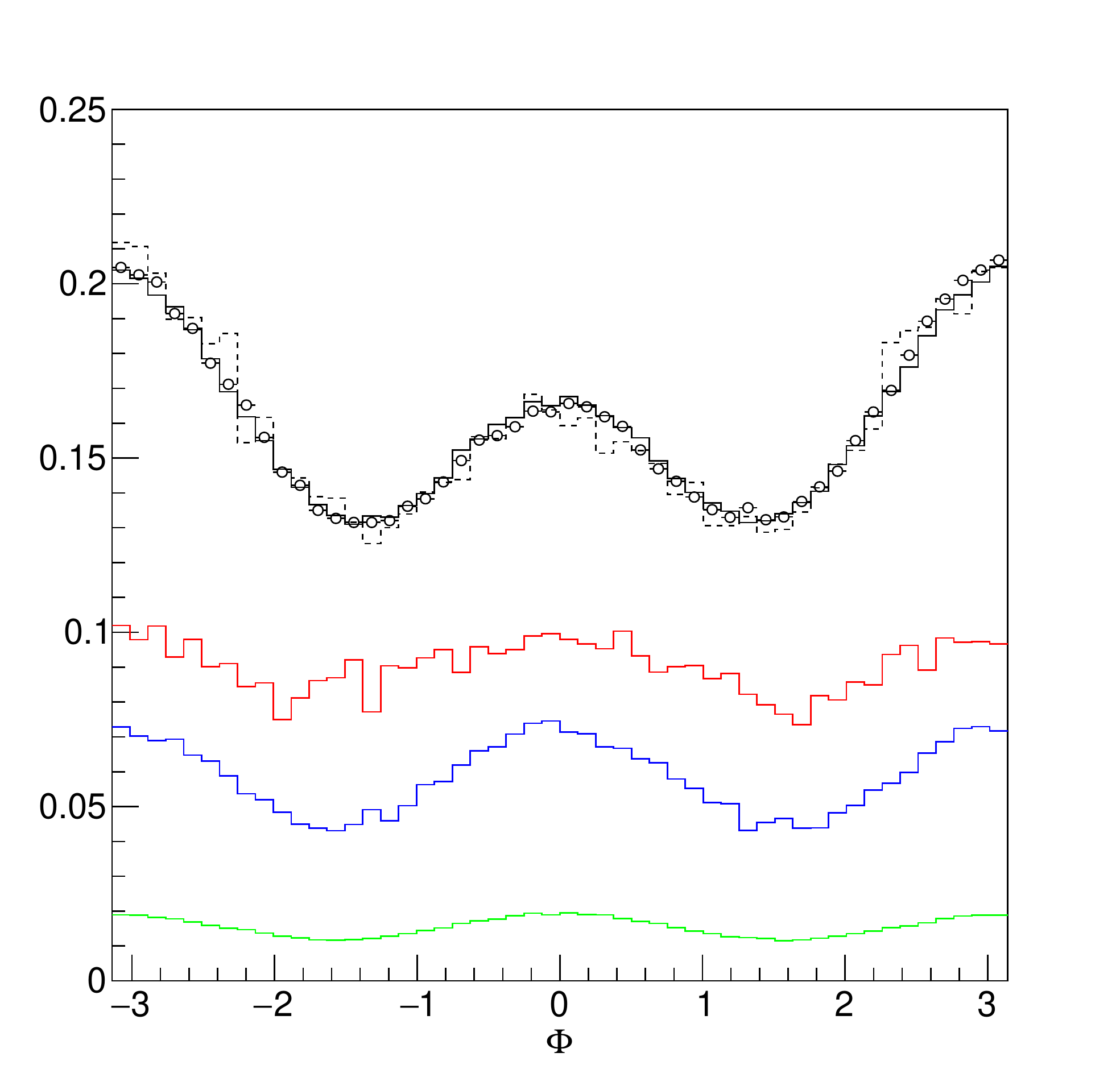}
}
\captionsetup{justification=centerlast}
\caption{
Distribution of the larger (left) and smaller (middle) dilepton invariant mass
in the $H\to4e / 4\mu$ (top) and $H\to2e2\mu$ (bottom) decay. Also shown are
the $\cos\theta_{1,2}$ (top right) and $\Phi$ (bottom right) distributions in the $H\to2e2\mu$ decay. 
Distributions are generated with \textsc{JHUGen} for $C_{HWB}=1$, with the three contributions due to the 
$HZZ$ (red), $HZ\gamma$ (green), and $H\gamma\gamma$ (blue) couplings shown separately. 
The \textsc{JHUGen} distributions are shown without (solid) and with (points) corrections to the
$Zff$ couplings, indicated with $\delta g^{Zff}$.
The comparison to \textsc{SMEFTsim} modeling (dashed) is also shown. 
}
\label{fig:Lexicon-H4l}
\end{figure}

\begin{figure}[tb]
\centering
\centerline{
\includegraphics[width=0.32\textwidth]{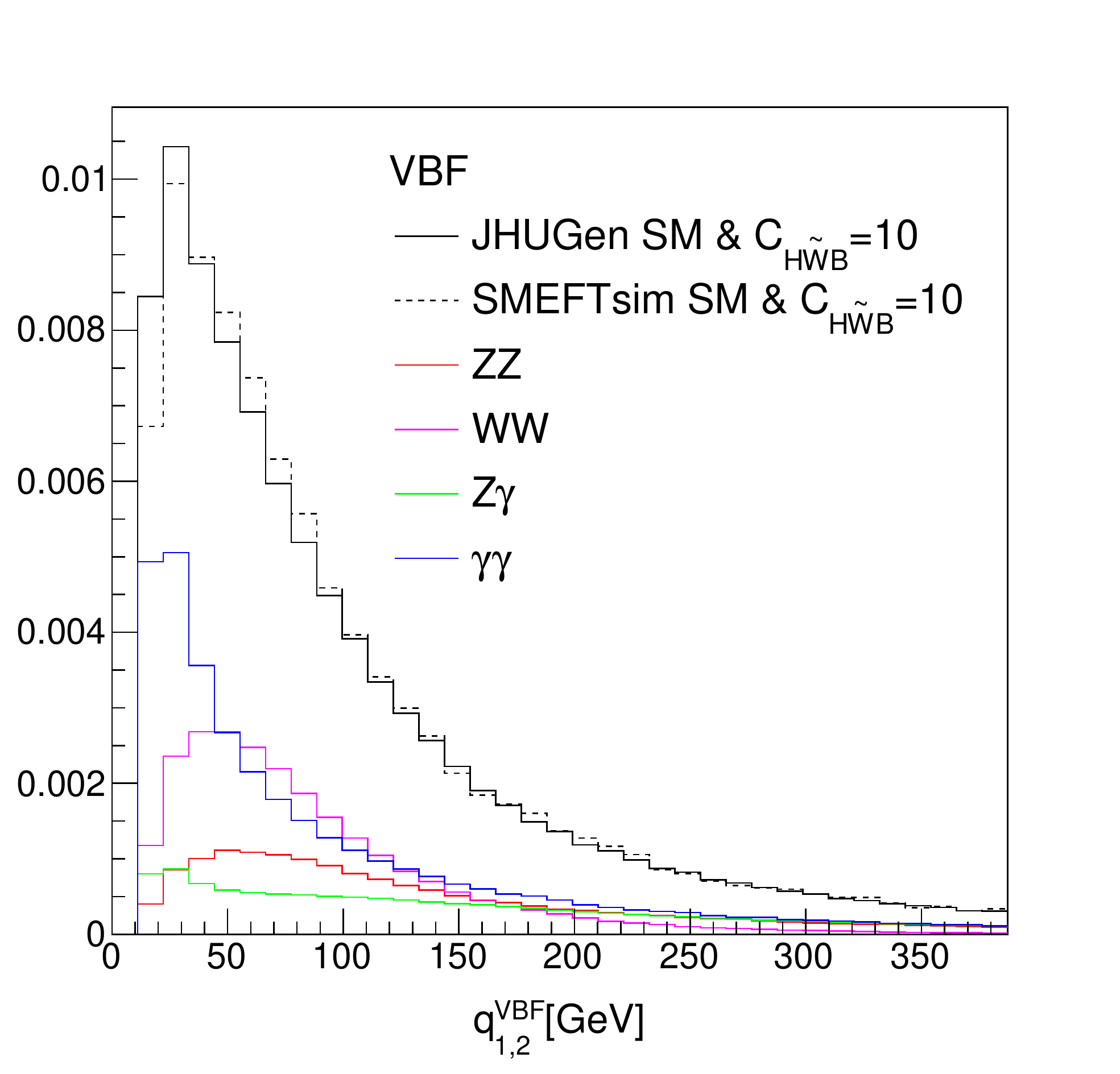}
\includegraphics[width=0.32\textwidth]{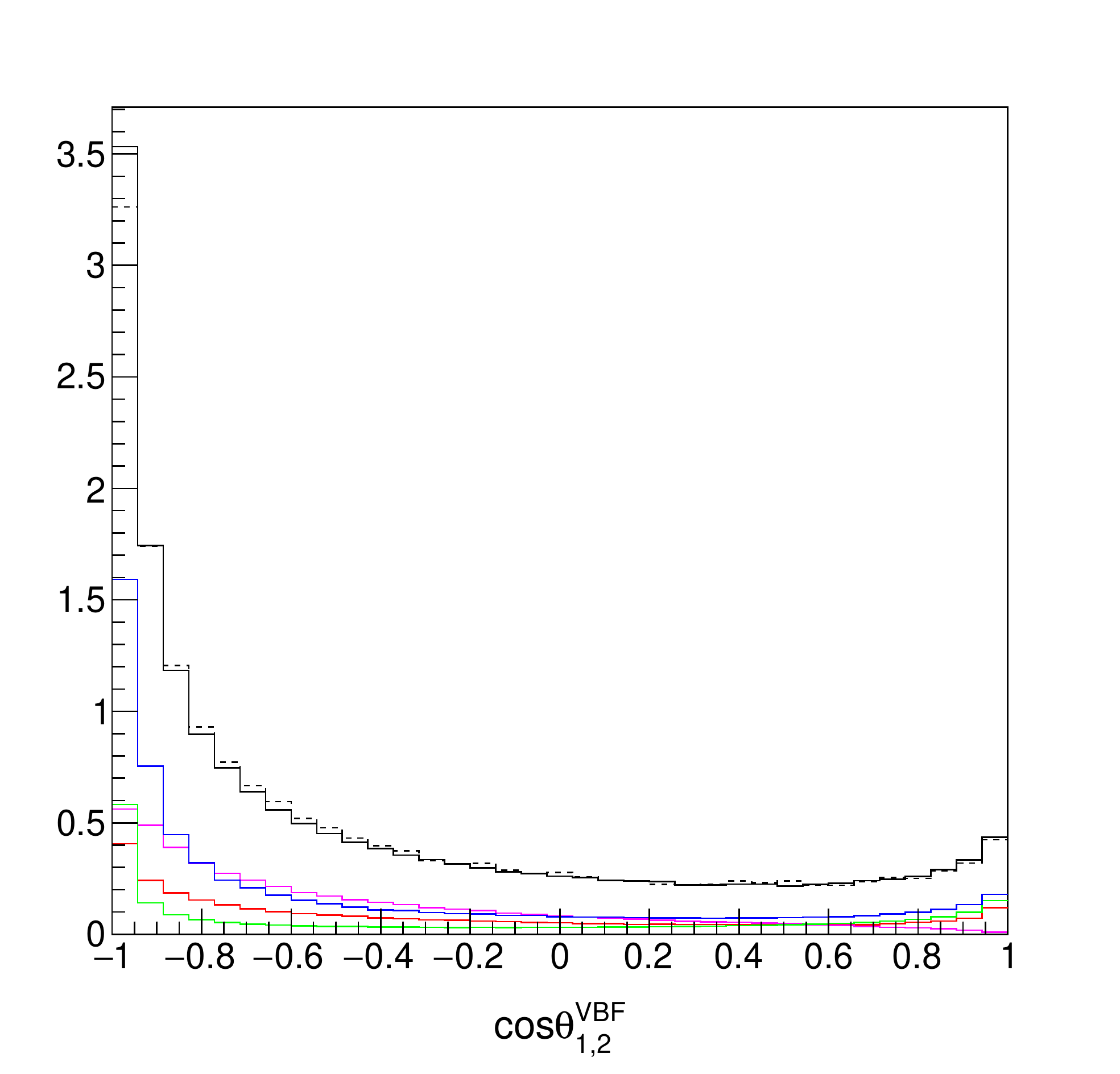}
\includegraphics[width=0.32\textwidth]{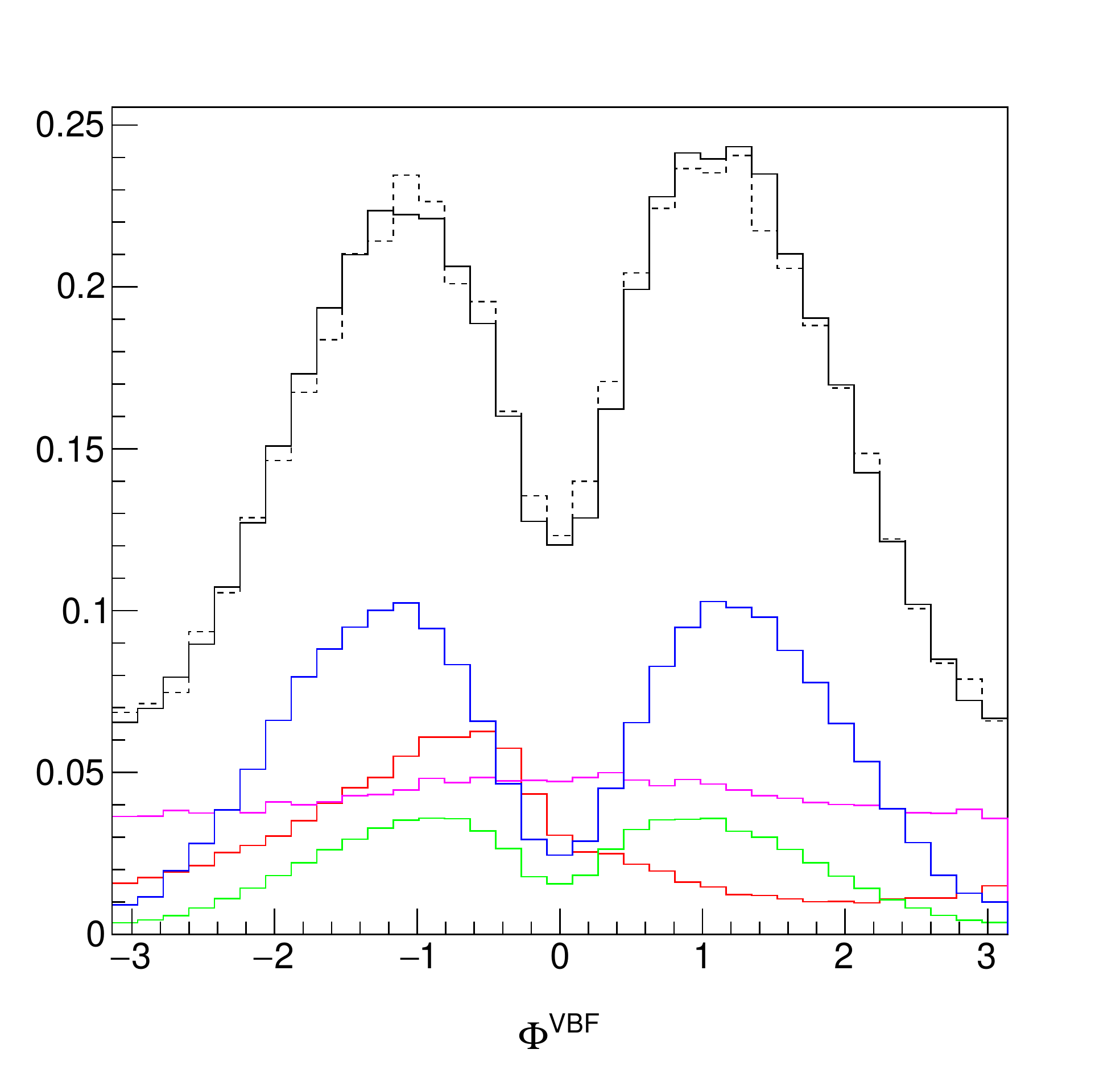}
}
\captionsetup{justification=centerlast}
\caption{
Distribution of $\sqrt{-q_{1,2}^2}$ (left), $\cos\theta_{1,2}^\mathrm{VBF}$ (middle), and $\Phi^\mathrm{VBF}$ (right)
for the intermediate vector boson in the VBF process generated with \textsc{JHUGen} for the 
$C_{\PH\widetilde{W}B}  =10$ with three contributions due to the 
$HZZ$ (red), $HWW$ (magenta), $HZ\gamma$ (green), and $H\gamma\gamma$ (blue) couplings shown separately. 
The comparison to \textsc{SMEFTsim} modeling (dashed) is also shown. 
}
\label{fig:Lexicon-VBF}
\end{figure}

\begin{figure}[tb]
\centering
\centerline{
\includegraphics[width=0.32\textwidth]{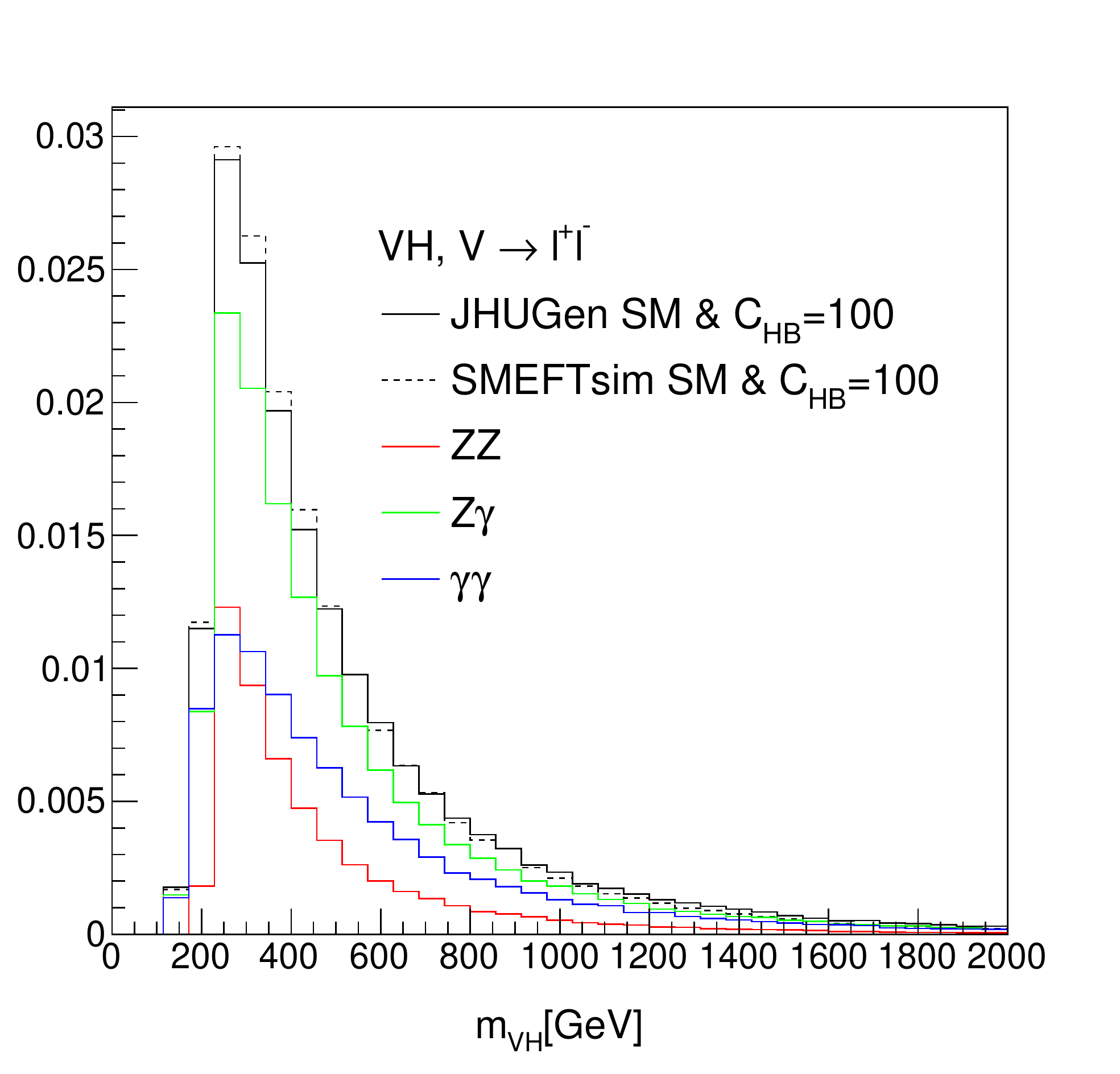}
\includegraphics[width=0.32\textwidth]{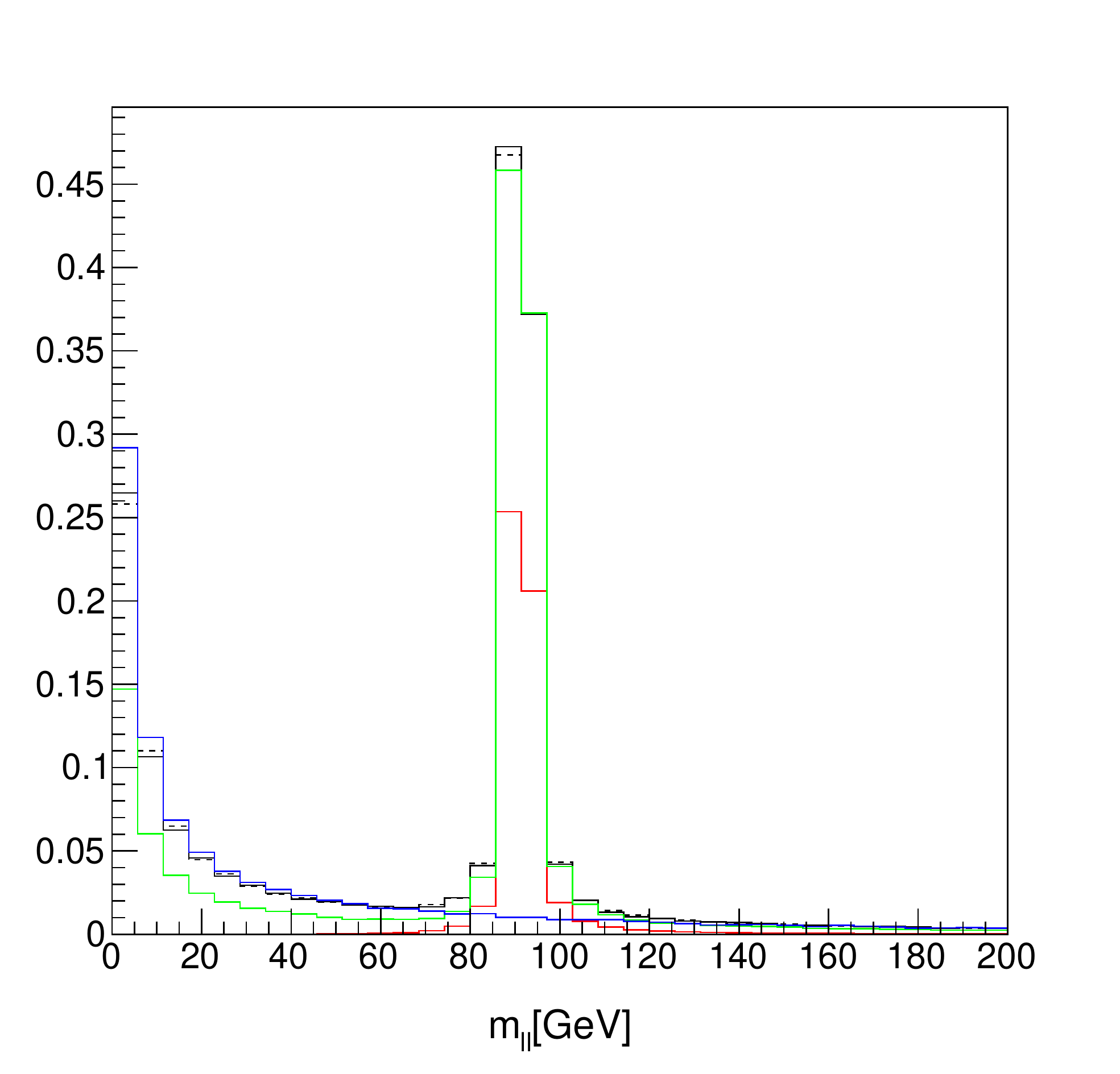}
}
\centerline{
\includegraphics[width=0.32\textwidth]{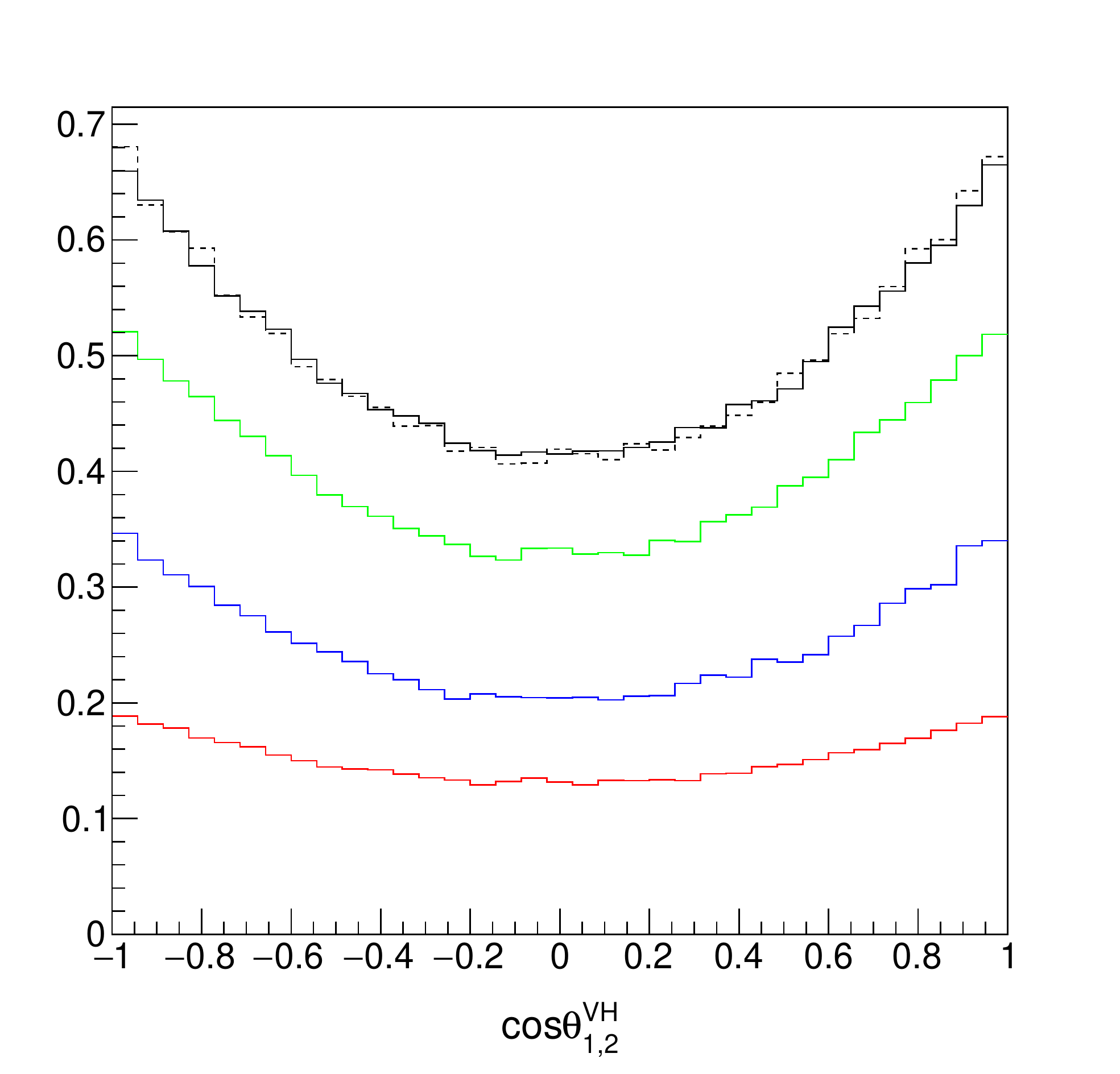}
\includegraphics[width=0.32\textwidth]{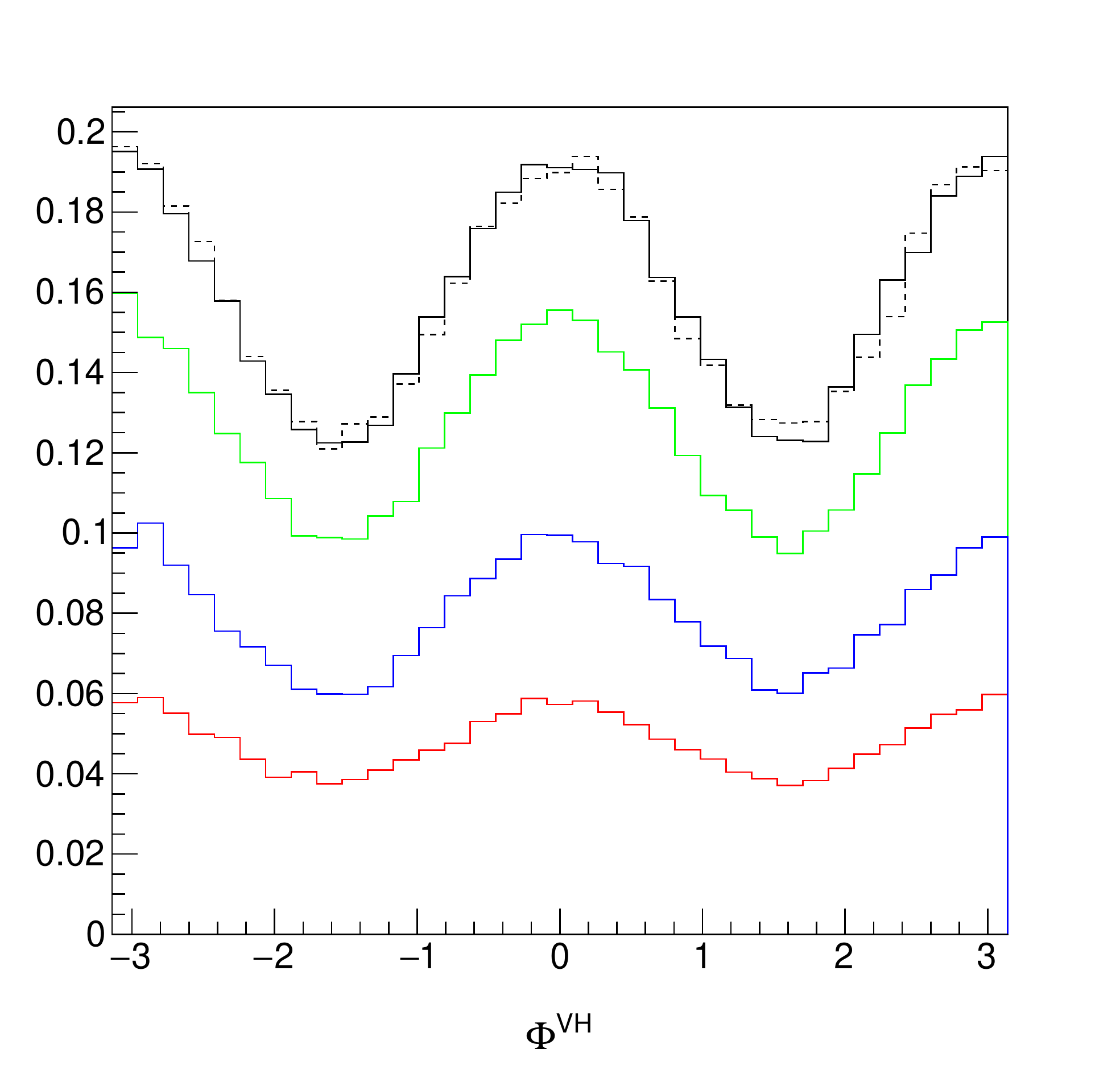}
}
\captionsetup{justification=centerlast}
\caption{
Distribution of the $m_{VH}$ (top left) and $m_{\ell\ell}$ (top right) invariant masses,
$\cos\theta_{1,2}^{VH}$ (bottom left), and $\Phi^{VH}$ (bottom right) 
in the $q\bar{q}\to VH\to \ell\ell H$ process
generated with \textsc{JHUGen} for the $C_{HB}=100$ with three contributions due to the 
$HZZ$ (red), $HZ\gamma$ (green), and $H\gamma\gamma$ (blue) couplings shown separately. 
The comparison to \textsc{SMEFTsim} modeling (dashed) is also shown. 
}
\label{fig:Lexicon-VH}
\end{figure}

\begin{figure}[h]
\centering
\centerline{
\includegraphics[width=0.32\textwidth]{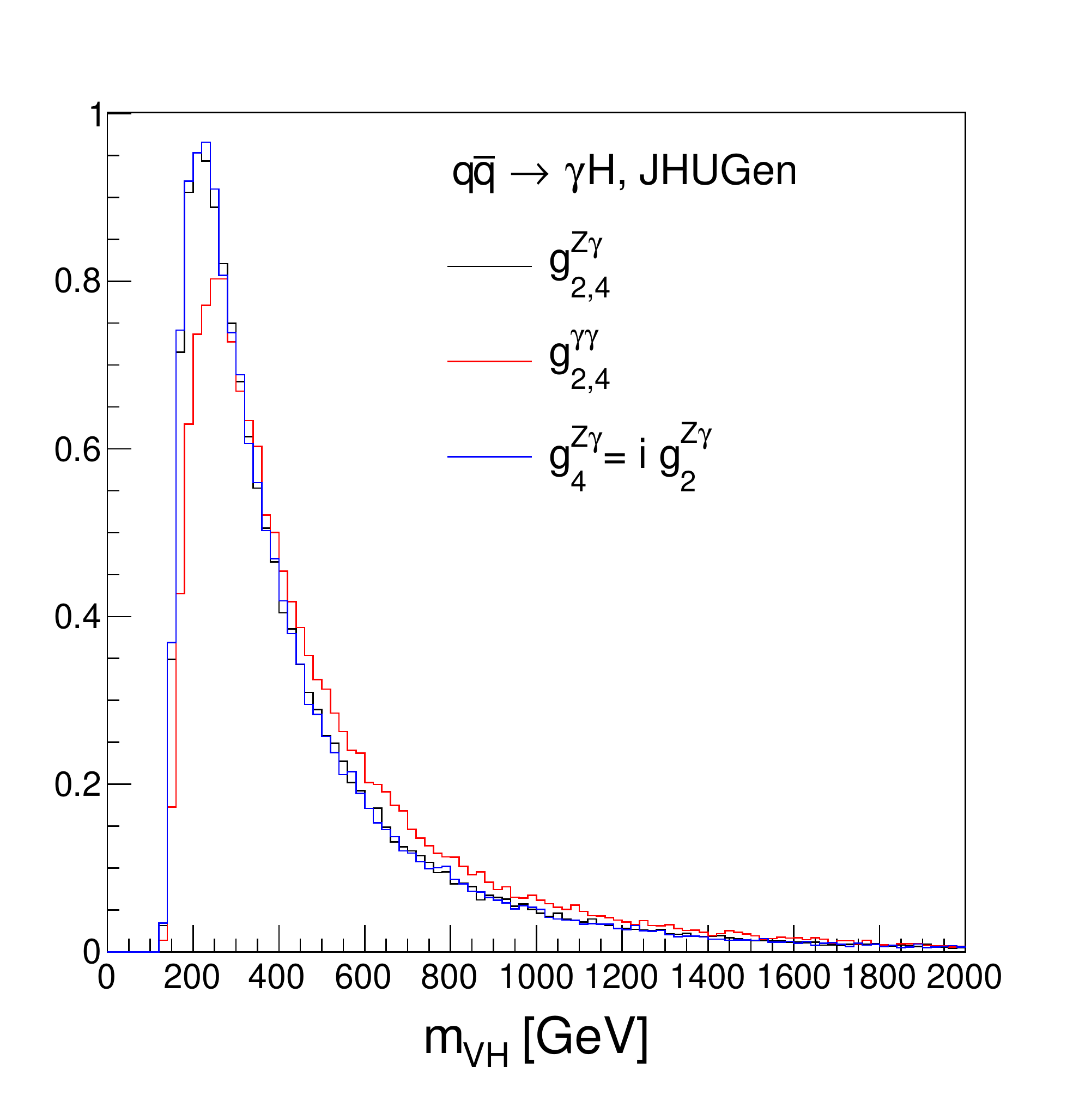}
\includegraphics[width=0.32\textwidth]{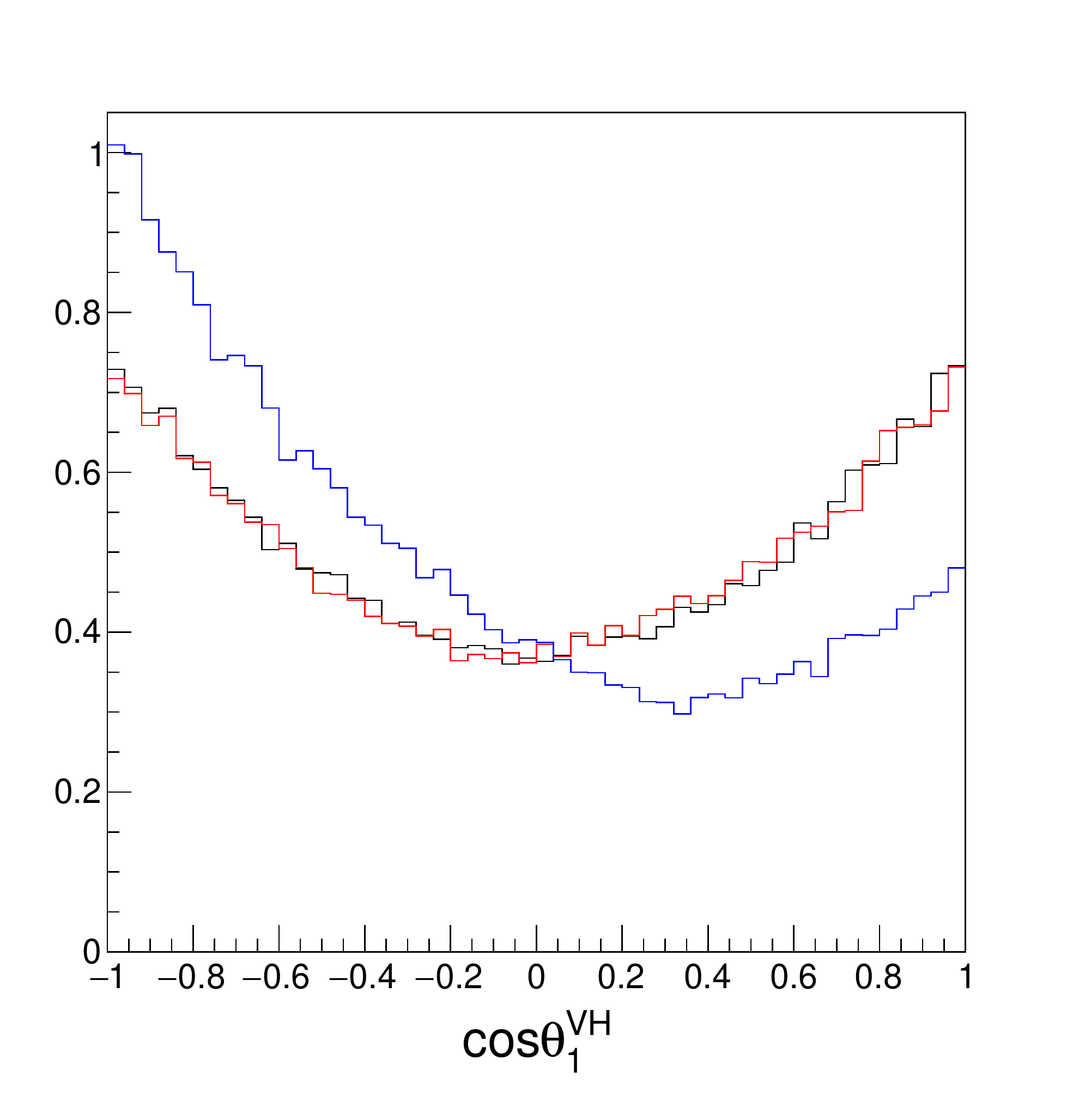}
}
\captionsetup{justification=centerlast}
\caption{
Distribution of $m_{\gamma H}$ (left) and $\cos\theta_{1}^{VH}$ (right) 
in the $q\bar{q}\to\gamma H$ process generated with \textsc{JHUGen} 
with the $g_2^{Z\gamma}$,  $g_2^{\gamma\gamma}$, $g_4^{Z\gamma}$, or $g_4^{\gamma\gamma}$ 
anomalous couplings. Distributions for individual couplings are shown in black ($HZ\gamma$) and 
red ($H\gamma\gamma$), and the mixture of $g_2^{Z\gamma}$ and $g_4^{Z\gamma}$ couplings
with the complex ratio is shown in blue. 
}
\label{fig:Lexicon-gammaH}
\end{figure}

Another approach to study anomalous \Hboson couplings involving photons is analysis of the $\gamma H$
process, which distinguishing feature is a high-momentum \onshell\ photon associated with the \Hboson. 
In the LO topology, where the $\gamma H$ system has no transverse boost, 
the transverse momentum of either photon $\gamma$ or the \Hboson is a dependent observable and the three primary 
measurements are the rapidity $y$ and the invariant mass $m_{\gamma H}$ of the $m_{\gamma H}$ system, 
and the angle $\theta_1$ formed by the outgoing photon with respect to the direction of incoming quark 
in the $\gamma H$ rest frame.
This angle is also defined in Fig.~\ref{fig:kinematics}, where $V_2=\gamma$ and which does not have 
a subsequent decay. While it is not possible to distinguish the incoming quark and antiquark on event-by-event 
basis, on average the boost direction of the $\gamma H$ provides the preferred direction of the quark, 
and we use this to define $\theta_1$. However, determination of the $\cos\theta_1$ sign becomes
important only in the special case of the forward-backward asymmetry discussed below. 
The ability to determine the $\cos\theta_1$ sign is a function of $y$ and has been discussed 
earlier~\cite{Chatrchyan:2011ya,Gritsan:2020pib}.

In Fig.~\ref{fig:Lexicon-gammaH}, the $m_{\gamma H}$ and $\cos\theta_{1}^{VH}$ distributions
are shown for the $g_2^{Z\gamma}$,  $g_2^{\gamma\gamma}$, $g_4^{Z\gamma}$, or $g_4^{\gamma\gamma}$ 
anomalous couplings, and for the mixture of $g_2^{Z\gamma}$ and $g_4^{Z\gamma}$ 
contributions with a complex phase of $g_4^{Z\gamma}$.
The $m_{\gamma H}$ distributions differ somewhat between the $HZ\gamma$ and $H\gamma\gamma$ couplings, 
due to the difference between the intermediate $Z^*$ and $\gamma^*$.
The $\cos\theta_{1}^{VH}$ distributions follow the $\left(1+\cos^2\theta_1\right)$ expectation for all real couplings.
This expectation can be traced back to Eq.\,(A2) in Ref.~\cite{Anderson:2013afp}, where $A_{00}=0$, which must be
averaged over $\Phi$ and $\cos\theta_2$. The situation becomes similar to the 
angular distribution in the $H\to Z\gamma$ decay, described with the same angular parameterization, 
as shown in Fig.~15 of Ref.\,\cite{Anderson:2013afp}, and where the forward-backward asymmetry 
may be generated with the mixture of $CP$-odd and $CP$-even couplings and in the presence of a complex phase. 

The size of the forward-backward asymmetry is proportional to the $A_f$ parameter defined in Ref.\,\cite{Anderson:2013afp}
for $Zff$ couplings, which is 0.15 for the lepton couplings, but is as large as 0.67 and 0.94 for the
up and down type quarks. Therefore, despite the sizable dilution in the measurement of the 
$\cos\theta_1$ sign, the forward-backward asymmetry is strongly pronounced in Fig.~\ref{fig:Lexicon-gammaH}
and could be measured in experiment once ${\gamma H}$ production is observed. 
The $A_f$ parameter is zero for the photon couplings $\gamma ff$, and such an effect is not possible 
in the mixture of couplings involving $g_2^{\gamma\gamma}$ and $g_4^{\gamma\gamma}$. 
Since non-trivial forward-backward asymmetry appears only in the special case of complex couplings,
we do not consider this asymmetry further in this work, but we point out that such a study is in principle possible.


\section{Parameterization of cross sections}
\label{sect:eft-xs}

In this Section, we discuss the relationship between the coupling constants and the cross section of a process involving the \Hboson. 
In Ref.~\cite{Gritsan:2020pib}, we calculated the scaling factors for the partial decay widths in the nine dominant \Hboson decay modes
as a function of anomalous couplings $a_i$, including the $H\to {\rm gg}$, $\gamma\gamma$, and $Z\gamma$ decays, by resolving the loop 
contributions. However, we omitted point-like contributions such as $g_{2,4}^{\gamma\gamma}$ and $g_{2,4}^{Z\gamma}$ due to their relatively 
lower importance in the VBF, $VH$, and $H\to 4f$ processes. Such couplings could be generated by a heavy quark ${\cal Q}$ with 
mass $m_{\cal Q}\gg M_H$. We assume that its couplings to the \Hboson are $\kappa_{\cal Q}$ and $\tilde\kappa_{\cal Q}$, 
the number of colors $N_c$, the electric charge $Q$, and the weak isospin projection $T^{3\mathrm{L}}$. This special model
allows us to derive the point-like interactions and relate those to the partial decay widths. While derivation applies to this special case, 
the final expression in terms of the $g_{2,4}^{\rm gg}$, $g_{2,4}^{\gamma\gamma}$, and $g_{2,4}^{Z\gamma}$ couplings becomes 
generic and remains valid for any new physics in the loop, generated by any combination of heavy fermions or bosons. 
Therefore, the resulting expressions are applicable to the general treatment of these loops in the EFT approach. 

First, we recall that in the narrow-width approximation for \onshell\ \Hboson production and decay, the cross section 
can be expressed as
\begin{eqnarray}
\sigma(i\to H\to f)\propto
\frac{\left(\sum  \alpha_{jk}^{(i)}a_ja_k\right)\left(\sum \alpha_{lm}^{(f)}a_la_m\right)}{\Gamma_{\rm tot}} \,,
\label{eq:diff-cross-section2}
\end{eqnarray}
where the total width $\Gamma_{\rm tot}= \Gamma_{\rm known} + \Gamma_{\rm other}$
representing decays to known particles and other unknown final states, either invisible or undetected in experiment. 
In the following we will focus on decay to the known SM particles which can be expressed as a sum of all partial decay widths as
\begin{eqnarray}
{\Gamma_{\rm known}}
  = {\Gamma_{\rm tot}^{\rm SM}} \times
\sum_{f} \left( \frac{\Gamma_f^{\rm SM}}{\Gamma_{\rm tot}^{\rm SM}} \times \frac{\Gamma_f }{ \Gamma_f^{\rm SM}} \right)
= \sum_{f}\Gamma_f^{\rm SM} R_f
\,,
\label{eq:width}
\end{eqnarray}
where $R_f$ is the scaling factor as function of the coupling constants $a_i$, 
and $\Gamma_f^{\rm SM}$ is the SM value of the partial decay width in the final state $f$.

In the following, we rely on \textsc{JHUGen} framework implementation, discussed in Section~\ref{sect:eft-basis}
and Ref.~\cite{Gritsan:2020pib}, to derive the loop contributions of the SM particles and the heavy quark ${\cal Q}$
to the scaling factor $R_{\rm gg}$, for both $CP$-even and $CP$-odd couplings. 
The $CP$-even coupling contributions of the quarks and $W$ boson to $R_{\gamma\gamma}$ and $R_{\Z\gamma}$ 
are derived with \textsc{HDECAY}~\cite{Djouadi:2018xqq}. 
The $CP$-odd contributions to $R_{\gamma\gamma}$ are calculated with the \textsc{JHUGen} framework 
in a manner analogous to $R_{\rm gg}$. 
The $CP$-odd contributions to $R_{Z\gamma}$ are calculated using \textsc{cHDECAY}~\cite{Fontes:2017zfn}.

The ratio of the decay width to the SM expectation in the $H\to{\rm gg}$ process~\cite{Gritsan:2020pib} is found to be
\begin{eqnarray}
\label{eq:ratio-4}
R_{\rm gg} = &&
1.1068\, \kappa_t^2 + 0.0082\, \kappa_b^2 - 0.1150\, \kappa_t\kappa_b
	+ 2.5717\, \tilde\kappa_t^2 + 0.0091\, \tilde\kappa_b^2 - 0.1982\, \tilde\kappa_t\tilde\kappa_b
	\nonumber  \\
    &&	+\, 1.0298\, (N_c/3)^2\kappa_{\cal Q}^{2} + 2.1357\, (N_c/3)\kappa_{\cal Q} \kappa_t - 0.1109\, (N_c/3)\kappa_{\cal Q} \kappa_b 
	\nonumber \\
    &&	+ 2.3170\, (N_c/3)^2\tilde\kappa_{\cal Q}^{2} + 4.8821\, (N_c/3)\tilde\kappa_{\cal Q} \tilde\kappa_t - 0.1880\, (N_c/3)\tilde\kappa_{\cal Q} \tilde\kappa_b
	\,.
\end{eqnarray}
The $\kappa_{\cal Q}$ and $\tilde\kappa_{\cal Q}$ couplings are connected to the $g_2^{\rm gg}$ and $g_4^{\rm gg}$ point-like 
interactions introduced in Eq.~(\ref{eq:HVV}) through
\begin{eqnarray}
\label{eq:g2gg}
g_{2}^{\rm gg, \cal Q} =-\alpha_s N_c\kappa_{\cal Q}/(18\pi)
\,,~~~~~~~~~
g_{4}^{\rm gg, \cal Q} =-\alpha_s N_c\tilde\kappa_{\cal Q}/(12\pi)
\,.
\end{eqnarray}
One can rewrite Eq.~(\ref{eq:ratio-4}) in terms of the $g_2^{\rm gg}$ and $g_4^{\rm gg}$ couplings in place 
of $N_c\kappa_{\cal Q}$ and $N_c\tilde\kappa_{\cal Q}$ by substituting Eq.~(\ref{eq:g2gg}). 
Even though Eq.~(\ref{eq:g2gg}) is derived in the special case of a heavy quark, the resulting expression of 
$R_{\rm gg}$ as a function of $g_2^{\rm gg}$ and $g_4^{\rm gg}$ and other terms is
valid for any heavy particles in the loop that generate these point-like interactions. 

The latter observation allows us to obtain the value of the effective $g_{2}^{\rm gg}$ coupling which leads
to the SM cross section in the gluons fusion process. By setting all couplings, other than $g_{2}^{\rm gg}$,
to zero and $R_{\rm gg} = 1$ in Eq.~(\ref{eq:ratio-4}), we obtain
\begin{eqnarray}
\label{eq:g2gg-effective}
g_{2}^{\rm gg, SM} =-0.00621
\,.
\end{eqnarray}
The $g_{2}^{\rm gg, SM}$ value differs by only 1.5\% from the value that one would obtain 
in the heavy top mass limit by setting $\kappa_{\cal Q}=1$ and $N_c=3$ in Eq.~(\ref{eq:g2gg}),
and the sign follows the prediction in this limit. 

An approximate way to express Eq.~(\ref{eq:ratio-4}) with the point-like interactions only 
in the case of SM couplings of fermions $\kappa_t=\kappa_b=1$ and $\tilde\kappa_t=\tilde\kappa_b=0$
would be to substitute the top and bottom quark contributions with an effective coupling $g_{2}^{\rm gg, SM}$
from Eq.~(\ref{eq:g2gg-effective}), substitute $\kappa_{\cal Q}$ and $\tilde\kappa_{\cal Q}$
for $g_2^{\rm gg}$ and $g_4^{\rm gg}$, and obtain
\begin{eqnarray}
\label{eq:ratio-gg-2}
R_{\rm gg} \simeq \frac{1}{\left(g_{2}^{\rm gg, SM}\right)^2}
\left[ \left(g_{2}^{\rm gg, SM} + g_{2}^{\rm gg}\right)^2 +  \left(g_{4}^{\rm gg}\right)^2 \right]
\,.
\end{eqnarray}

For the $H\to\gamma\gamma$  final states, 
we include the $W$ boson in addition to the top, bottom, and heavy ${\cal Q}$ quarks in the loop
and obtain\footnote{
Due to updated EW parameters, there is a small change in the numerical values of coefficients in Eqs.~(\ref{eq:ratio-gammagamma})
and (\ref{eq:ratio-zgamma}) that are in common with Ref.~\cite{Gritsan:2020pib}.
}
\begin{eqnarray}
\label{eq:ratio-gammagamma}
R_{\gamma\gamma} = && 
 1.60932\left(\frac{g_1^{WW}}{2}\right)^2
 -0.69064\left(\frac{g_1^{WW}}{2}\right)\kappa_t
  +0.00912\, \left(\frac{g_1^{WW}}{2}\right)\kappa_b 
  -0.49725\, \left(\frac{g_1^{WW}}{2}\right)(N_c \, Q^2  \, \kappa_{\cal Q})
       \nonumber 
         \\
&&     +0.07404\,  \kappa_t^2 
         +0.00002\,  \kappa_b^2  
         -0.00186\, \kappa_t\kappa_b
         \nonumber 
         \\
&&  	+0.03841\,\left(N_c \, Q^2  \, \kappa_{\cal Q}\right)^2
        +0.10666\, \kappa_t\left(N_c \, Q^2  \, \kappa_{\cal Q}\right) 
	-0.00136\, \kappa_b\left(N_c \, Q^2  \, \kappa_{\cal Q}\right) \, 
        \nonumber 
          \\
&&      +0.20533\,  \tilde\kappa_t^2 
          +0.00006\,  \tilde\kappa_b^2 
          -0.00300\, \tilde\kappa_t\tilde\kappa_b
         \nonumber 
          \\
&& 	 + 0.10252\, \left(N_c \, Q^2  \, \tilde\kappa_Q\right)^2
         +0.29018\, \tilde\kappa_t\left(N_c \, Q^2  \, \tilde\kappa_{\cal Q}\right)
         -0.00202\, \tilde\kappa_b\left(N_c \, Q^2  \, \tilde\kappa_{\cal Q}\right)
       \,.
\end{eqnarray}

For the contribution of a heavy quark in the loop we find
\begin{eqnarray} \label{eq:g2AA}
	g_2^{\gamma\gamma,\cal Q} = -\frac{\alpha}{3\pi} N_c \, Q^2  \, \kappa_{\cal Q}
	\,,~~~~~~~~~
	g_4^{\gamma\gamma,\cal Q} = -\frac{\alpha}{2\pi} N_c \, Q^2  \, \tilde\kappa_{\cal Q}.
\end{eqnarray}

Following the idea described above for $R_{\rm gg}$,
one can rewrite Eq.~(\ref{eq:ratio-gammagamma}) in terms of the $g_2^{\gamma\gamma}$ and $g_4^{\gamma\gamma}$ couplings in place 
of $N_cQ^2\kappa_{\cal Q}$ and $N_cQ^2\tilde\kappa_{\cal Q}$ by substituting Eq.~(\ref{eq:g2AA}). 
The final expression of $R_{\gamma\gamma}$ as a function of $g_2^{\gamma\gamma}$ and $g_4^{\gamma\gamma}$ 
and other terms is again valid for any heavy particles in the loop, fermions or bosons, that generate these point-like interactions. 
By setting all couplings other than $g_{2}^{\gamma\gamma}$ to zero and $R_{\gamma\gamma} = 1$ in Eq.~(\ref{eq:ratio-gammagamma}), 
we obtain the effective coupling which leads to the SM cross section 
\begin{eqnarray}
\label{eq:g2AA-effective}
g_{2}^{\gamma\gamma,\rm SM} = 0.00423
\,.
\end{eqnarray}

The $g_{2}^{\gamma\gamma,\rm SM}$ value differs slightly from $0.00400$ obtained from the general expression 
of the SM loops derived from Refs.~\cite{Low:2012rj,Contino:2014aaa} and shown in Eq.~(\ref{eq:g2AAanalytic}). 
The difference could be explained by the higher-order effects incorporated in Eq.~(\ref{eq:ratio-gammagamma})
and the fact that in our approach we match the SM rate $R_{\gamma\gamma} = 1$.
The sign in Eq.~(\ref {eq:g2AA-effective}) follows Eq.~(\ref{eq:g2AAanalytic}).
\begin{eqnarray} \label{eq:g2AAanalytic}
	g_2^{\gamma\gamma} &=& \left( -\frac{\alpha}{4\pi} \right) \left[ \left(\frac{g_1^{WW}}{2}\right) \times A_1^{\gamma\gamma}\!\left(\tau_W\right) 
	                                                              + \kappa_t N_c Q_t^2 \times A_{1/2}^{\gamma\gamma}\!\left(\tau_t\right) \right] \nonumber \\
	                   &=& 0.00516 \left(\frac{g_1^{WW}}{2}\right) - 0.00116 \kappa_t,
\end{eqnarray}
where the one-loop functions are given by 
\begin{eqnarray} \label{eq:AloopFunctions1}
	A_1^{\gamma\gamma}\!\left(\tau_W\right) =  \begin{cases}  -8.32  &\quad\text{for}\quad  \tau_W=M_W^2/M_H^2 \\ -7   &\quad\text{for}\quad  \tau_W \rightarrow\infty \end{cases}
\end{eqnarray}
and
\begin{eqnarray} \label{eq:AloopFunctions2}
	A_{1/2}^{\gamma\gamma}\!\left(\tau_t\right) =  \begin{cases}  +1.38  &\quad\text{for}\quad  \tau_t=m_t^2/M_H^2 \\ +4/3   &\quad\text{for}\quad  \tau_t \rightarrow\infty \end{cases}.
\end{eqnarray}

An approximate way to express Eq.~(\ref{eq:ratio-gammagamma}) with point-like interactions only would be
to follow the idea used to create Eq.~(\ref{eq:ratio-gg-2}) and
substitute the SM couplings with $g_{2}^{\gamma\gamma, \rm SM}$
from Eq.~(\ref{eq:g2AA-effective}), substitute $\kappa_{\cal Q}$ and $\tilde\kappa_{\cal Q}$
for $g_2^{\gamma\gamma}$ and $g_4^{\gamma\gamma}$, and obtain
\begin{eqnarray}
\label{eq:ratio-gammagamma-2}
R_{\gamma\gamma} \simeq
\frac{1}{\left(g_{2}^{\gamma\gamma, \rm SM}\right)^2}
\left[
 \left(g_{2}^{\gamma\gamma, \rm SM} + g_{2}^{\gamma\gamma}\right)^2
+ \left(g_{4}^{\gamma\gamma}\right)^2
\right]
\,.
\end{eqnarray}

For the $H\to Z\gamma$  final states, for the coupling of the heavy ${\cal Q}$ quark to the $Z$ boson,
we introduce the following parameter
\begin{eqnarray} 
\label{eq:RQ}
	{\cal R}_{\cal Q} = Q\, \frac{T^{3\mathrm{L}}_{\cal Q}-2s_w^2 \, Q}{s_w c_w}  \,,
\end{eqnarray}
which corresponds to the following values for the SM parameters of the 
top ($T^{3\mathrm{L}}_{t}=+1/2$, $Q_{t}=+2/3$) and bottom ($T^{3\mathrm{L}}_{b}=-1/2$, $Q_{b}=-1/3$) quarks
\begin{eqnarray} 
\label{eq:RQtop}
	{\cal R}_{t} =   0.3032 \,,~~~~~
	{\cal R}_{b} =   0.2735 \,.
\end{eqnarray}

We obtain
\begin{eqnarray}
\label{eq:ratio-zgamma}
R_{Z\gamma} = && 
   1.11965\,  \left(\frac{g_1^{WW}}{2}\right)^2 
 -  0.12652\,   \left(\frac{g_1^{WW}}{2}\right)\kappa_t
+ 0.00348\, \left(\frac{g_1^{WW}}{2}\right)\kappa_b 
- 0.13021\, \left(\frac{g_1^{WW}}{2}\right) (N_c \, {\cal R}_{\cal Q}\kappa_{\cal Q})
          \nonumber 
       \\
&&    + 0.00357\, \kappa_t^2
        + 0.000003\, \kappa_b^2 
        - 0.00018\, \kappa_t\kappa_b
        \nonumber 
        \\
&&    + 0.00377\, \left(N_c \, {\cal R}_{\cal Q}\kappa_{\cal Q}\right)^2
	+ 0.00734\, \kappa_t\left(N_c \, {\cal R}_{\cal Q}\kappa_{\cal Q}\right) 
        - 0.00019\, \kappa_b \left(N_c \, {\cal R}_{\cal Q}\kappa_{\cal Q}\right)
        \nonumber 
        \\
&&      + 0.00849\, \tilde\kappa_t^2
           + 0.000004\, \tilde\kappa_b^2
           -0.00025\, \tilde\kappa_t\tilde\kappa_b 
       \nonumber 
        \\
&&        + 0.00883\, \left(N_c \, {\cal R}_{\cal Q}\tilde\kappa_{\cal Q}\right)^2
            + 0.01723\, \tilde\kappa_t\left(N_c \, {\cal R}_{\cal Q}\tilde\kappa_{\cal Q}\right) 
            - 0.00024\, \tilde\kappa_b\left(N_c \, {\cal R}_{\cal Q}\tilde\kappa_{\cal Q}\right)
        \,.
\end{eqnarray}

For the contribution of a heavy forth generation quarks in the loop we find
\begin{eqnarray} \label{eq:g2ZA}
	g_2^{Z\gamma,\cal Q} = - \frac{\alpha}{6\pi} N_c \, {\cal R}_{\cal Q} \, \kappa_{\cal Q}
	\,,~~~~~~~~~
	g_4^{Z\gamma,\cal Q} = - \frac{\alpha}{4\pi} N_c \, {\cal R}_{\cal Q}  \, \tilde\kappa_{\cal Q} \,.
\end{eqnarray}

We note that the effective value of $g_{2}^{Z\gamma}$ for a heavy quark ${\cal Q}$ 
which reproduces the SM partial width, is
\begin{eqnarray}
\label{eq:g2ZA-effective}
g_{2}^{Z\gamma, \rm SM} =  0.00675 .
\end{eqnarray}

The $g_{2}^{Z\gamma,\rm SM}$ value differs slightly from $0.00724$ obtained from the general expression 
of the SM loops derived from Refs.~\cite{Low:2012rj,Contino:2014aaa}\footnote{We thank Ian Low for updating 
the results in Eq.~(7) of Ref.~\cite{Low:2012rj}.}
and shown in Eq.~(\ref{eq:g2ZAanalytic}). 
As before, the difference could be explained by the higher-order effects incorporated in Eq.~(\ref{eq:ratio-zgamma})
and the fact that in our approach we match the SM rate $R_{Z\gamma} = 1$.
The sign in Eq.~(\ref {eq:g2ZA-effective}) follows Eq.~(\ref{eq:g2ZAanalytic}).
\begin{eqnarray} \label{eq:g2ZAanalytic}
	g_2^{Z\gamma} &=& \frac{\alpha}{4\pi}  \left[ \left(\frac{g_1^{WW}}{2}\right) \frac{c_w}{s_w} \times A_1^{Z\gamma}\!\left(\tau_W\right) 
	                                                              + \kappa_t N_c \mathcal{R}_t \times A_{1/2}^{Z\gamma}\!\left(\tau_t\right) \right] \nonumber \\
	              &=& 0.00747 \left(\frac{g_1^{WW}}{2}\right) - 0.00023 \kappa_t,
\end{eqnarray}
where the one-loop functions are given by $A_1^{Z\gamma}\!\left( M_W^2/M_H^2\right) = 6.58$ and $A_{1/2}^{Z\gamma}\!\left( m_t^2/M_H^2 \right)=-0.35$.

An approximate way to express Eq.~(\ref{eq:ratio-zgamma}) with point-like interactions only would be
to substitute the SM contributions with an effective coupling $g_{2}^{Z\gamma, \rm SM}$
from Eq.~(\ref{eq:g2ZA-effective}), substitute $\kappa_{\cal Q}$ and $\tilde\kappa_{\cal Q}$
for $g_2^{Z\gamma}$ and $g_4^{Z\gamma}$, and obtain
\begin{eqnarray}
\label{eq:ratio-zgamma-2}
R_{Z\gamma} \simeq
\frac{1}{\left(g_{2}^{Z\gamma, \rm SM}\right)^2}
\left[
 \left(g_{2}^{Z\gamma, \rm SM} + g_{2}^{Z\gamma}\right)^2
+\left(g_{4}^{Z\gamma}\right)^2
\right]
\,.
\end{eqnarray}

In the above calculation, the  $H\to{\gamma^*\gamma}$ process is not included, 
for which the full loop calculation with anomalous couplings is not available. 
For the $H\to ZZ/Z\gamma^*/\gamma^*\gamma^*\rightarrow$\,four-fermion final state, 
the full one-loop calculation with anomalous couplings is not available either. 
The EW loop corrections under the SM assumption are discussed in Section~\ref{sect:eft-sm}. 
A more careful treatment of the singularities appearing in the presence of anomalous couplings 
in both of the above cases is discussed in Section~\ref{sect:eft-lowq}.
For the leading tree-level contributions, we derived the $R_{ZZ/Z\gamma^{*}/\gamma^{*}\gamma^{*}}$ parameterization
in Ref.~\cite{Gritsan:2020pib}, in which case we set $g_2^{Z\gamma}=g_4^{Z\gamma}=g_2^{\gamma\gamma}=g_4^{\gamma\gamma}=0$
to avoid collinear singularities. In the following, we introduce these four couplings
and avoid singularities in the $\gamma^*\to 2f$ transition with the finite fermion mass threshold $q^2>(2m_f)^2$. 
We set $\Lambda_1^{Z\gamma}=\Lambda_1^{ZZ}=100$\,GeV in Eq.~(\ref{eq:HVV}) 
and rely on the $\kappa_2^{Z\gamma}$ and $\kappa_1^{ZZ}=\kappa_2^{ZZ}$ parameters to express the
scaling factor as\footnote{There is a sign change of the $\kappa_{2}^{Z\gamma}$ coupling when compared
to coefficients in common with Ref.~\cite{Gritsan:2020pib}, because here we use the convention 
$D_\mu = \partial_\mu -\mathrm{i} \frac{e}{2 s_w} \sigma^i W_\mu^i - \mathrm{i} \frac{e}{2 c_w} B_\mu$,
as discussed in Section~\ref{sect:eft_couplings}. 
}
\begin{eqnarray}
\label{eq:ratio-3}
\nonumber
R_{ZZ/Z\gamma^{*}/\gamma^{*}\gamma^{*}} = 
&& \left(\frac{g_{1}^{ZZ}}{2}\right)^{2}
+ 0.17\left( \kappa_{1}^{ZZ}\right)^{2}
+ 0.09\left( g_{2}^{ZZ}\right)^{2}
+ 0.04\left( g_{4}^{ZZ}\right)^{2}
+ 0.10\left( \kappa_{2}^{Z\gamma}\right)^{2}
\\ \nonumber
&&
+ 79.95\left( g_{2}^{Z\gamma}\right)^{2}
+ 75.23\left( g_{4}^{Z\gamma}\right)^{2}
+ 29.00\left( g_{2}^{\gamma\gamma}\right)^{2}
+ 29.47\left( g_{4}^{\gamma\gamma}\right)^{2}
\\ \nonumber
&&
+ 0.81 \frac{g_{1}^{ZZ}}{2} \kappa_{1}^{ZZ}
+ 0.50 \frac{g_{1}^{ZZ}}{2} g_{2}^{ZZ}
+ 0 \times \frac{g_{1}^{ZZ}}{2} g_{4}^{ZZ}
- 0.19 \frac{g_{1}^{ZZ}}{2} \kappa_{2}^{Z\gamma}
\\ \nonumber
&&
- 1.56 \frac{g_{1}^{ZZ}}{2} g_{2}^{Z\gamma}
+ 0 \times \frac{g_{1}^{ZZ}}{2} g_{4}^{Z\gamma}
+ 0.06 \frac{g_{1}^{ZZ}}{2} g_{2}^{\gamma\gamma}
+ 0 \times \frac{g_{1}^{ZZ}}{2} g_{4}^{\gamma\gamma}
\\ \nonumber
&&
+ 0.21 \kappa_{1}^{ZZ} g_{2}^{ZZ}
+ 0 \times \kappa_{1}^{ZZ} g_{4}^{ZZ}
- 0.07 \kappa_{1}^{ZZ} \kappa_{2}^{Z\gamma}
- 0.64 \kappa_{1}^{ZZ} g_{2}^{Z\gamma}
\\ \nonumber
&&
+ 0 \times \kappa_{1}^{ZZ} g_{4}^{Z\gamma}
+ 0.00 \kappa_{1}^{ZZ} g_{2}^{\gamma\gamma}
+ 0 \times \kappa_{1}^{ZZ} g_{4}^{\gamma\gamma}
+ 0 \times g_{2}^{ZZ} g_{4}^{ZZ}
\\ \nonumber
&&
- 0.05 g_{2}^{ZZ} \kappa_{2}^{Z\gamma}
- 0.51 g_{2}^{ZZ} g_{2}^{Z\gamma}
+ 0 \times g_{2}^{ZZ} g_{4}^{Z\gamma}
- 0.02 g_{2}^{ZZ} g_{2}^{\gamma\gamma}
\\ \nonumber
&&
+ 0 \times g_{2}^{ZZ} g_{4}^{\gamma\gamma}
+ 0 \times g_{4}^{ZZ} \kappa_{2}^{Z\gamma}
+ 0 \times g_{4}^{ZZ} g_{2}^{Z\gamma}
+ 0.36 g_{4}^{ZZ} g_{4}^{Z\gamma}
\\ \nonumber
&&
+ 0 \times g_{4}^{ZZ} g_{2}^{\gamma\gamma}
-0.57 g_{4}^{ZZ} g_{4}^{\gamma\gamma}
+ 1.80 \kappa_{2}^{Z\gamma} g_{2}^{Z\gamma}
+ 0 \times \kappa_{2}^{Z\gamma} g_{4}^{Z\gamma}
\\ \nonumber
&&
- 0.05 \kappa_{2}^{Z\gamma} g_{2}^{\gamma\gamma}
+ 0 \times \kappa_{2}^{Z\gamma} g_{4}^{\gamma\gamma}
+ 0 \times g_{2}^{Z\gamma} g_{4}^{Z\gamma}
- 1.84 g_{2}^{Z\gamma} g_{2}^{\gamma\gamma}
\\ 
&&
+ 0 \times g_{2}^{Z\gamma} g_{4}^{\gamma\gamma}
+ 0 \times g_{4}^{Z\gamma} g_{2}^{\gamma\gamma}
- 2.09 g_{4}^{Z\gamma} g_{4}^{\gamma\gamma}
+ 0 \times g_{2}^{\gamma\gamma} g_{4}^{\gamma\gamma}
\end{eqnarray}

Equation~(\ref{eq:ratio-3}) covers all final states with $Z/\gamma^*\to q\bar{q}$ and $\ell^+\ell^-$ with quarks and charged leptons,
while neutrinos are included with $Z\to\nu\bar\nu$. The treatment of $q\bar{q}$ hadronization with the low-mass resonances is 
not included here, and is discussed in more detail in Section~\ref{sect:eft-lowq}.
The interference between the $CP$-odd and $CP$-even contribution integrates out to zero,
as reflected in the zero terms in Eq.~(\ref{eq:ratio-3}).


Let us conclude this Section by discussing the cross section of the $\qqbar\to\gamma H$ process as a function 
of the anomalous couplings summarized in Table~\ref{tab:warsaw-gammaH}. 
Detecting or setting limits on this process will be of interest for constraining the following couplings, 
as discussed in Section~\ref{sect:eft-basis}:
\begin{eqnarray}
&&
\frac{\sigma(\qqbar\to\gamma H)}{\sigma_{\rm ref}^{\gamma\PH}} = 
  \left(g_2^{Z\gamma}\right)^2
+ \left(g_4^{Z\gamma}\right)^2
+0.553 \left(g_2^{\gamma\gamma}\right)^2
+0.553 \left(g_4^{\gamma\gamma}\right)^2
-0.578 \, g_2^{Z\gamma}g_2^{\gamma\gamma}
-0.578 \, g_4^{Z\gamma}g_4^{\gamma\gamma}
\label{eq:cross-section-gammaH}
\end{eqnarray}
where the reference cross section is $\sigma_{\rm ref}^{\gamma\PH} = 1.33\times 10^4$\,fb.
We will investigate this channel further in Section~\ref{sect:eft-analysis}.



\section{Loop-induced standard model contributions}
\label{sect:eft-sm}

The NLO EW corrections from the \textsc{Prophecy4f} and \textsc{HAWK} generators have been widely 
used in calculations of the \Hboson production and decay cross sections at the LHC and included in the
LHC Higgs Working Group recommendations~\cite{deFlorian:2016spz}. The corrections are generally positive in the 
$H\to4\ell$ process~\cite{Bredenstein:2006rh} and negative in the VBF and $VH$ processes~\cite{Denner:2011id}. 
Differential distributions also show growth of these effects at higher energy, such as at high transverse momentum $p_T^H$ 
of the \Hboson in the case of VBF and $VH$ production, as expected for the well-known EW Sudakov enhancement. 
Our goal here is to reexamine some of these effects, focus on certain kinematic distributions, 
and compare the NLO EW effects to those generated by the EFT operators. In particular, 
we also produce kinematic distributions with \textsc{JHUGen} at LO, and introduce effective 
$g_2^{\gamma\gamma,\rm SM}$ and $g_2^{Z\gamma,\rm SM}$ couplings to model what 
one can call pseudo-EW corrections. 
Both \textsc{Prophecy4f} and \textsc{HAWK} include the interference of the loop-induced contributions 
with the Born process as dictated at NLO accuracy, but do not include squared contributions, which are 
formally of higher order. Nonetheless, these squared terms may be comparable to or larger than the
interference contributions, and we examine this with the effective $g_2^{\gamma\gamma,\rm SM}$ and 
$g_2^{Z\gamma,\rm SM}$ couplings by keeping or excluding their squared contributions. 

\begin{figure}[t]
\centering
\centerline{
\includegraphics[width=0.32\textwidth]{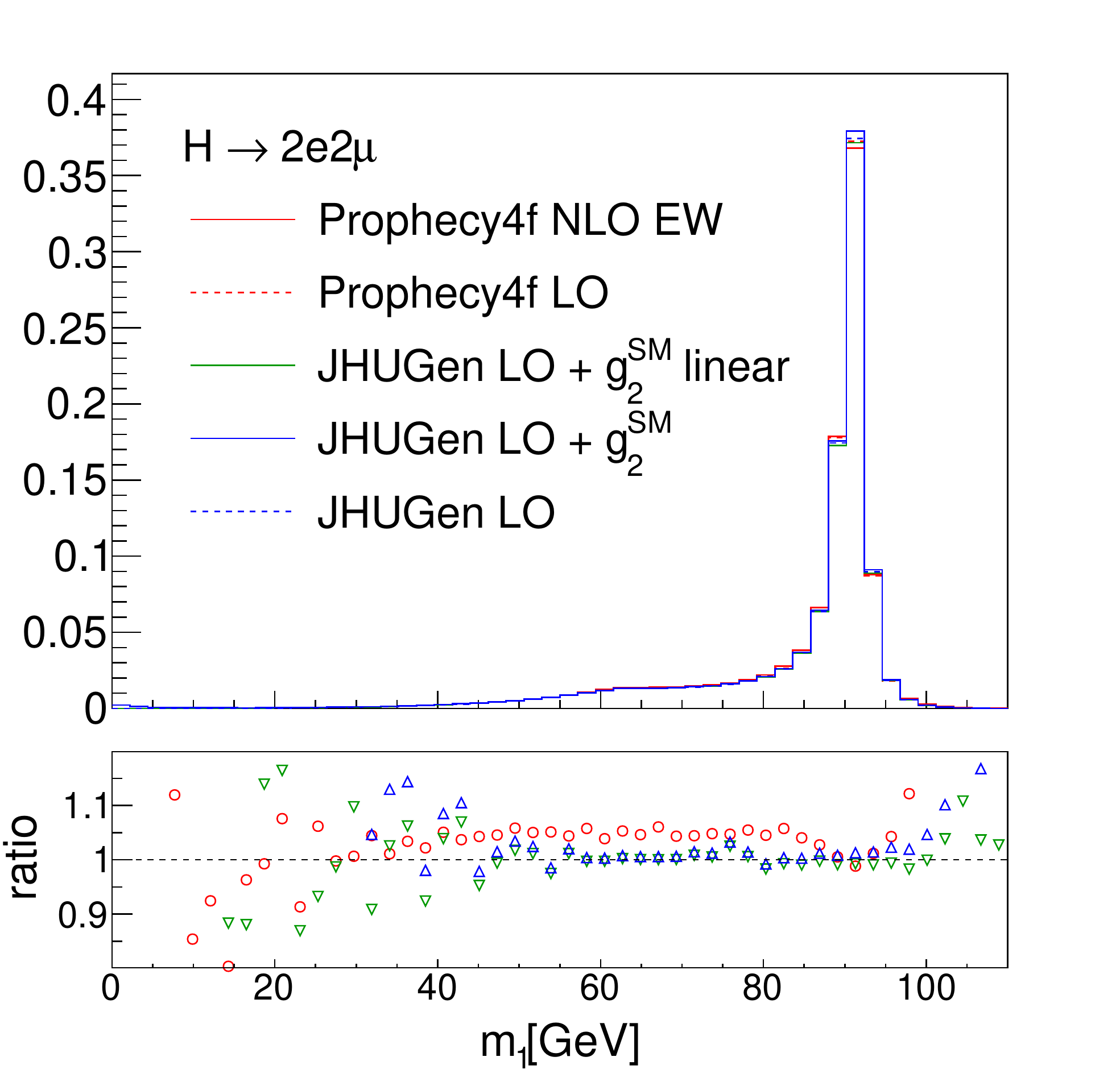}
\includegraphics[width=0.32\textwidth]{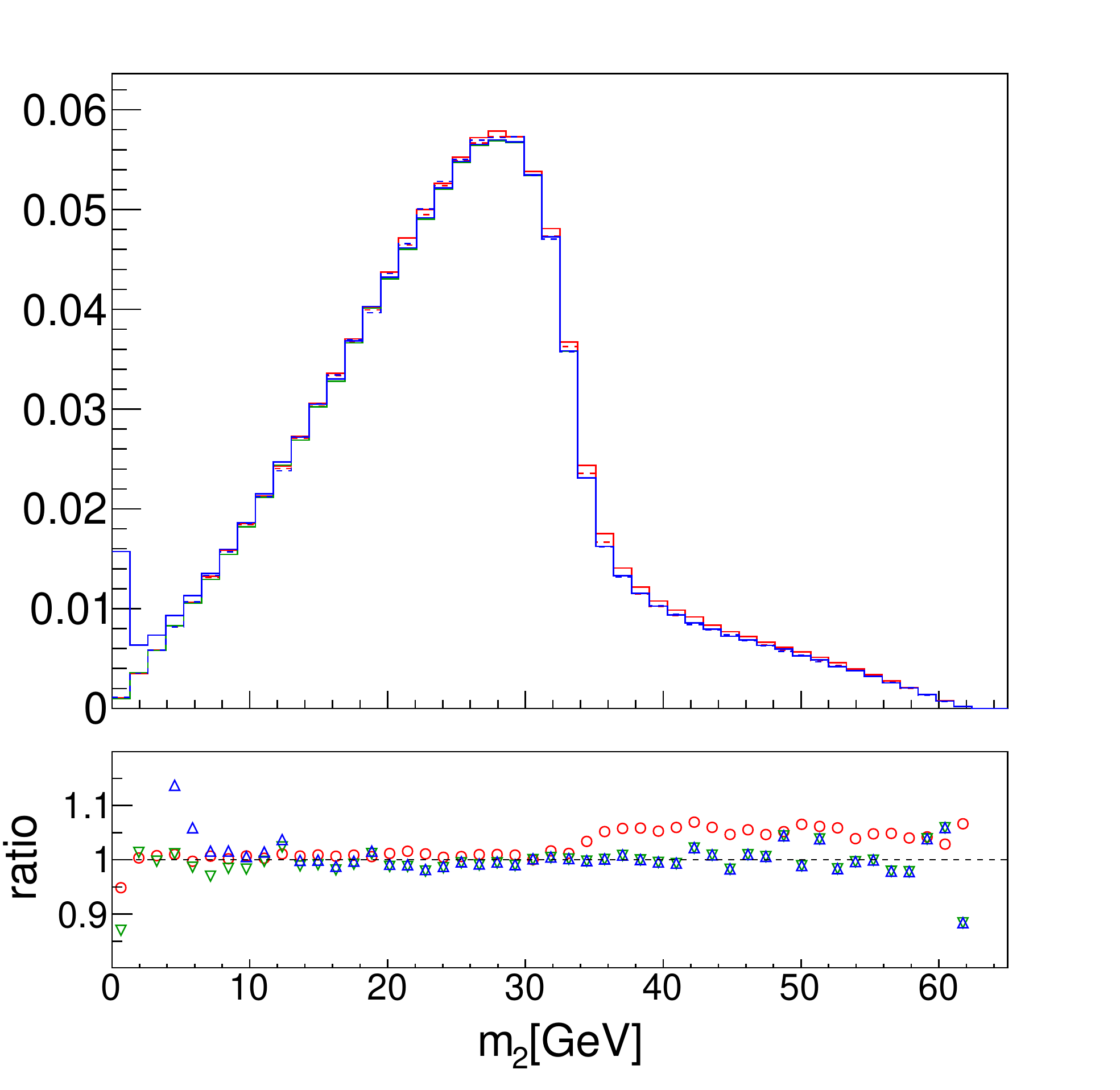}
}
\centerline{
\includegraphics[width=0.32\textwidth]{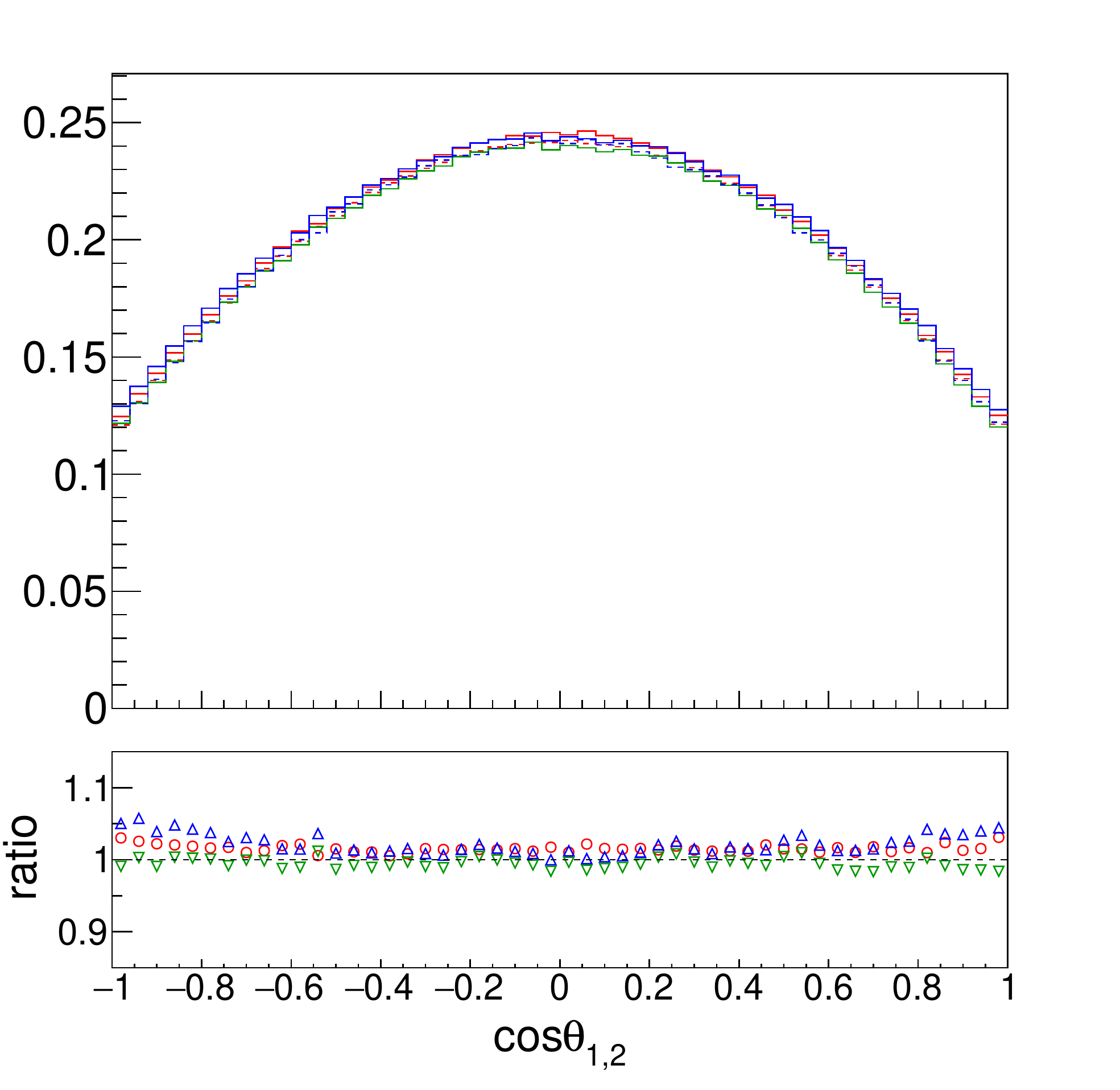}
\includegraphics[width=0.32\textwidth]{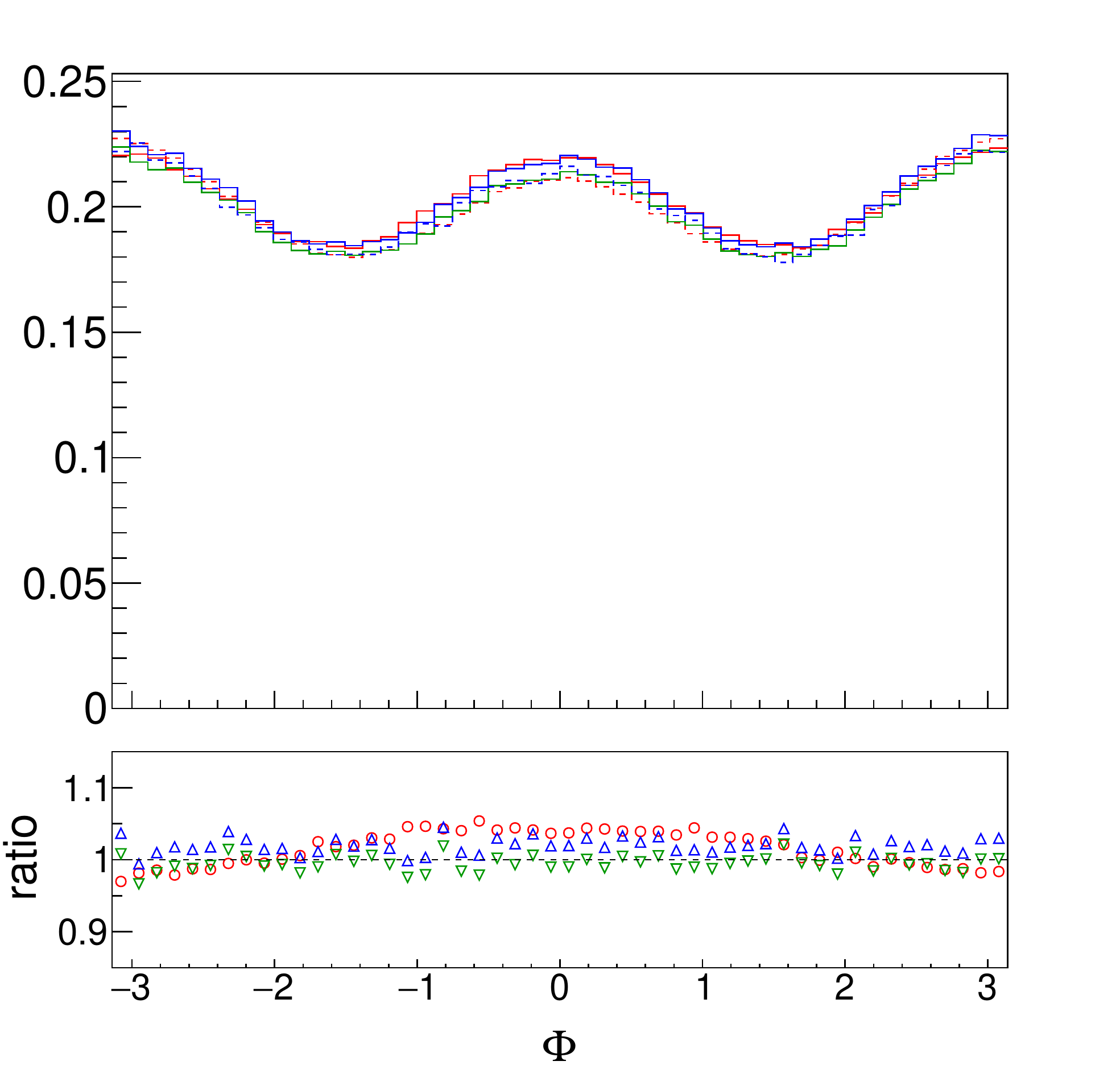}
}
\captionsetup{justification=centerlast}
\caption{
Distributions of kinematic observables in the $H\to2e2\mu$ decay: $m_1$, $m_2$, $\cos\theta_{1,2}$, $\Phi$. 
Five distributions are shown in each case: LO simulation (dashed red), NLO EW (solid red) with \textsc{Prophecy4f}, 
LO (dashed blue) and ad-hoc loop correction with $g_2^{\gamma\gamma,\rm SM}$ and $g_2^{Z\gamma,\rm SM}$ 
with (solid blue) and without (solid green) quadratic terms with \textsc{JHUGen}. 
Ratio of distributions with and without corrections are also shown.
}
\label{fig:EW-Profecy4f}
\end{figure}

\begin{figure}[t]
\centering
\centerline{
\includegraphics[width=0.32\textwidth]{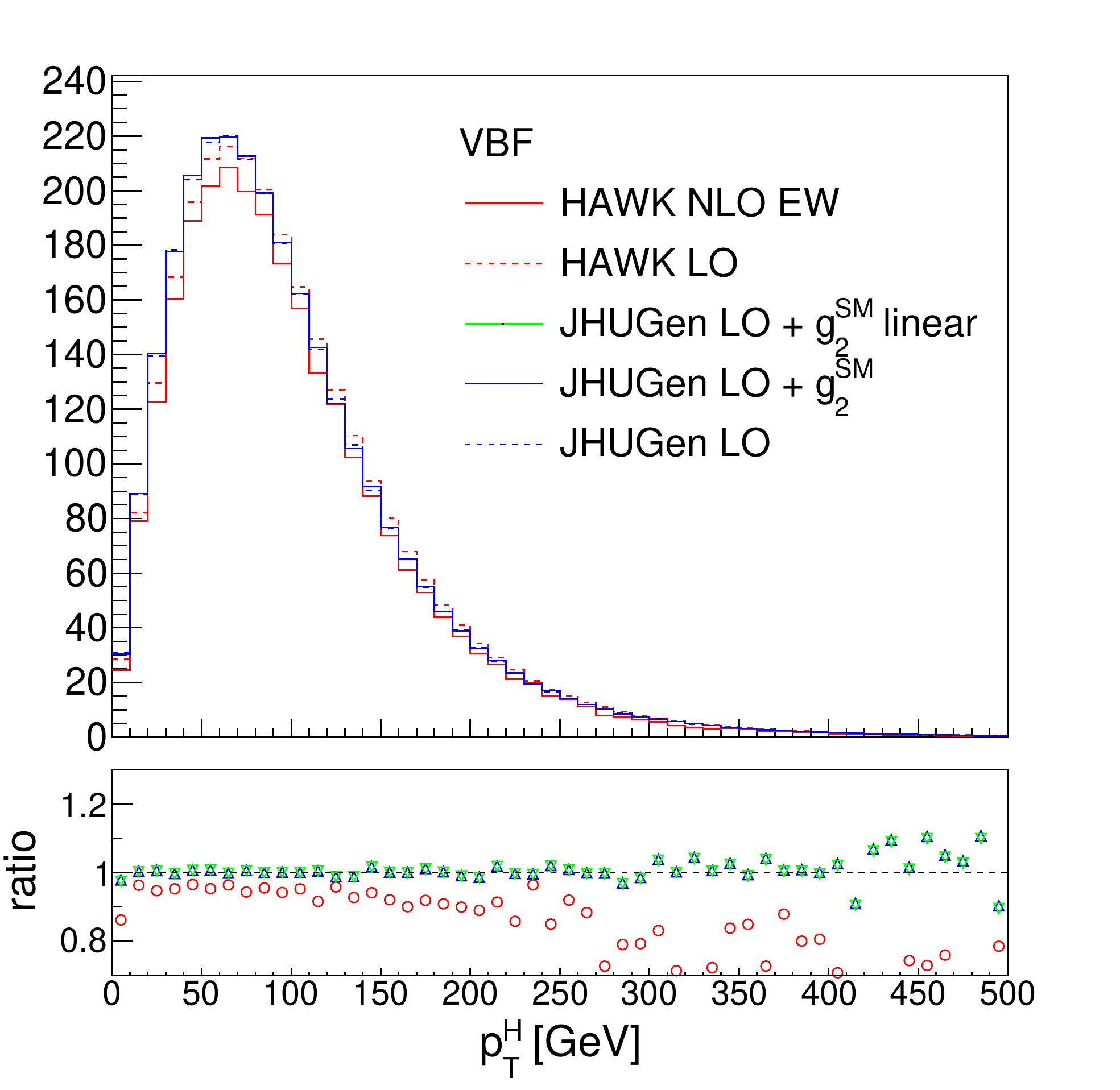}
\includegraphics[width=0.32\textwidth]{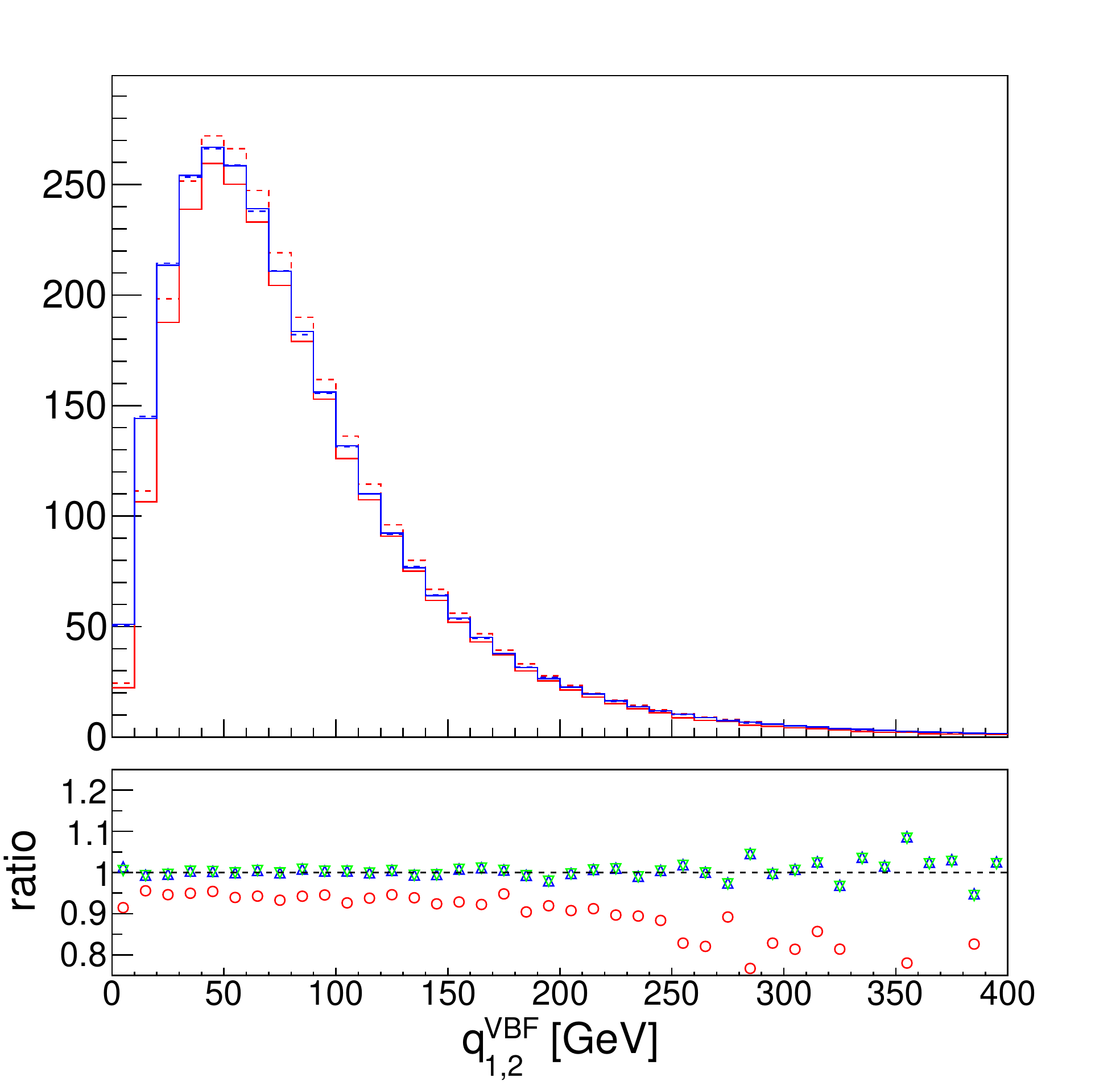}
}
\centerline{
\includegraphics[width=0.32\textwidth]{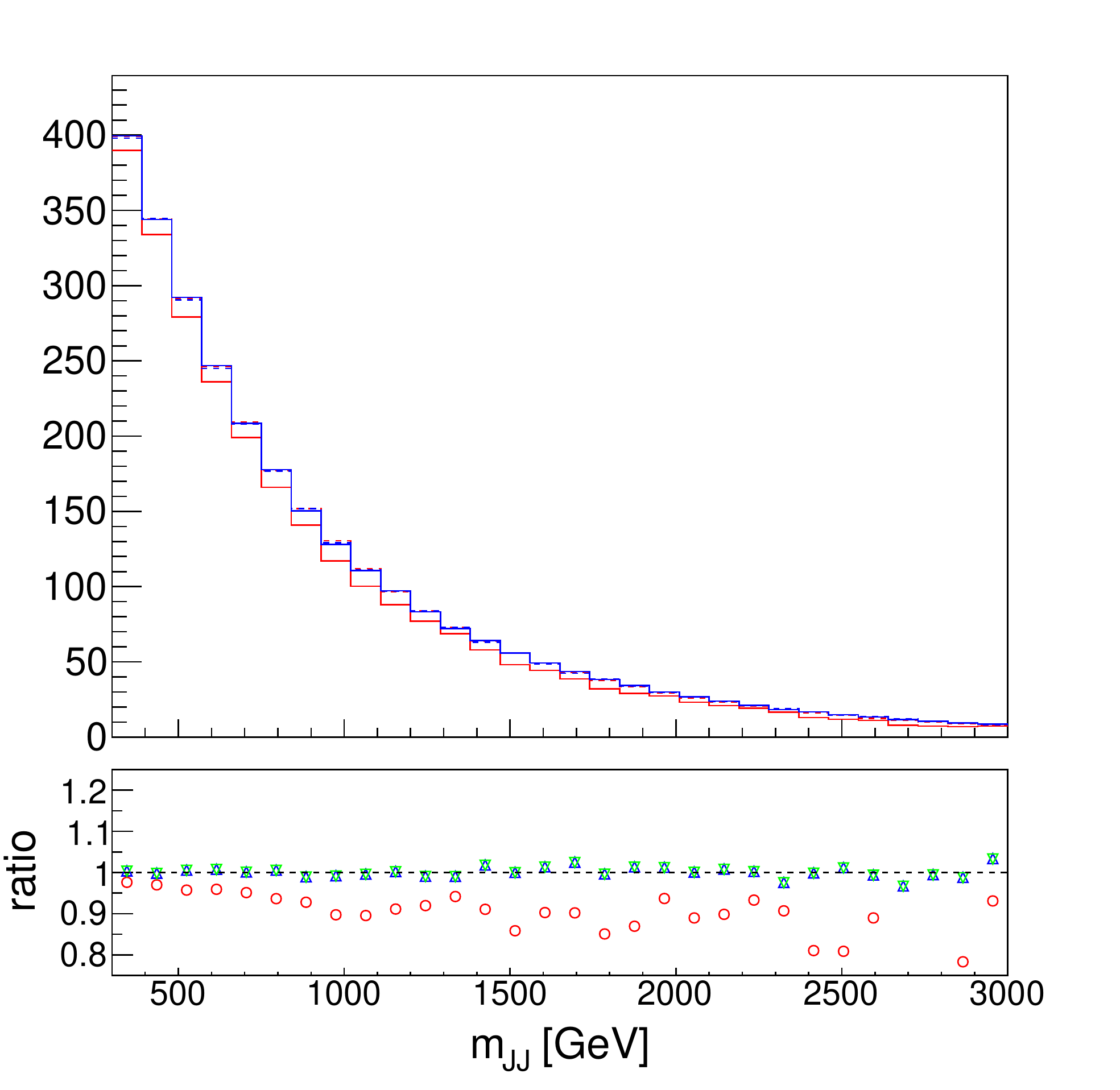}
\includegraphics[width=0.32\textwidth]{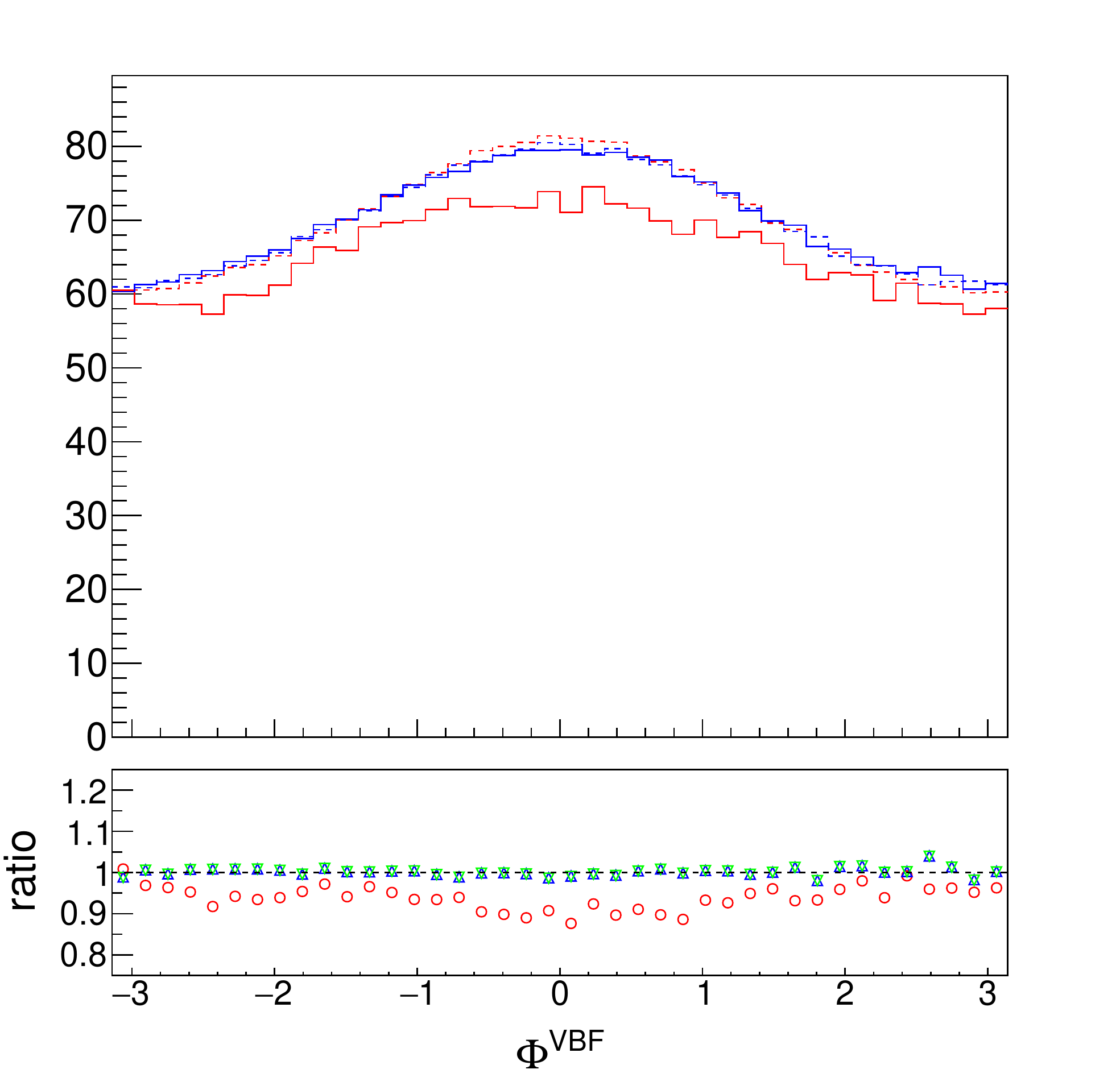}
}
\captionsetup{justification=centerlast}
\caption{
Distributions of kinematic observables in the VBF production: $p_T(H)$, $\sqrt{-q_{1,2}^2}$, $m_{jj}$, and $\Phi^{\rm VBF}$. 
Five distributions are shown in each case: LO simulation (dashed red), NLO EW (solid red) with \textsc{HAWK}, 
LO (dashed blue) and ad-hoc loop correction with $g_2^{\gamma\gamma,\rm SM}$ and $g_2^{Z\gamma,\rm SM}$ 
with (solid blue) and without (solid green) quadratic terms with \textsc{JHUGen}. 
Ratio of distributions with and without corrections are also shown.
}
\label{fig:EW-HAWK-VBF}
\end{figure}

\begin{figure}[t]
\centering
\centerline{
\includegraphics[width=0.32\textwidth]{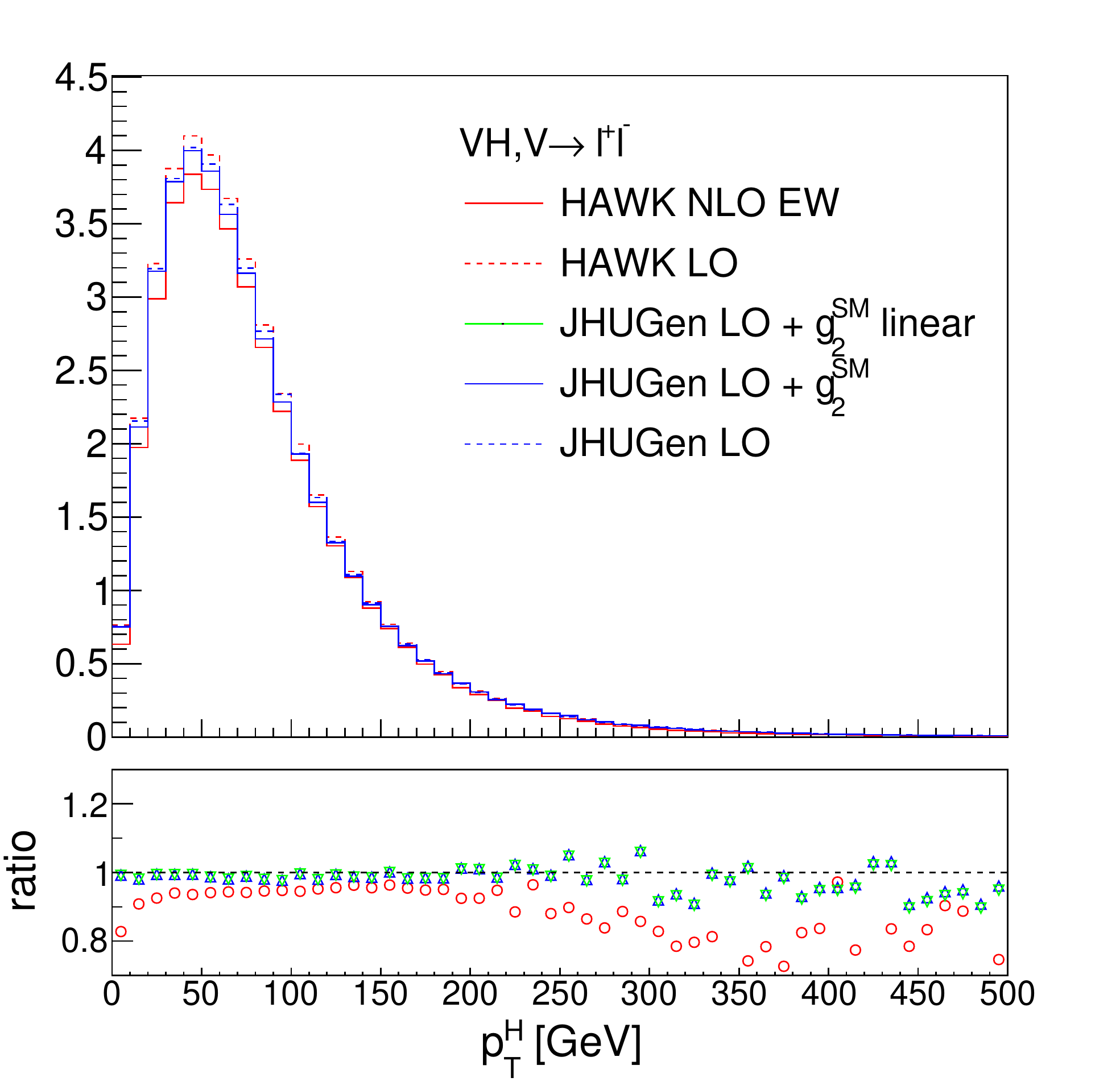}
\includegraphics[width=0.32\textwidth]{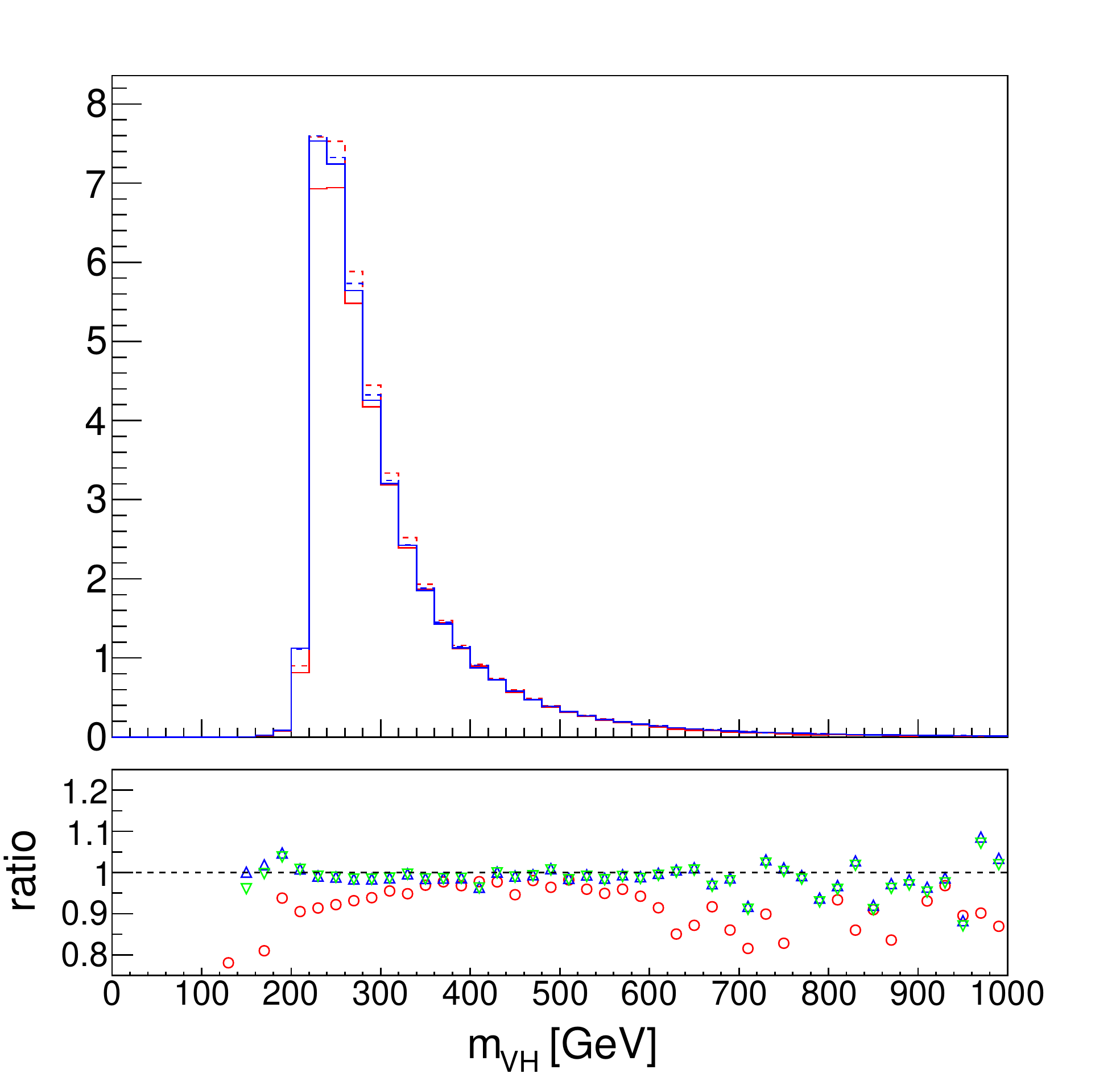}
}
\centerline{
\includegraphics[width=0.32\textwidth]{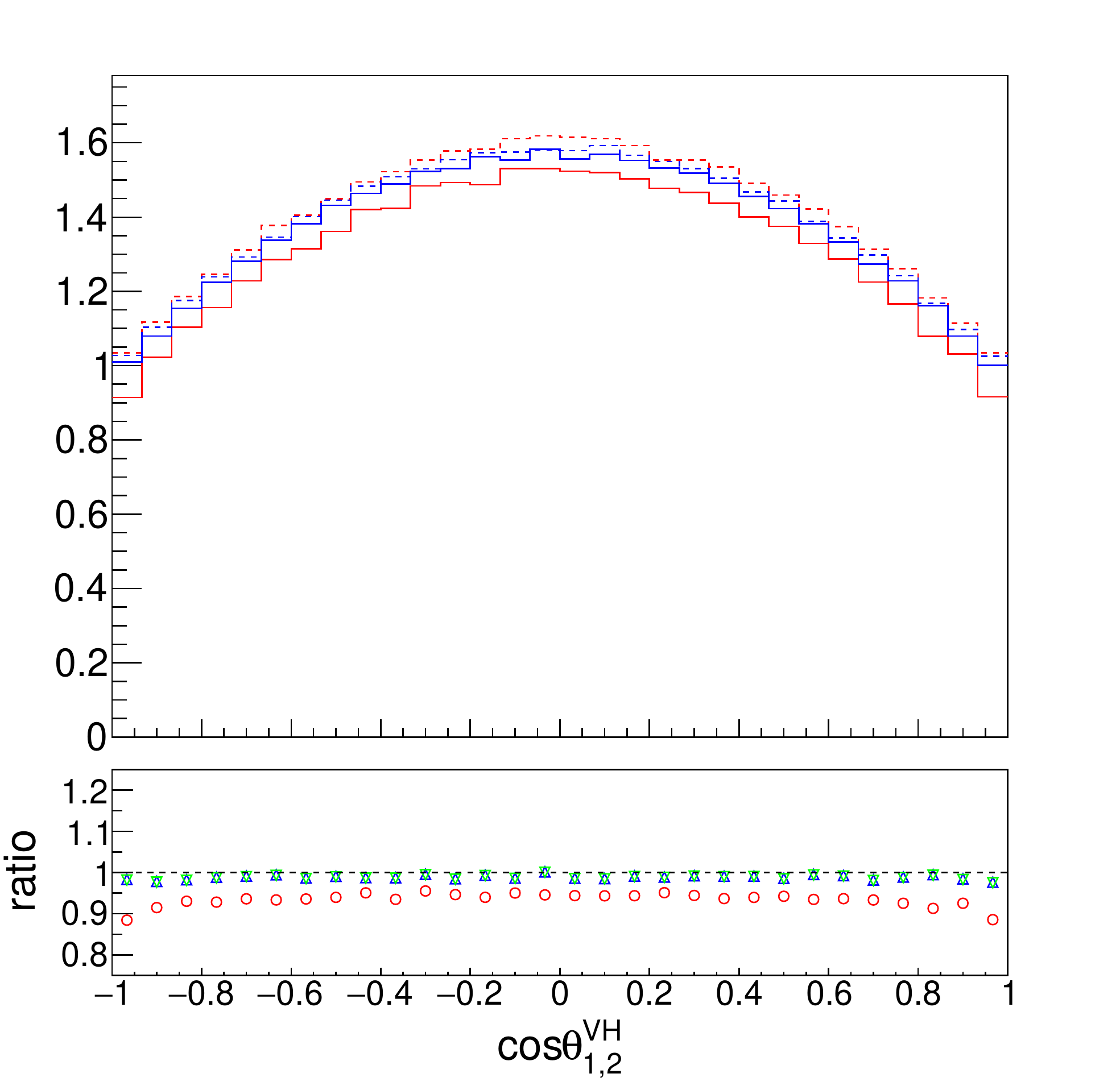}
\includegraphics[width=0.32\textwidth]{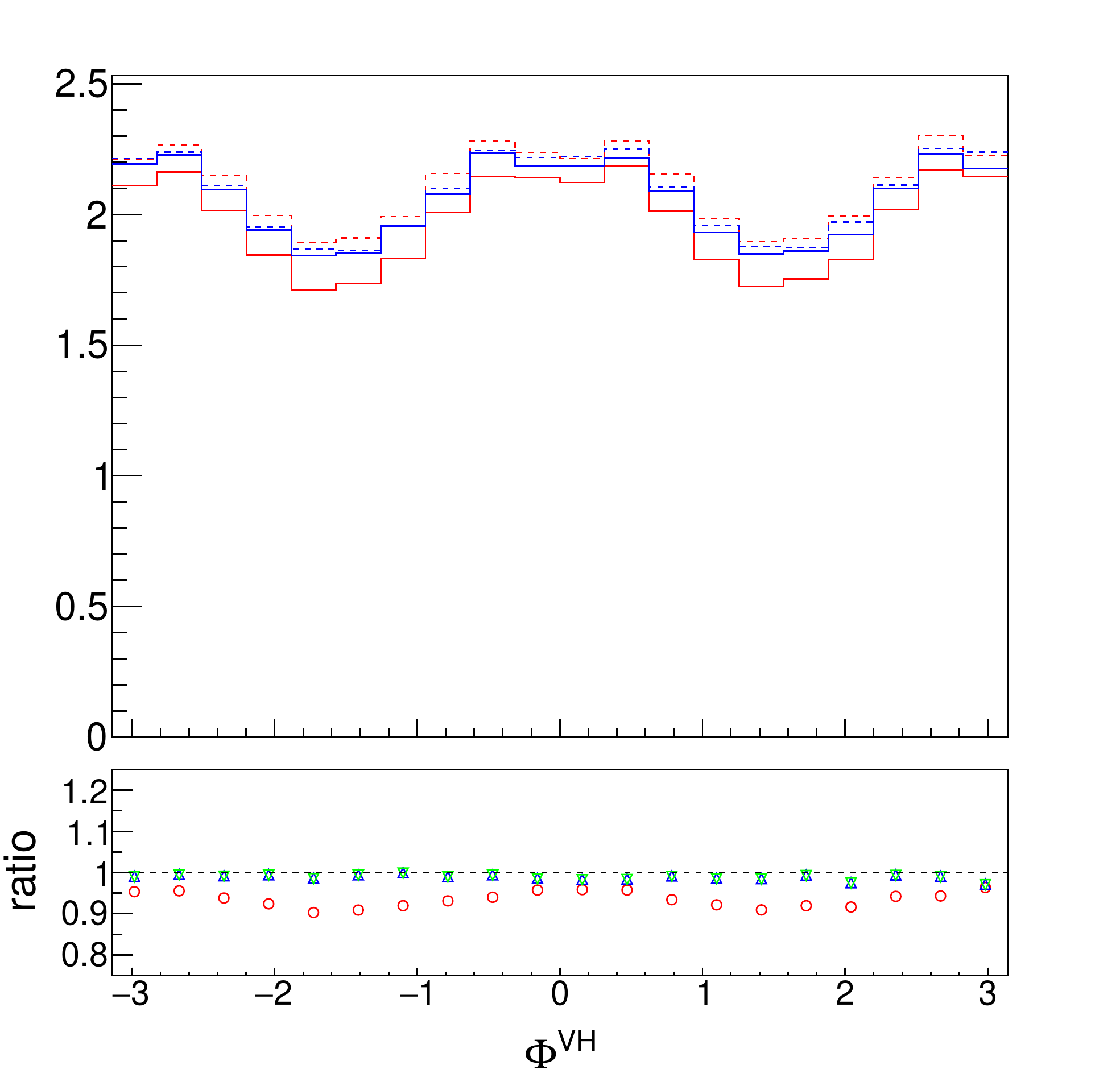}
}
\captionsetup{justification=centerlast}
\caption{
Distributions of kinematic observables in $VH$ production: $p_T(H)$, $m_{VH}$, $\cos\theta^{VH}_{1,2}$, and $\Phi^{VH}$. 
Five distributions are shown in each case: LO simulation (dashed red), NLO EW (solid red) with \textsc{HAWK}, 
LO (dashed blue) and ad-hoc loop correction with $g_2^{\gamma\gamma,\rm SM}$ and $g_2^{Z\gamma,\rm SM}$ 
with (solid blue) and without (solid green) quadratic terms with \textsc{JHUGen}. 
Ratio of distributions with and without corrections are also shown.
}
\label{fig:EW-HAWK-VH}
\end{figure}

The effective point-like couplings, such as  $g_2^{\rm gg, SM}$, $g_2^{\gamma\gamma,\rm SM}$, 
and $g_2^{Z\gamma,\rm SM}$, can model SM loop effects only in decay (or production) with \onshell\ particles,
such as $H\to\mathrm{gg}, \gamma\gamma, Z\gamma$.
The numerical values of these couplings from Eq.~(\ref{eq:HVV}) can be found in Eqs.~(\ref{eq:g2gg-effective}), 
(\ref{eq:g2AA-effective}), and (\ref{eq:g2ZA-effective}), respectively.
As we illustrate below, these couplings are inadequate for modeling loop effects in decays to 
virtual vector bosons, such as $H\to \gamma^*\gamma/Z\gamma \to f\bar{f}\gamma$ or
$H\to \gamma^*\gamma^*/Z\gamma^*/ZZ \to f\bar{f}f^\prime\bar{f^\prime}$.
The non-trivial $q^2$ dependence of the effective $HVV$ vertex cannot be described this way 
and the $g_2^{ZZ}$ and other tensor structures appearing in Eq.~(\ref{eq:HVV}) are not represented. 
Similar considerations apply to the VBF and $VH$ production, and also to the $\gamma H$ production,
as discussed in Section~\ref{sect:eft-basis}.
To make these statements in a quantitative way, we compare a simluation of these
effective couplings to a complete modeling of the NLO EW effects.

In order to model the SM loop corrections in the $H\to f\bar{f}f^\prime\bar{f^\prime}$ process, we employ the \textsc{Prophecy4f} 
generator, and in VBF and $ZH$ production we use \textsc{HAWK}.
In both cases, the NLO EW corrections can be applied to the process of interest
and compared to the LO simulation. 
We note that \textsc{HAWK} provides all results in the form of binned distributions, since unweighted events are not available
and events are not stored in LHE format~\cite{Alwall:2006yp}. 
This may complicate analysis and comparison of generated events, since different code would have to be employed 
in calculating the observables. Therefore, we have introduced a software interface which writes weighted events
from \textsc{HAWK} simulation in the LHE format, and all further analysis is performed in a unified way. 
Moreover, photon bremsstrahlung leads to smearing of kinematic distributions. In order to disentangle photon radiation 
from purely EW effects in kinematic distributions in both the $H\to4\ell$ and $VH$ with $V\to\ell^+\ell^-$ processes, 
we have introduced a recombination algorithm in analysis of events written in LHE format. 
The four-momentum of the associated photon is added to the nearest lepton in this algorithm. 
We note that NLO QCD+EW predictions for $ZH$ production have been recently implemented in the \textsc{POWHEG}
framework~\cite{Granata:2017iod}, but are not included in this study. 

We start with the study of NLO EW corrections in the $H\to2e2\mu$ decay process. In Fig.~\ref{fig:EW-Profecy4f},
the LO and NLO EW modeling of the process is shown, as generated with the \textsc{Prophecy4f} generator, 
and compared to ad-hoc loop correction with $g_2^{\gamma\gamma,\rm SM}$ and $g_2^{Z\gamma,\rm SM}$ with \textsc{JHUGen}. 
The overall correction to the decay width is $+1.5\%$, as shown in Table~\ref{tab:ew-nlo}.
The size of the effect with the pseudo-EW correction of $-0.6\%$ is similar, but does not reproduce the sign. 
However, including the quadratic terms with the pseudo-EW corrections appears important, as there is 
a growing importance of these effects at $q^2\to0$, which increases the correction by $+2.6\%$. 
The effect of linear terms appearing with the proper NLO EW corrections is most pronounced in the 
intermediate $m_1$ and $m_2$ ranges, away from the pole of \onshell\ $Z$. This is where the effect
of interference is most pronounced. Overall, we conclude that the pseudo-EW corrections only roughly
model the effect in the $H\to2e2\mu$ process of an order of magnitude, but are not adequate to describe 
the proper EW corrections. Nonetheless, they also indicate that the quadratic terms may become sizable 
and more important than the linear terms at the values of $m_2$ of a few GeV or below. 

\begin{table}[!b]
\centering
\captionsetup{justification=centerlast}
\caption{
The effect of NLO EW corrections calculated with the \textsc{Prophecy4f} and \textsc{HAWK} programs in the 
three processes with the selection requirements discussed in Section~\ref{sect:eft-basis}. Also shown are the effects 
of the $g_2^{\gamma\gamma,\rm SM}$ and $g_2^{Z\gamma,\rm SM}$ couplings with and without (linear) using their
squared contributions calculated with the \textsc{JHUGen} program. 
}
\label{tab:ew-nlo}
\begin{tabular}{lccc}
\vspace{-0.3cm} \\
   \hline
   &  EW NLO/LO  &   (LO + $g_2^{\rm SM}$)/LO  &    (LO + $g_2^{\rm SM}$ linear)/LO     \\
   \hline
   $H\to4\ell$               & $+1.5\%$  & $+2.0\%$ & $-0.6\%$ \\
   VBF                         & $-6.7\%$  & $+0.2\%$ & $+0.1\%$ \\
   $Z(\to\ell^+\ell^-)H$ & $-6.4\%$   & $-1.2\%$ & $-1.2\%$ \\
   \hline
\end{tabular}
\end{table}

The study of NLO EW corrections in the VBF process is shown in Fig.~\ref{fig:EW-HAWK-VBF}
and in the $VH$ process in Fig.~\ref{fig:EW-HAWK-VH}, where the LO and NLO EW modeling of the process 
is shown with the \textsc{HAWK} generator, and compared to ad-hoc loop correction 
with $g_2^{\gamma\gamma,\rm SM}$ and $g_2^{Z\gamma,\rm SM}$ with \textsc{JHUGen}. 
The selection requirements are similar to those in Section~\ref{sect:eft-basis}, except that in VBF we 
do not place a requirement on $q^2_V$ and in $VH$ we apply a looser requirement $m_{\ell\ell}> 0.1$ GeV
and a tighter requirement $p_T^\ell> 5$ GeV.
The overall NLO EW correction is negative in the range of $6-7\%$, as shown in Table~\ref{tab:ew-nlo},
and grows in size with energy represented by transverse momentum $p_T^H$ in both cases. 
In the case of VBF, the momentum of the intermediate vector bosons $\sqrt{-q_{1,2}^2}$ also
shows this feature. The pseudo-EW corrections show about $-1.2\%$ correction in the $VH$ process
and small growth of the effect with $p_T^H$, but no sizable effect in the VBF process. In both VBF
and $VH$, there is no evidence of importance of the quadratic terms with the $g_2^{\rm SM}$ expansion. 
We conclude that the pseudo-EW corrections are not adequate in the VBF and $VH$ processes,
even if in the $VH$ process cross section modifications appear in the same direction. 

Given the inadequacy of the pseudo-EW corrections, we have investigated an approximate approach 
of re-weighting the LO EW simulation with a dynamic k factor, which is a ratio of the NLO and LO EW 
kinematic distributions. This approach can capture the main features of the correction, such as 
its growth with $p_T^H$, if this quantity is used as the kinematic distribution in re-weighting. 
However, this approach does not guarantee adequate modeling of the other kinematic distributions 
simultaneously, if those are not also used in re-weighting. For example, we found that re-weighting 
based on $p_T^H$ does not bring angular distributions to an agreement. 

The importance of the NLO EW corrections will become evident in the actual analysis of LHC data. 
The precision of existing LHC constraints~\cite{Chatrchyan:2012jja,Aad:2013xqa,Chatrchyan:2013mxa,Khachatryan:2014kca,
Khachatryan:2015mma,Aad:2015mxa,Aad:2016nal,Khachatryan:2016tnr,Sirunyan:2017tqd,Aaboud:2017oem,Aaboud:2017vzb,
Aaboud:2018xdt,Sirunyan:2019twz,Sirunyan:2019nbs,Aad:2020mnm,Sirunyan:2021fpv}
is not sufficient for reaching the NLO EW effects appearing in the SM. Therefore, accurate modeling
of such effects may not appear as critical at present. However, with growing precision of experimental measurements
a careful investigation of NLO EW effects on kinematic distributions will become important. 


\section{Calculation of the width in the presence of intermediate photons}
\label{sect:eft-lowq}

The presence of intermediate photons in \Hboson decays to fermions via $\gamma^* \to f^+f^-$ 
can lead to sharp peaks in the spectrum when $q^2=(p_{f^+}+p_{f^-})^2$ is small.
For decays into leptons, these peaks are cut off by the physical lepton masses. 
In the case of quarks, non-perturbative effects wash out this peak structure and introduce hadronic resonances instead. 
In the following, we introduce a procedure, based on matching amplitudes in the collinear limit, 
to handle these singularities in a way which allows their efficient numerical evaluation. 
We draw parts of this description from Ref.~\cite{Denner:2019zfp}.
Technical details can be found in Appendix~\ref{app:eft-lowq}.

We write the partial \Hboson decay widths as 
\begin{eqnarray} \label{eq:hwidth4f}
	\Gamma_{H\to 2f2f'} &=& \frac{1}{2m_H} \frac1{(4_{ff'})}  \int \! \mathrm{dPS}^{(4)} \, |\mathcal{M}_{H\to 2f2f'}|^2,
	\\ \label{eq:hwidth2fga}
	\Gamma_{H\to 2f\gamma} &=& \frac1{2m_H} \int \! \mathrm{dPS}^{(3)} \, |\mathcal{M}_{H\to 2f\gamma}|^2,
	\\ \label{eq:hwidthgaga}
	\Gamma_{H\to \gamma\gamma} &=& \frac{1}{2m_H} \frac1{(2_{\gamma\gamma})} \int \! \mathrm{dPS}^{(2)} \, |\mathcal{M}_{H\to \gamma\gamma}|^2,
\end{eqnarray}
where we introduced the symbols $(j_{pp'})=j^{\, \delta_{p p'}}$ for symmetry factors for identical particles.
Explicit parameterization of the phase spaces, using appropriate variables, yields 
\begin{eqnarray}
	 \int \! \mathrm{dPS}^{(3)} &=& 
	 \int_{4m_1^2}^{M_H^2}\! \frac{\mathrm{d}q_{12}^2}{2\pi} \, 
	 \int\! \mathrm{dPS}^{(2)}\!\!\left(M_H^2,q_{12}^2,0\right) \, \mathrm{dPS}^{(2)}\!\!\left(q_{12}^2,m_f^2,m_f^2\right),
	 \label{eq:finalps3}
\end{eqnarray}
and 
\begin{eqnarray}
	 \int \! \mathrm{dPS}^{(4)} 
	 &=&
	 \int_{4m_f^2}^{M_H^2}\! \frac{\mathrm{d}q_{12}^2}{2\pi} \, 
	 \int_{4m_{f'}^2}^{(M_H^2-\sqrt{q_{12}})^2}\! \frac{\mathrm{d}q_{34}^2}{2\pi} \, 
	 \int\! \mathrm{dPS}^{(2)}\!\!\left(M_H^2,q_{12}^2,q_{34}^2\right) \, \mathrm{dPS}^{(2)}\!\!\left(q_{12}^2,m_f^2,m_f^2\right) \, \mathrm{dPS}^{(2)}\!\!\left(q_{34}^2,m_{f'}^2,m_{f'}^2\right),
	 \nonumber\\ \label{eq:finalps4}
\end{eqnarray}
where simply $\mathrm{dPS}^{(2)}(q^2,m_1^2,m_2^2) = \mathrm{d}\!\cos\theta \, \mathrm{d}\phi /(2q^2)$.
We split the invariant mass integrations in Eqs.~(\ref{eq:finalps3},\ref{eq:finalps4}) into a low and high virtuality region 
\begin{eqnarray}\label{eq:pssplit}
	\int_{4 m^2}^{M^2}\! \mathrm{d}q^2_\gamma = \int_{4m^2}^{\mu^2}\! \mathrm{d} q^2_\gamma + \int_{\mu^2}^{M^2}\! \mathrm{d}q^2_\gamma
\end{eqnarray}
separated by an arbitary parameter $\mu^2 \ll M^2$. In this form, we can apply the collinar approximation to the squared matrix elements 
$|\mathcal{M}_{X+\gamma^*\to X+2f}|^2 \xrightarrow[]{\,q_{\gamma^*}^2 \ll \mu^2 \,}  P_{ff}(z)\big/q_{\gamma^*}^2 \times |\mathcal{M}_{X+\gamma}|^2$
in the low virtuality region and analytically integrate over $q^2_\gamma$ \cite{Denner:2019zfp}.
The high virtuality region does not contain sharp peaks and can be treated numerically in a standard manner. 
As a result, the partial decay width for $H \to 2\ell\gamma$ can be written as   
\begin{eqnarray}
	\Gamma_{H\to 2\ell\gamma} &=& \Gamma_{H\to 2\ell\gamma} \Big|_{q_{2\ell}^2 \ge \mu^2}
 	+ \frac{\alpha}{2\pi} \left[ \frac{2}{3} \log\left(\frac{\mu^2}{m_\ell^2}\right) - \frac{10}{9} \right] 2\, \Gamma_{H\to \gamma\gamma} + \mathcal{O}(\mu^2\big/ M_H^2),
\label{eq:result2lgamma}
\end{eqnarray}	
where the left-hand side is independent of $\mu^2$.
We note that the low virtuality region is conveniently expressed in terms of an analytic function containing a potentially large $\log(m_\ell^2)$
times the width $\Gamma_{H\to \gamma\gamma}$, which can be straight-forwardly obtained using numerical methods for any combination of anomalous couplings (cfg. Eq.~(\ref{eq:ratio-gammagamma})).  
Contributions from low virtuality $Z$ bosons decaying into $2\ell$ are parametrically suppressed by $\mu^2\big/M_H^2$. 
In a similar fashion, we obtain the results for the \Hboson decays into four leptons
\begin{eqnarray}
	\Gamma_{H\to 2\ell 2\ell'} &=& \Gamma_{H\to 2\ell 2\ell'} \Big|_{q_{2\ell}^2 \ge \mu^2,q_{2\ell'}^2 \ge \mu^2}
	\nonumber \\
	&&+ \frac{\alpha}{2\pi} \left[ \frac{2}{3} \log\left(\frac{\mu^2_\ell}{m_\ell^2}\right) - \frac{10}{9} \right] \; \Gamma_{H \to  2\ell' \gamma} \Big|_{q_{2\ell'}^2 \ge \mu^2}
	+ \frac{\alpha}{2\pi}  \left[ \frac{2}{3} \log\left(\frac{\mu^2_{\ell'}}{m_{\ell'}^2}\right) - \frac{10}{9} \right] \; \Gamma_{H \to  2\ell \gamma} \Big|_{q_{2\ell}^2 \ge \mu^2}
	\nonumber \\
	&&+ \left(\frac{\alpha}{2\pi}\right)^2 \left[ \frac{2}{3} \log\left(\frac{\mu^2_\ell}{m_\ell^2}\right) - \frac{10}{9} \right] 
	 \left[ \frac{2}{3} \log\left(\frac{\mu^2_{\ell'}}{m_{\ell'}^2}\right) - \frac{10}{9} \right] \; 2\,\Gamma_{H\to \gamma\gamma} + \mathcal{O}(\mu^2\big/ M_H^2),
\label{eq:result2l2l}
	\\ 
	\Gamma_{H\to 4\ell} &=& \Gamma_{H\to 4\ell } \Big|_{q_{2\ell}^2 \ge \mu^2}
	+ \frac{\alpha}{2\pi} \left[ \frac{2}{3} \log\left(\frac{\mu^2}{m_{\ell}^2}\right) - \frac{10}{9} \right] \; \Gamma_{H \to  2\ell \gamma} \Big|_{q_{2\ell}^2 \ge \mu^2}
	\nonumber \\
	&&+ \left(\frac{\alpha}{2\pi}\right)^2 \left[ \frac{2}{3} \log\left(\frac{\mu^2}{m_\ell^2}\right) - \frac{10}{9} \right]^2 
       \; \Gamma_{H\to \gamma\gamma} + \mathcal{O}(\mu^2\big/ M_H^2).
\label{eq:result4l}
\end{eqnarray}	
\\
If the \Hboson decay occurs into quark final states, the low virtuality region is affected by sizable non-perturbative effects, which can be related to the experimentally 
measured quantity $\Delta \alpha_\mathrm{had}^{(5)}(M_Z^2)$ via a dispersion relation and unitarity \cite{Denner:2019zfp}.
The value of $\Delta \alpha_\mathrm{had}^{(5)}(M_Z^2)=(276.11 \pm 1.11) \times 10^{-4}$ \cite{Eidelman:1995ny,Keshavarzi:2018mgv} has been extracted from the low energy region of the 
ratio $\sigma(e^+e^- \to \text{hadrons}) \big/ \sigma(e^+e^- \to \mu^+\mu^-)$
and is related to the fine structure constant via $\alpha(s) = \alpha(0) \big/ (1-\Delta \alpha(s))$.
Summing over the five light quark flavors, labelling $\sum_{q} \Gamma_{H\to 2q\gamma}= \Gamma_{H\to 2j\gamma}$, and choosing $\mu^2 \gg 4m_b^2 $, we find 
\begin{eqnarray}
	\Gamma_{H\to 2j\gamma} &=& \Gamma_{H\to 2j\gamma} \Big|_{q_{2q}^2 \ge \mu^2}
 	+ \left[ \Delta \alpha_\mathrm{had}^{(5)}(M_Z^2) + \frac{\alpha}{\pi} \frac{11}{9}  \log\left(\frac{\mu^2}{M_Z^2}\right)  \right] 2\, \Gamma_{H\to \gamma\gamma} + \mathcal{O}(\mu^2\big/ M_Z^2),
\label{eq:result2lgamma-had}
\end{eqnarray}	
and similar for $\Gamma_{H\to 4j}$.

\begin{figure}[t]
\captionsetup{justification=centerlast}
\centerline{
\includegraphics[width=0.32\textwidth]{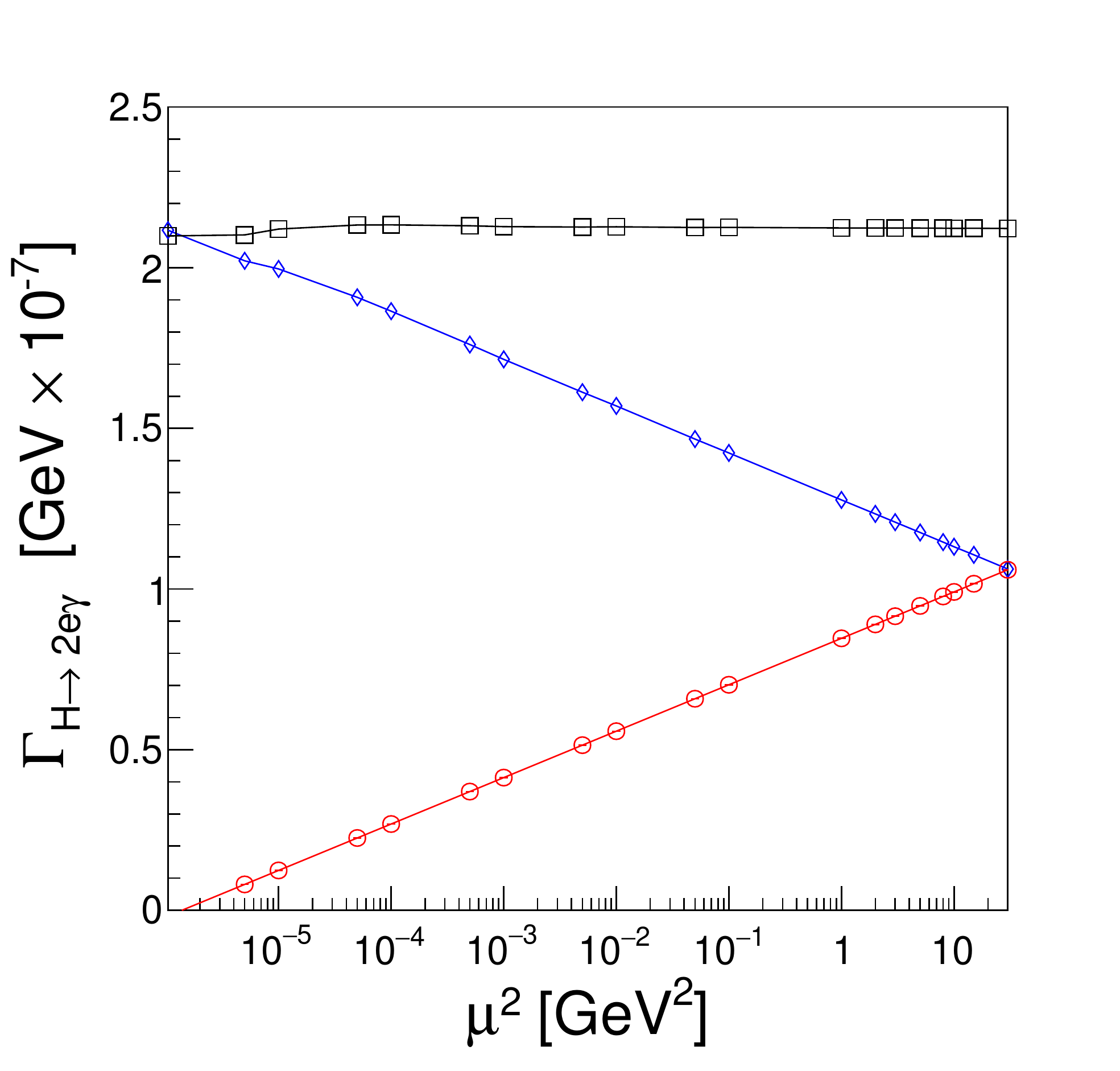}
\includegraphics[width=0.32\textwidth]{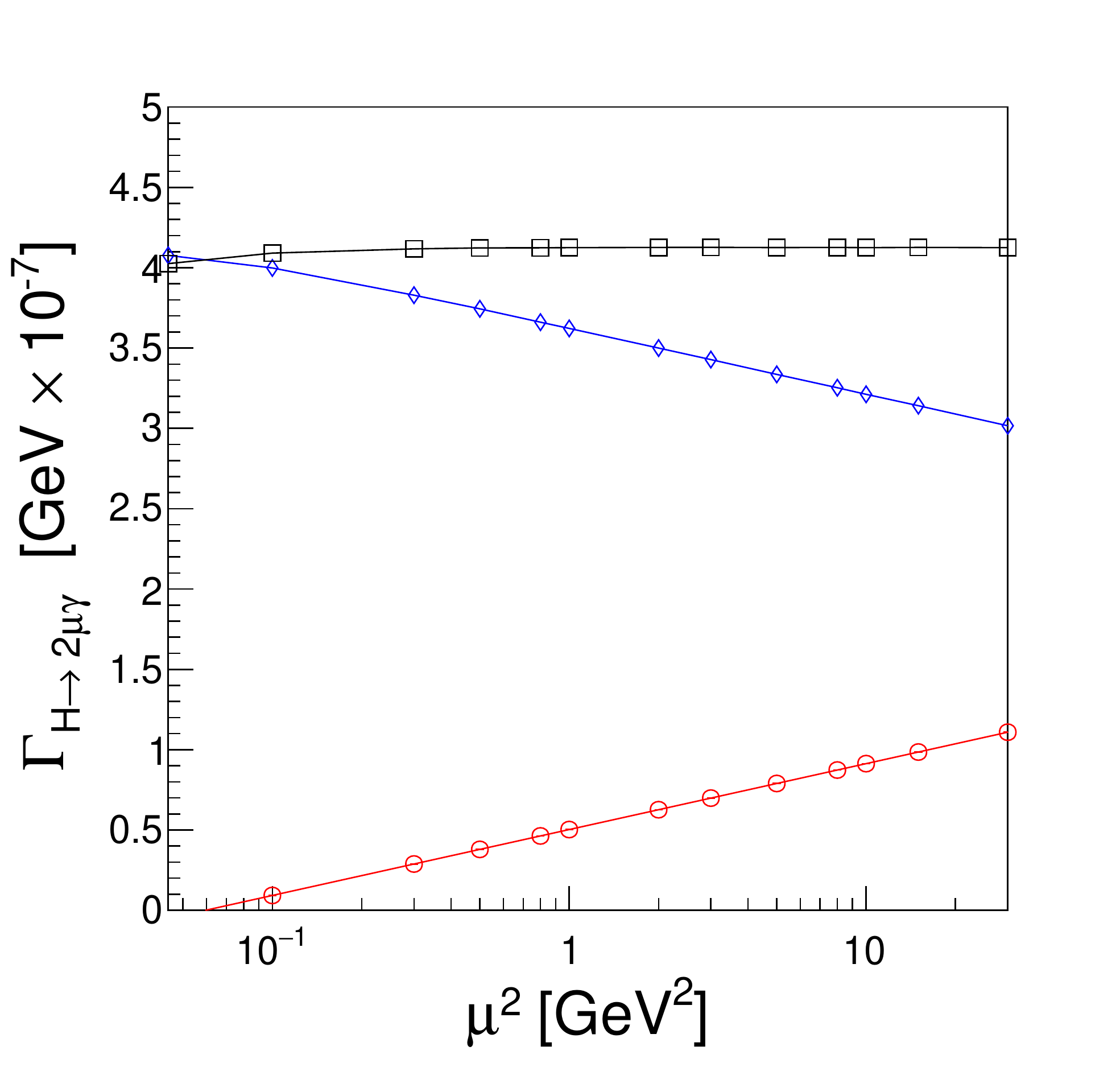}
}
\caption{
The partial decay width of the process $H\to2\ell\gamma$ calculated using Eq.~(\ref{eq:result2lgamma}) 
for the $2e\gamma$ (left) and  $2\mu\gamma$ (right) final states with the couplings discussed in text. 
The first two terms appearing in Eq.~(\ref{eq:result2lgamma}) are shown for several values of $\mu^2$:
$\Gamma_{H\to 2\ell\gamma}$ with the requirement $q_{2\ell}^2 \ge \mu^2$ (blue diamonds, $\diamond$)
and $\Gamma_{H\to \gamma\gamma}$ multiplied by the $\mu^2$-dependent factor (red circles, $\circ$). 
The sum of the cross sections is also shown (black squares, $\square$). 
}
\label{fig:test-lowq2-2lgamma}
\end{figure}

In order to illustrate the performance of Eq.~(\ref{eq:result2lgamma}), we model the $H\to2\ell\gamma$ decay 
with the $HVV$ couplings corresponding to $g_{2}^{\gamma\gamma,\rm SM}$ and $g_{2}^{Z\gamma, \rm SM}$ 
as defined in Eqs.~(\ref{eq:g2AA-effective}) and~(\ref{eq:g2ZA-effective}).
We scan the value of cutoff $\mu^2$ in Eq.~(\ref{eq:result2lgamma}) from the 
threshold value of $(2m_\ell)^2$ up to $(5\,{\rm GeV})^2$ for both electrons ($\ell=e$) and muons ($\ell=\mu$). 
Figure~\ref{fig:test-lowq2-2lgamma} shows the first two terms appearing in Eq.~(\ref{eq:result2lgamma}),
the value of $\Gamma_{H\to 2\ell\gamma}$ with the requirement $q_{2\ell}^2 \ge \mu^2$
and the value of $\Gamma_{H\to \gamma\gamma}$ multiplied by the $\mu^2$-dependent factor. 
The other terms in Eq.~(\ref{eq:result2lgamma}) can be neglected in this comparison. 
With the $\mu^2$ increasing, the former cross section falls while the latter rises, leading to a constant 
cross section of the $H\to 2\ell\gamma$ process, as one would expect to observe for the proper modeling 
of the effect. 

In order to illustrate the performance of Eqs.~(\ref{eq:result2l2l}) and~(\ref{eq:result4l}), we model the $H\to 2e2\mu$, $4e$, and $4\mu$ 
decays with the $g_{4}^{\gamma\gamma}$ and $g_{4}^{Z\gamma}$ couplings set to the values of $g_{2}^{\gamma\gamma,\rm SM}$ 
and $g_{2}^{Z\gamma, \rm SM}$, respectively, as used in the previous test. Both in this test and in the previous test of Eq.~(\ref{eq:result2lgamma}),
the rationale is to model the processes involving virtual photons which exhibit growth with $q^2\to 0$. 
We choose to illustrate the performance with $CP$-even couplings in one
case and $CP$-odd couplings in the other, but the procedure has been validated to work in any combination of couplings. 
The $\kappa_2^{Z\gamma}$ coupling formally involves a virtual photon. However, due to appearance of $q_\gamma^2$ 
in the numerator of the tensor structure in Eq.~(\ref{eq:HVV}), this coupling does not
exhibit divergence with $q^2\to 0$.  Its behavior is similar to the terms with the virtual $Z$
and can be absorbed in the corresponding terms. 

\begin{figure}[t]
\captionsetup{justification=centerlast}
\centerline{
\includegraphics[width=0.32\textwidth]{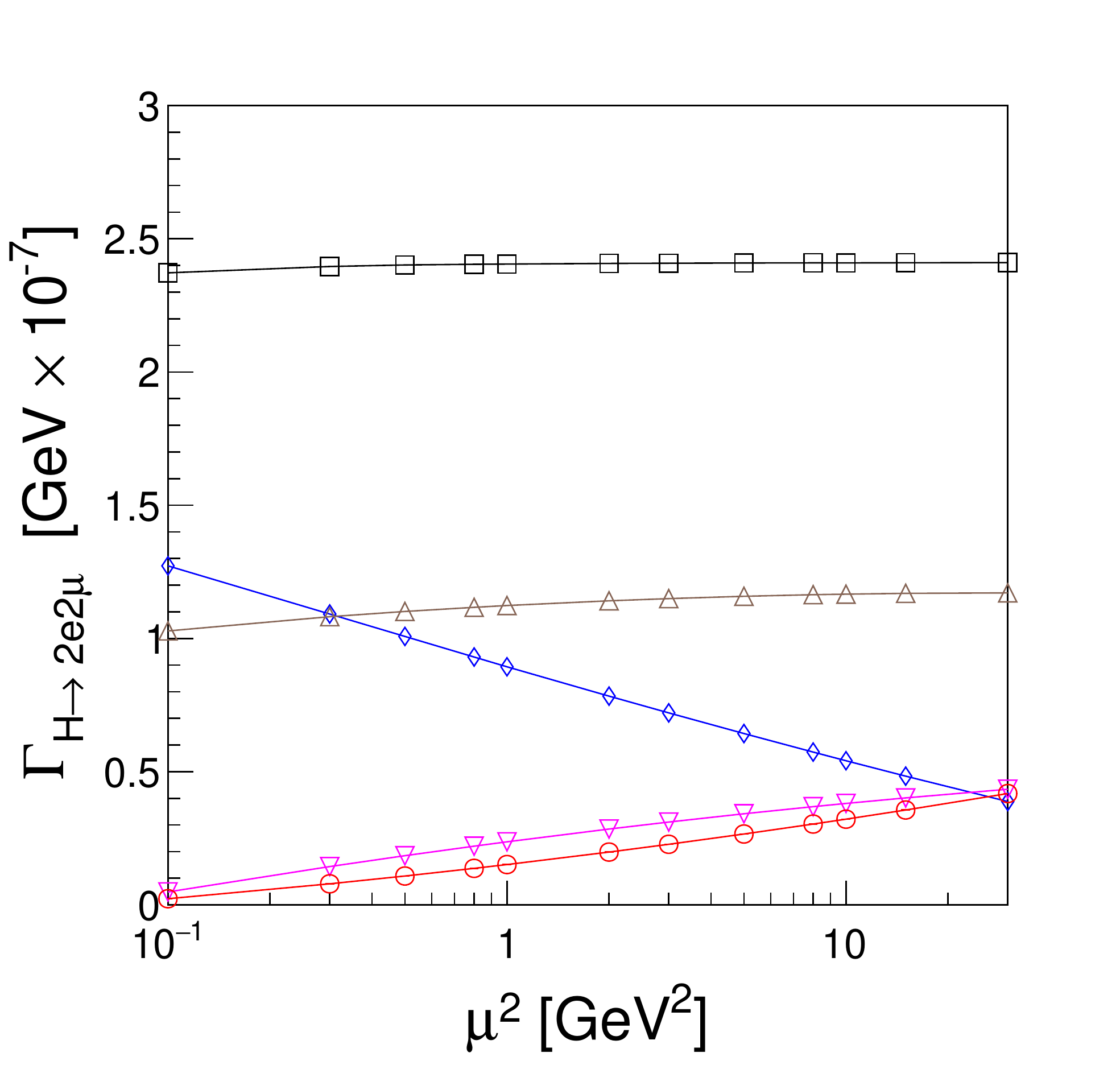}
\includegraphics[width=0.32\textwidth]{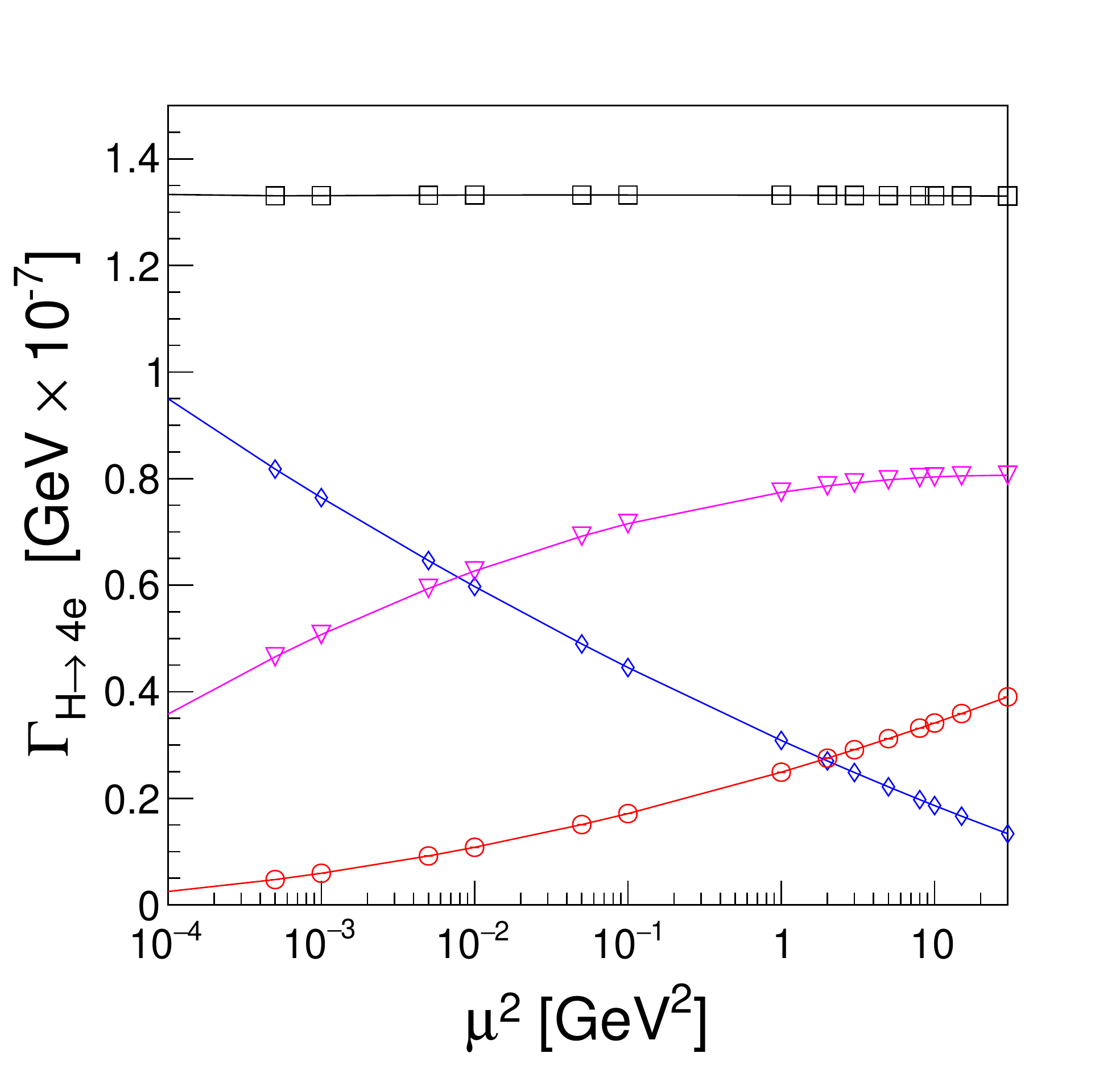}
\includegraphics[width=0.32\textwidth]{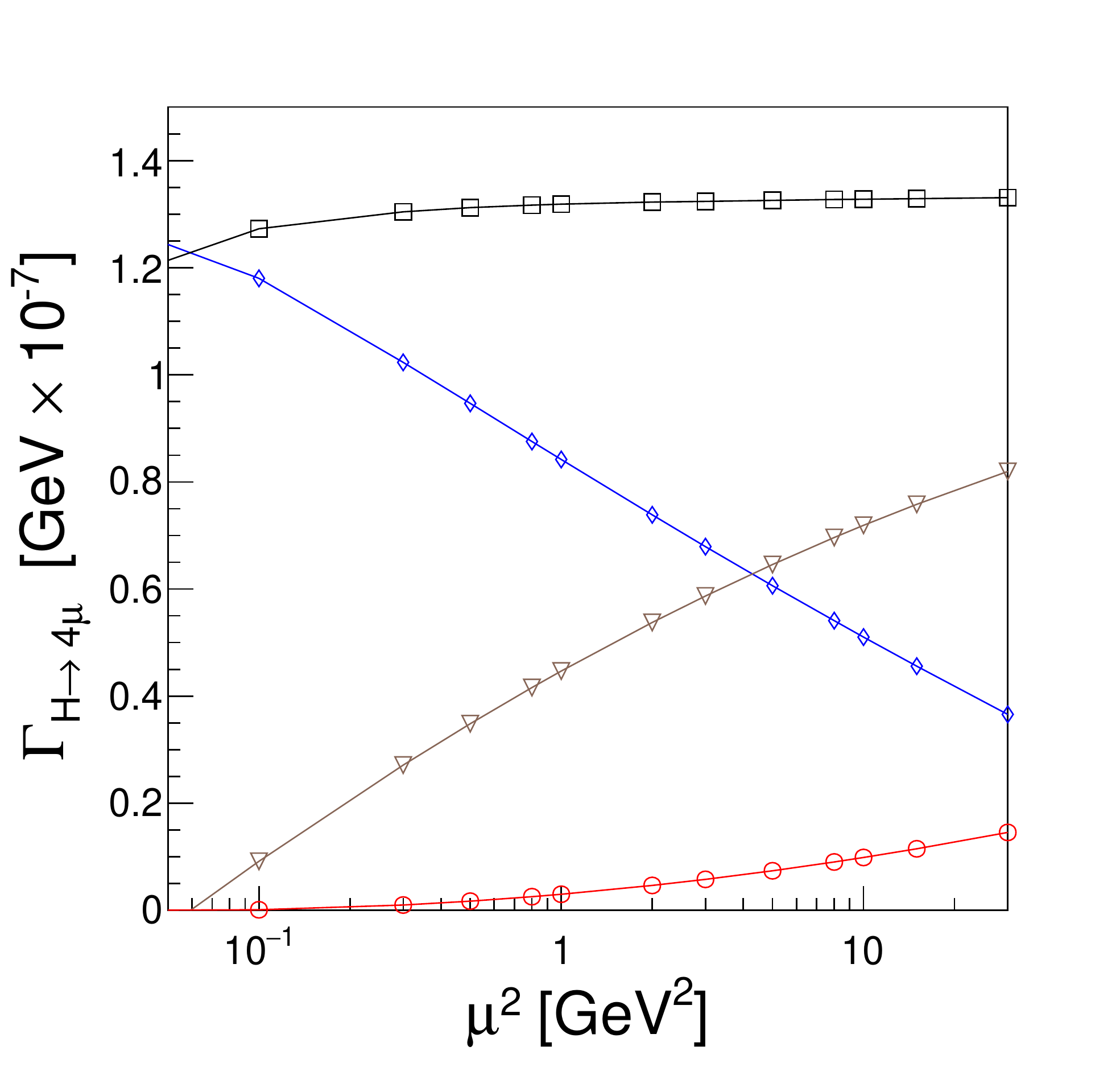}
}
\caption{
The partial decay width of the process $H\to2e2\mu$ calculated using Eq.~(\ref{eq:result2l2l}) (left), 
$H\to4e$ (middle) and $H\to4\mu$ (right) calculated using Eq.~(\ref{eq:result4l}) with the couplings discussed in text. 
The first term in each equation is shown with the requirement $q_{2\ell}^2 \ge \mu^2$
common for electrons and muons (blue diamonds, $\diamond$). 
The following terms multiplied by the $\mu^2$-dependent factors are also shown:
the $\Gamma_{H\to \gamma\gamma}$  (red circles, $\circ$), 
the $\Gamma_{H\to 2e\gamma}$ with the requirement $q_{2e}^2 \ge \mu^2$ (magenta triangles pointing down, $\bigtriangledown$), and 
the $\Gamma_{H\to 2\mu\gamma}$ with the requirement $q_{2\mu}^2 \ge \mu^2$ (brown triangles pointing up, $\bigtriangleup$).
The sum of the cross sections is also shown (black squares, $\square$). 
}
\label{fig:test-lowq2-4l}
\end{figure}

As before, we scan the value of cutoff $\mu^2$ in Eq.~(\ref{eq:result2l2l}) for the $H\to 2e2\mu$ process and
in Eq.~(\ref{eq:result4l}) for the $H\to 4e$ and $4\mu$ processes from the 
threshold value of $(2m_\ell)^2$ up to $(5\,{\rm GeV})^2$. The threshold value is defined for muons 
in the case of $H\to 4\mu$ and electrons in the other two cases. 
Figure~\ref{fig:test-lowq2-4l} shows the partial decay width of the process $H\to2e2\mu$ calculated using Eq.~(\ref{eq:result2l2l}),
and $H\to4e$ and $4\mu$ calculated using Eq.~(\ref{eq:result4l}). Several terms in the corresponding equations are isolated:
the first term in each equation is shown with the requirement $q_{2\ell}^2 \ge \mu^2$ common for electrons and muons; 
the $\Gamma_{H\to \gamma\gamma}$ multiplied by the $\mu^2$-dependent factor, 
the $\Gamma_{H\to 2e\gamma}$ with the requirement $q_{2e}^2 \ge \mu^2$, and 
the $\Gamma_{H\to 2\mu\gamma}$ with the requirement $q_{2\mu}^2 \ge \mu^2$, 
where both $\Gamma_{H\to 2\ell\gamma}$ are multiplied by the $\mu^2$-dependent factor as well. 
With the $\mu^2$ increasing, the first term falls while the other terms generally rise. Again, we obtain a constant 
cross section of the four-lepton process, as one would expect to observe for the proper modeling of the effect. 

The combination of formulas in 
Eqs.~(\ref{eq:result2lgamma}) and~\ref{eq:result2lgamma-had} for $\Gamma_{H\to 2f\gamma}$
and Eqs.~(\ref{eq:result2l2l}), (\ref{eq:result4l}), and similar ones involving hadronic jets
for $\Gamma_{H\to 4f}$ should allow one to handle low-$q^2$ singularities and hadronic 
structure with efficient numerical evaluation. 


\section{Expected constraints on the couplings with photons}
\label{sect:eft-analysis}

We continue by investigating the on-shell production and decay of the \Hboson with its couplings to weak vector bosons 
in the VBF, $VH$, and $H\to VV\to 4\ell$ processes. There has already been extensive study of the $HVV$ couplings, 
both by experimental collaborations and in phenomenological work. However, there was no conclusive study 
on the effects of the photon contribution in the production topology. In our previous work in Ref.~\cite{Gritsan:2020pib}, 
we pointed out that the decays $H\to \gamma\gamma$ and $Z\gamma$ with on-shell photons provide constraints 
on the $g_2^{\gamma\gamma}$, $g_4^{\gamma\gamma}$, $g_2^{Z\gamma}$, and $g_4^{Z\gamma}$ couplings 
which are stronger than those that can be obtained from the VBF, $VH$, and $H\to 4\ell$ processes. For this reason, 
the analysis of multiple operators was simplified by setting those four couplings to zero. The constraints from
the decays with on-shell photons can be illustrated with the simplified partial decay width expressions in 
Eqs.~(\ref{eq:ratio-gammagamma-2}, \ref{eq:ratio-zgamma-2}). This effect in the $H\to 4\ell$ was studied
with LHC data~\cite{Khachatryan:2014kca} and with phenomenological tools~\cite{Chen:2012jy,Chen:2014gka}. 
In the following we re-examine the $H\to 4\ell$ decay and investigate the VBF and $VH$ processes. 

First, we would like to point to the effect already observed in Tables~\ref{tab:warsaw-4l}, \ref{tab:warsaw-VBF}, and~\ref{tab:warsaw-VH}.
Let us focus on any of the three operators  $C_{\PH\PW}$, $C_{\PH\PW{B}}$, or $C_{\PH{B}}$   
where the $g_2^{ZZ}$, $g_2^{Z\gamma}$, and  $g_2^{\gamma\gamma}$ couplings contribute, 
or equivalently any of $C_{\PH\widetilde{\PW}}$,  $C_{\PH\widetilde{\PW}B}$, or $C_{\PH\widetilde{B}}$,
where $g_4^{ZZ}$, $g_4^{Z\gamma}$, and  $g_4^{\gamma\gamma}$  contribute. 
While the $g_{2,4}^{ZZ}$ contributions to the VBF and $VH$ processes are comparable and sometimes
even dominant for a given $C_{\PH X}$ operator, their contributions to the $H\to 4\ell$ decay appear to be
negligible in comparison.  For photon couplings, the reverse is the case: their contribution to decay
is much larger than to production.
Therefore, the photon couplings have relatively higher importance in decay compared
to the production processes. This still does not tell us if the photon couplings are better constrained in 
production or decay, and this is what we investigate below. 

In the following, prospects with either 3000\,fb$^{-1}$ (HL-LHC) or 300\,fb$^{-1}$ (full LHC) are studied at a 13 TeV collision energy. 
We use the \textsc{JHUGen} simulation to model the VBF, $WH$, $ZH/\gamma^*H$, $\gamma H$,  
and gluon fusion production with the decay $H\to 4\ell$.
We include the effective background with the \textsc{POWHEG}~\cite{Frixione:2007vw}
simulation of the $q\bar{q}\to 4\ell$ process, which we scale to match the contributions of other processes 
as found in experiment, such as gg~$\to 4\ell$ and Drell-Yan $Z$ production~\cite{Sirunyan:2021rug}. 
The detector effects are modeled with ad-hoc acceptance selection and empirical efficiency corrections, 
and the lepton and hadronic jet momenta are smeared to achieve realistic resolution effects. 
The following selection requirements are applied: 
$p_T^e > 5$\,GeV, $p_T^\mu > 7$\,GeV, $p_T^\mathrm{jet} > 30$\,GeV, 
$|\eta^e|<2.5$, $|\eta^\mu|<2.4$, $|\eta^\mathrm{jet} |<4.7$, 
$|m_{\ell\ell}|>12$\,GeV, $|m_{4\ell} - m_H| < 3.5$\,GeV. 

We target the optimal analysis of four anomalous couplings expressed through the fractional contributions to the $\PH\to 2e2\mu$ 
process $f_{g2}^{Z\gamma}$, $f_{g2}^{\gamma\gamma}$, $f_{g4}^{Z\gamma}$, and $f_{g4}^{\gamma\gamma}$, 
with the approach similar to Ref.~\cite{Gritsan:2020pib}. 
The cross section fractions are defined following 
\begin{eqnarray}
 f_{gn} =  \frac{g_{n}^2 \alpha_{nn}^{(f)}}{\sum_{j}{g_{j}^2 \alpha_{jj}^{(f)} }}
 ~\mathrm{sign}\left(\frac{g_n}{g_{1}} \right),
\label{eq:fgn}
\end{eqnarray}
where the $\alpha_{nn}^{(f)}$ coefficients are introduced in Eq.~(\ref{eq:diff-cross-section2}) for the final state $(f)$. 
The numerical values of these coefficients are given in Table~\ref{tab:ratios}, where they are normalized
with respect to the $\alpha_{11}$ coefficient, corresponding to the cross section calculated for $g_{1}=1$. 
The $\alpha_{nn}$ are the cross sections for $g_{n}=1$. In Table~\ref{tab:ratios}, we also quote the cross section
ratios defined in VBF and $VH$ production for comparison. As noted above and evident from this table, 
the relative importance of photon couplings is higher in decay compared to production. 

\begin{table}[!t]
\captionsetup{justification=centerlast}
\caption{
List of anomalous $HZ\gamma$ and $H\gamma\gamma$ couplings $g_n$, cross section fractions $f_{gn}$, 
and the cross section ratio $\alpha_{nn}/\alpha_{11}$ defined in $H\to2e2\mu$ decay, VBF, and $q\bar{q}\to ZH/\gamma^*H$ production. 
For comparison, the $g_4^{ZZ}$ coupling is also shown. 
The requirements $m_{\ell\ell}>4$ GeV and $p_T^\mathrm{jet} >1$ GeV are introduced in 
$\alpha_{nn}^{(f)}$ and $\alpha_{nn}^{(i)}$ calculation by convention. 
}
\label{tab:ratios}
\begin{tabular}{ccccc}
\vspace{-0.3cm} \\
\hline
  \multicolumn{1}{c}{Coupling}         &  \multicolumn{1}{c}{Fraction}             
            &  \multicolumn{1}{c}{$H\to2e2\mu$} &  \multicolumn{1}{c}{VBF} &  \multicolumn{1}{c}{$ZH/\gamma^*H$}  \\
  \multicolumn{1}{c}{$g_n$}    &  \multicolumn{1}{c}{$f_{gn}$}   
            & \multicolumn{1}{c}{$\alpha_{nn}^{(f)}/\alpha_{11}$} & \multicolumn{1}{c}{$\alpha_{nn}^{(i)}/\alpha_{11}$} & \multicolumn{1}{c}{$\alpha_{nn}^{(i)}/\alpha_{11}$}   \\
  \hline
 $g_2^{\gamma\gamma}$ & $f_{g2}^{\gamma\gamma}$ & $355.1$ & $65.04$  & $2.330$ \\
 $g_2^{Z\gamma}$ & $f_{g2}^{Z\gamma}$                    & $438.5$ & $24.89$  & $50.51$ \\
 $g_4^{\gamma\gamma}$ & $f_{g4}^{\gamma\gamma}$ & $348.0$ & $64.28$  & $1.790$ \\
 $g_4^{Z\gamma}$ & $f_{g4}^{Z\gamma}$                     & $356.7$ & $23.44$  & $32.50$ \\
 $g_4^{ZZ}$ & $f_{g4}^{ZZ}$                     & $0.153$ & $11.27$  & $47.94$ \\
  \hline
 \end{tabular}
\end{table}

For simplicity, we set the contributions of the other anomalous contributions to zero: 
$f_{g2}^{ZZ}=f_{g4}^{ZZ}=f_{\Lambda1}^{ZZ}=f_{\Lambda1}^{Z\gamma}=0$. 
This assumption will provide tighter constraints than one could achieve otherwise, but this 
is sufficient for comparison to the precision obtained from $H\to\gamma\gamma$ and $Z\gamma$
following Eqs.~(\ref{eq:ratio-gammagamma-2}) and~(\ref{eq:ratio-zgamma-2}).
We use the relationship with $\Delta{M}_W=0$ in Eq.~(\ref{eq:deltaMW}) for the SM-like contribution, 
but we do not enforce the $SU(2)\times U(1)$ symmetry in Eqs.~(\ref{eq:g2WW}--\ref{eq:kappa2Zgamma})
and instead set $g_2^{WW}=g_4^{WW}=\kappa_1^{WW}=\kappa_2^{Z\gamma}=0$. This is done to isolate the contributions with 
genuine photon couplings, which exhibit the features of virtual photons. These are the contributions which also appear
and are constrained in the $H\to Z\gamma$ and $\gamma\gamma$ processes. 

\subsection{Expected constraints on photon couplings from $H\to VV\to4\ell$ decay}
\label{subsec:decay}

We build the analysis following the MELA approach with the two types of optimal discriminants, using the 
full kinematic information in the four-lepton decay. 
There are four discriminants ${\cal D}_{g2}^{Z\gamma}$, ${\cal D}_{g2}^{\gamma\gamma}$, ${\cal D}_{g4}^{Z\gamma}$, 
and ${\cal D}_{g4}^{\gamma\gamma}$ which are designed to separate the pure anomalous contributions from the SM-like,
and there are four interference  discriminants ${\cal D}_{\rm int}^{Z\gamma}$, ${\cal D}_{\rm int}^{\gamma\gamma}$, 
${\cal D}_{CP}^{Z\gamma}$, ${\cal D}_{CP}^{\gamma\gamma}$ which isolate interference of the SM with the same four
anomalous contributions. The full available information is used in calculating the discriminants,
and further details on the MELA approach can be found in Ref.~\cite{Gritsan:2020pib} and references therein. 
In this and further analyses discussed below, events are split into the $\PH\to2e2\mu$, $4e$, and $4\mu$ categories, 
which is an important aspect because the relative fractions of events between these types change with anomalous 
couplings, due to interference of diagrams with identical leptons in the $4e$ and $4\mu$ final states. 
In the end, for each event in a category $j$, a set of observables ${\bf x}$ is defined. 

Since analysis of decay information is essentially independent from the production mechanism, we model kinematic
distributions using simulation of the gluon fusion process. However, in Section~\ref{subsec:vbf} we will also show 
how the full production and decay information can be taken into account. 
The probability density function for the signal decay process in gluon fusion production, before proper normalization, 
is defined as
\begin{eqnarray}
{\rm ggH}: ~~~ &&
\mathcal{P} \left({\bf x}; \vec{f} \,\right) 
\propto\sum_{\substack{k,l=1\\k\le l}}^K
\mathcal{P}_{kl}\left({\bf x}\right)
\sqrt{|f_{gk} \cdot f_{gl}|} ~\mathrm{sign}(f_{gk} \cdot f_{gl})\,,
\label{eq:psignalshorter}
\end{eqnarray}
where ${\bf x}$ are the observables and $f_{gn}$ are the cross-section fractions of the couplings,
$K=5$ for the four anomalous couplings and one SM coupling. 
Equation~(\ref{eq:psignalshorter}) is obtained from Eq.~(\ref{eq:diff-cross-section2}), 
where the width and $H\mathrm{gg}$ couplings are absorbed into the overall normalization. 
In gluon fusion production, the electroweak $HVV$ couplings appear only in decay.
Therefore, there are 15 terms in Eq.~(\ref{eq:psignalshorter}).
To populate the probability distributions, we use a simulation of unweighted events of several samples that adequately 
cover the phase space and re-weight those samples using the MELA package to parameterize the other terms.
As in Ref.~\cite{Gritsan:2020pib}, we implement a cutting planes algorithm~\cite{cuttingplanes} using the
Hom4PS~\cite{HomotopyContinuation,Hom4PS1,Hom4PS2} and Gurobi~\cite{gurobi} programs to ensure that the
probability density function remains positive definite for all possible values of $f_{gn}$.
With 3000\,fb$^{-1}$ data at 13 TeV, we expect about 4500 events reconstructed in the $H\to4\ell$ channel. 
In Fig.~\ref{fig:fg_scan} we show the expected constraints on the four parameters of interest
expressed through effective factions
$f_{g2}^{\gamma\gamma}$, $f_{g4}^{\gamma\gamma}$, $f_{g2}^{Z\gamma}$, and $f_{g4}^{Z\gamma}$.
However, before discussing the results, we should turn to analysis of production information. 

\begin{figure}[t]
\centering
\includegraphics[width=0.24\textwidth]{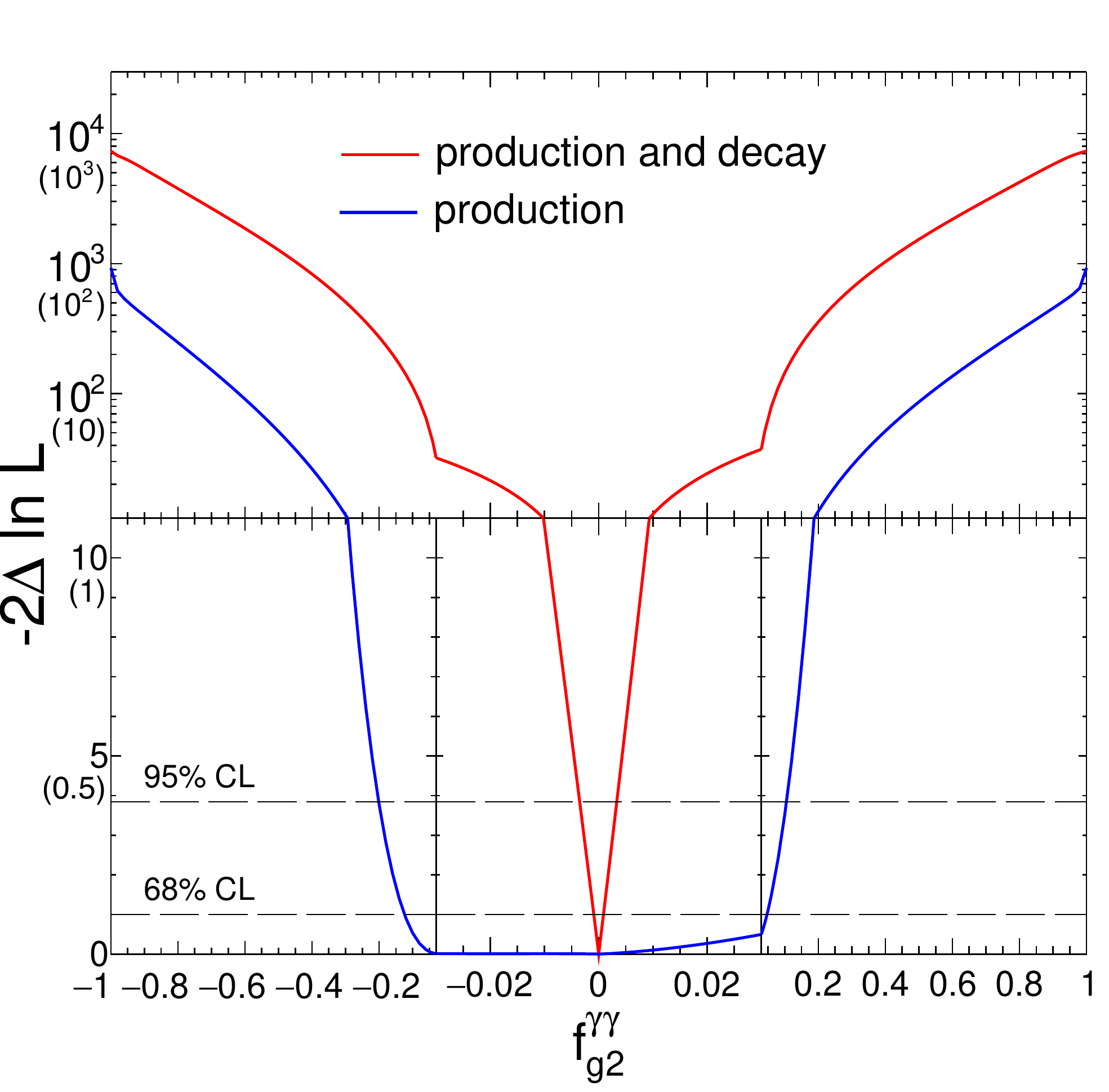}
\includegraphics[width=0.24\textwidth]{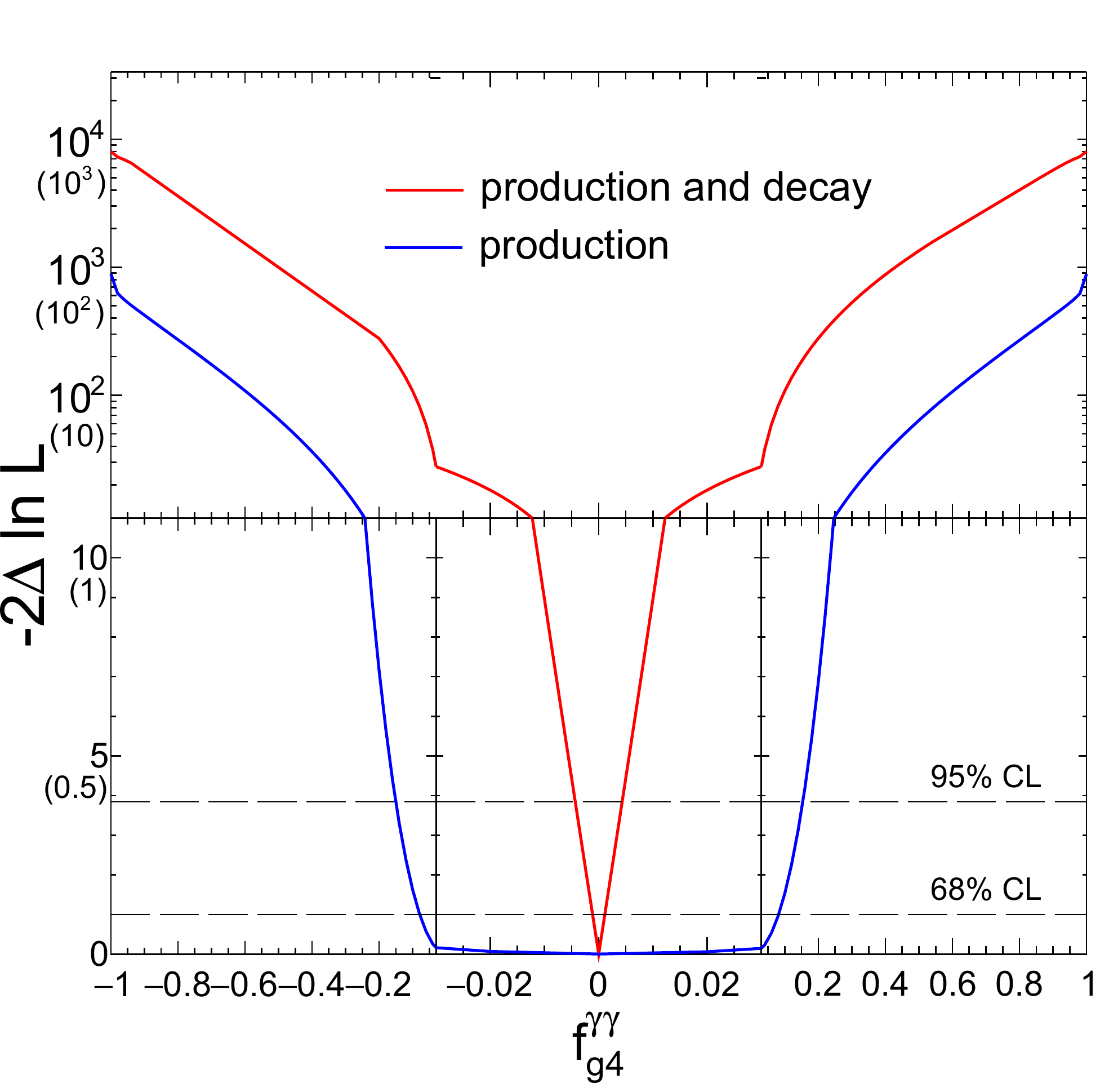}
\includegraphics[width=0.24\textwidth]{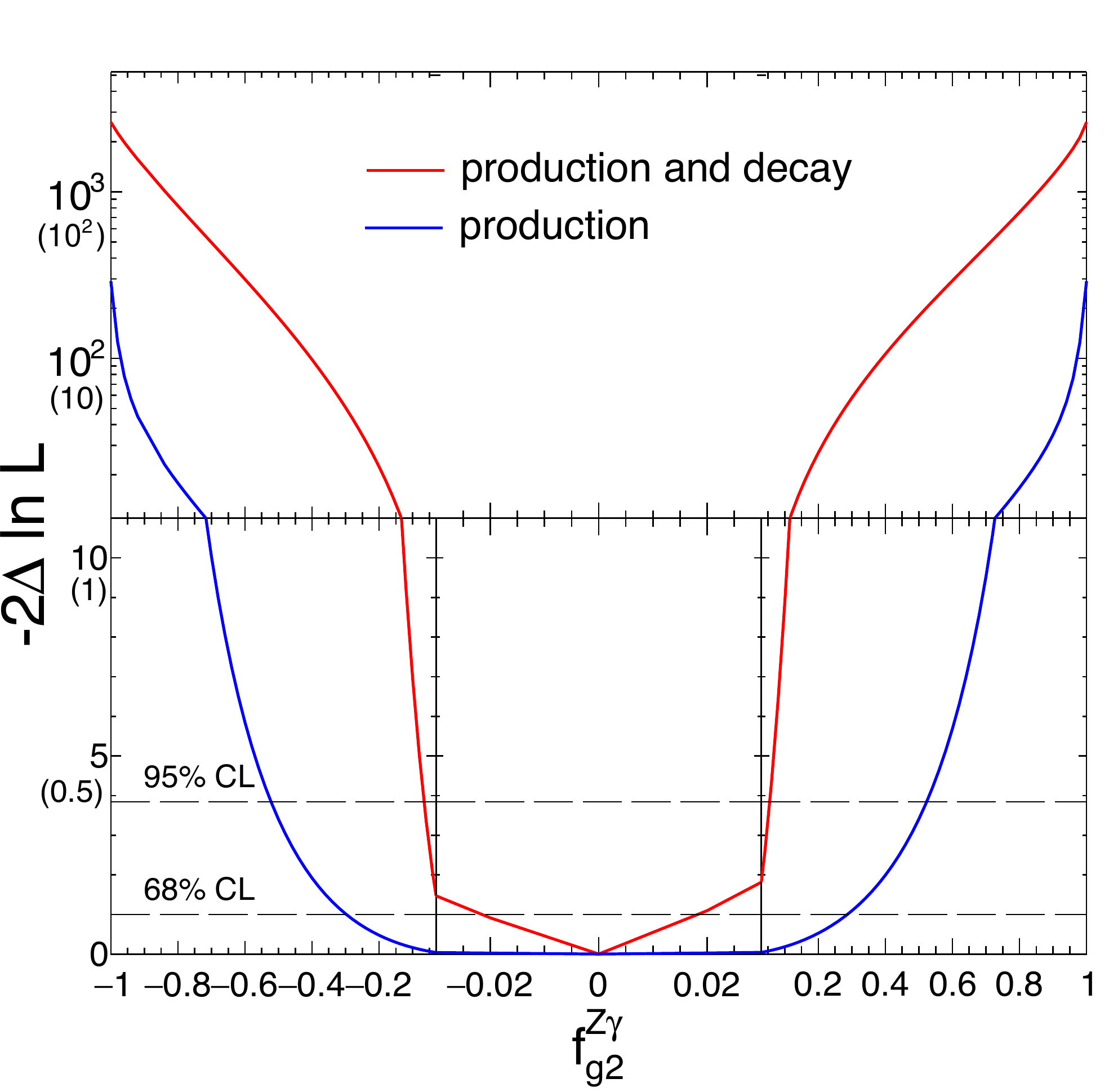}
\includegraphics[width=0.24\textwidth]{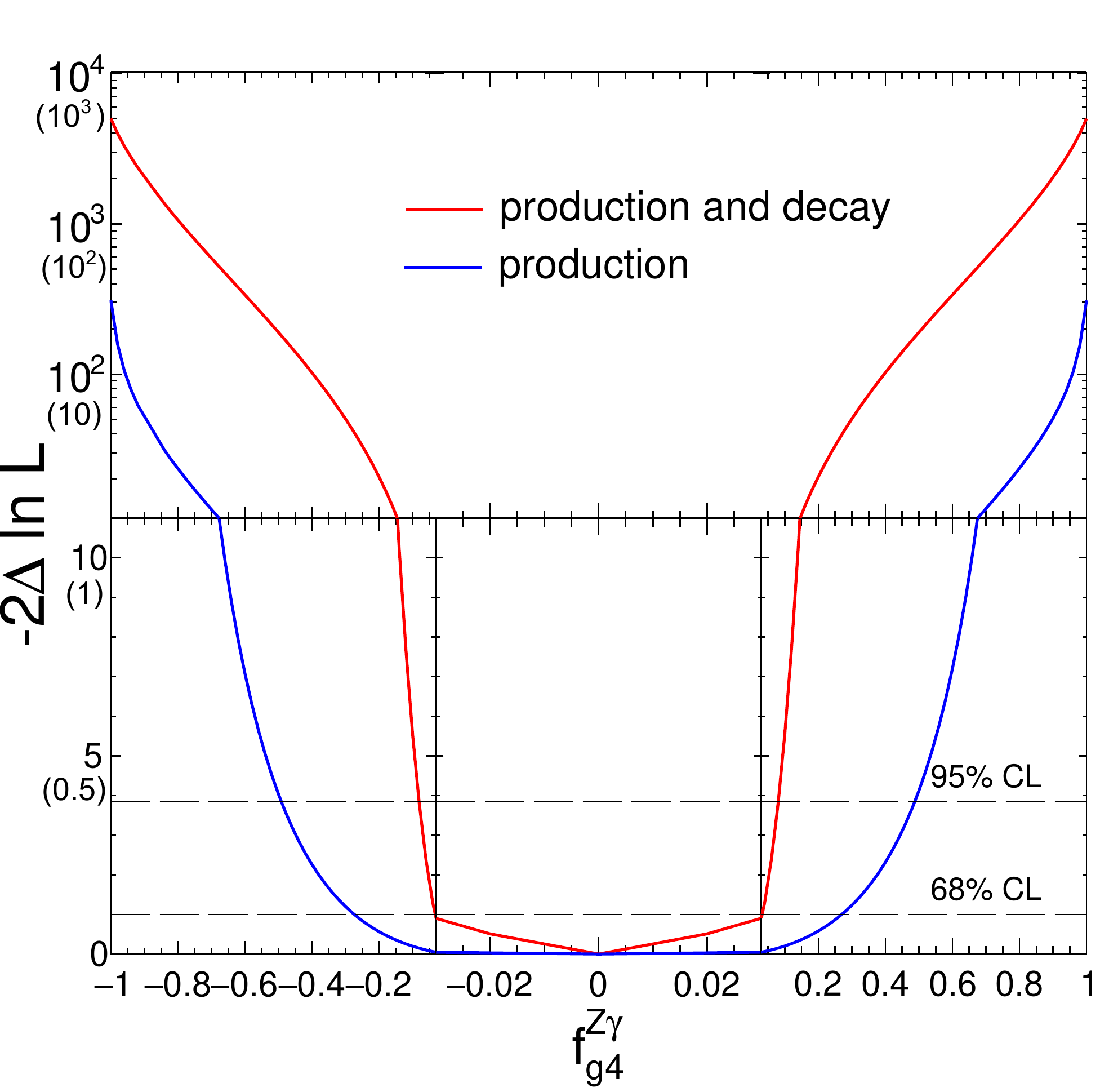}
\captionsetup{justification=centerlast}
\caption{
Expected constraints from a simultaneous fit of 
$f_{g2}^{\gamma\gamma}$, $f_{g4}^{\gamma\gamma}$, $f_{g2}^{Z\gamma}$, and $f_{g4}^{Z\gamma}$
using associated production and $\PH\to4\ell$ decay with 3000 (300)\,fb$^{-1}$ data at 13 TeV.
Two scenarios are shown: using MELA observables with production and decay (red) or production only (blue) information.
The dashed horizontal lines show the 68 and 95\% confidence level (CL) regions.
}
\label{fig:fg_scan}
\end{figure}

\begin{figure}[t]
\centering
\includegraphics[width=0.32\textwidth]{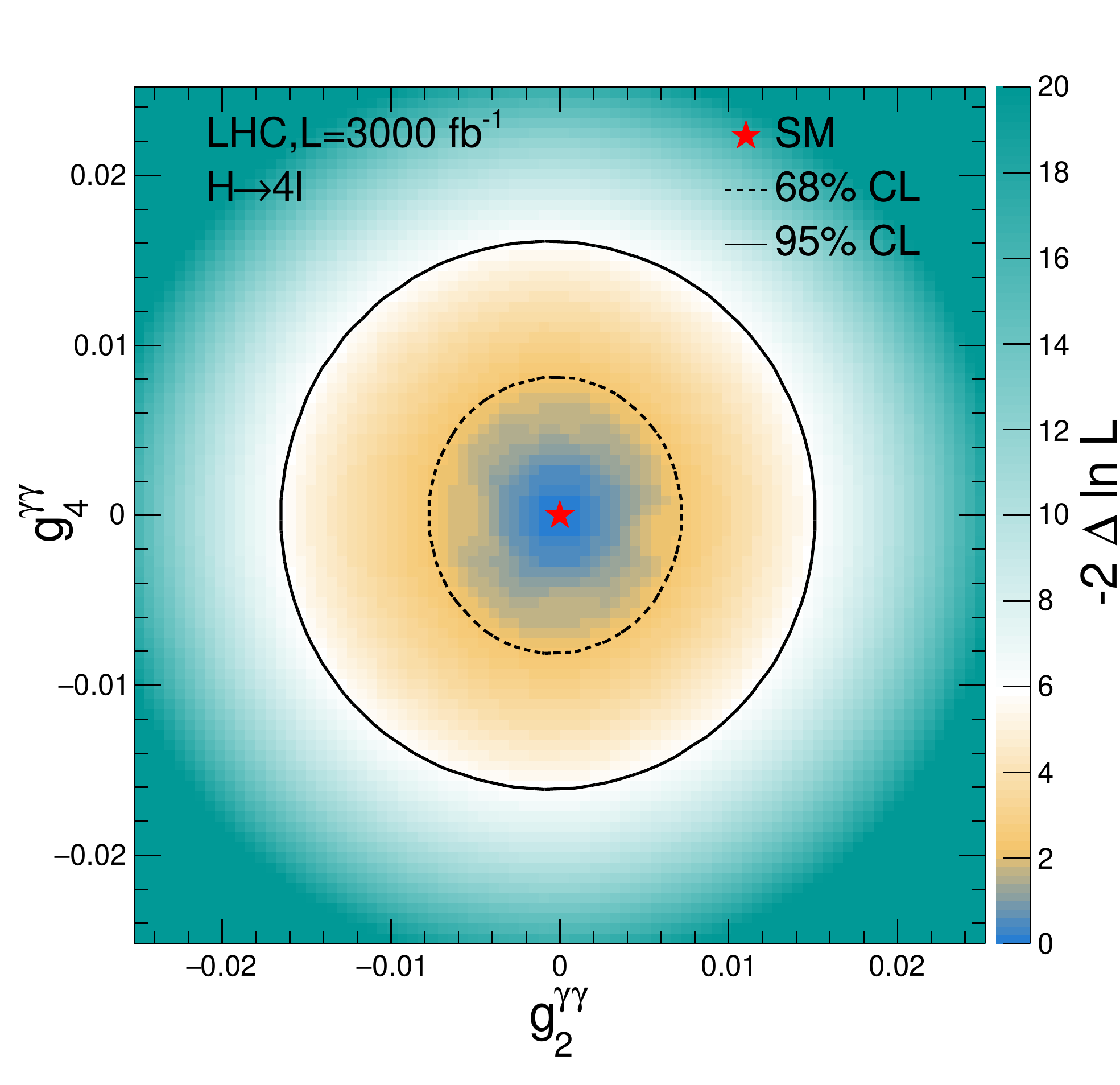}
\includegraphics[width=0.32\textwidth]{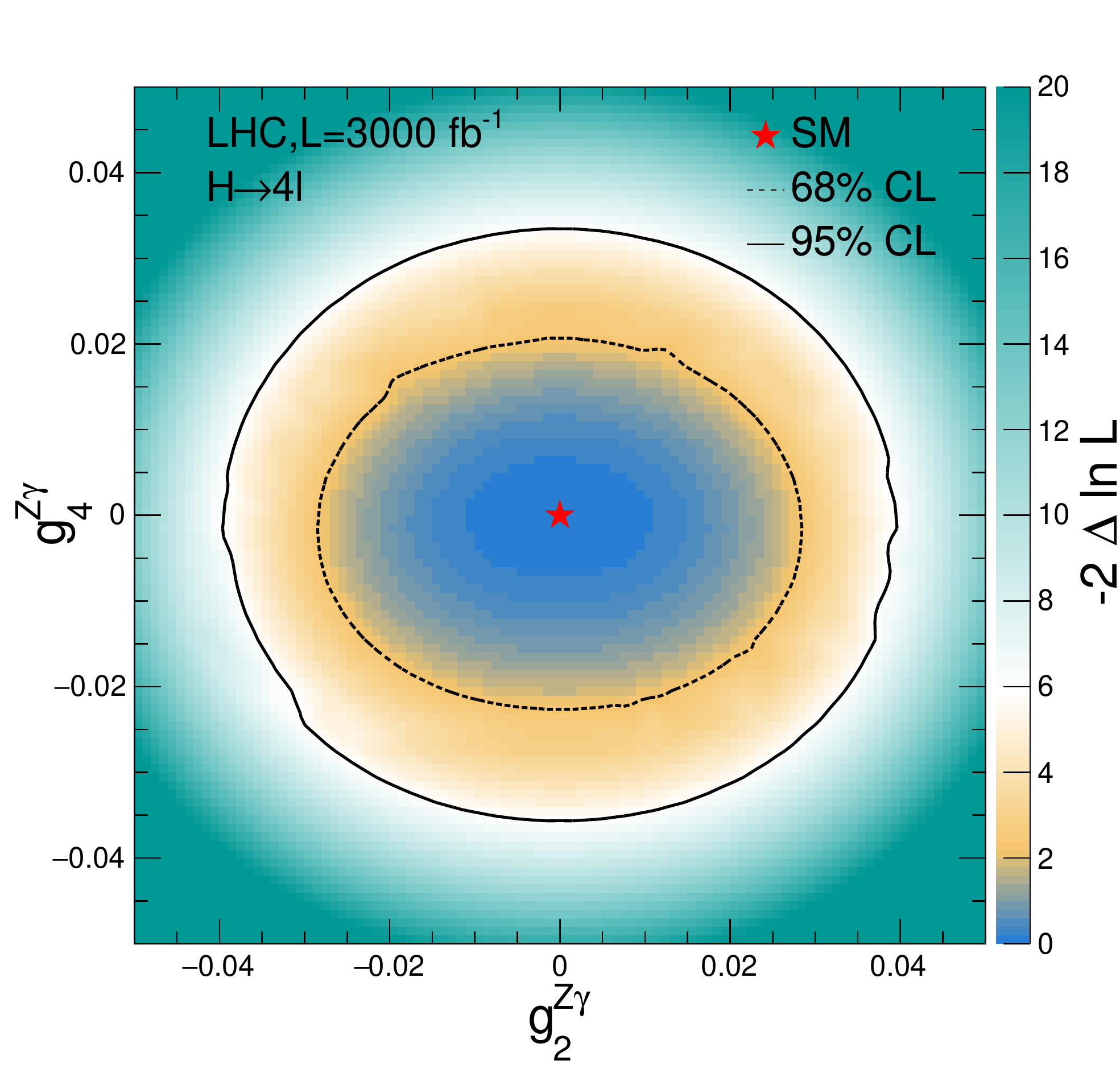}
\includegraphics[width=0.32\textwidth]{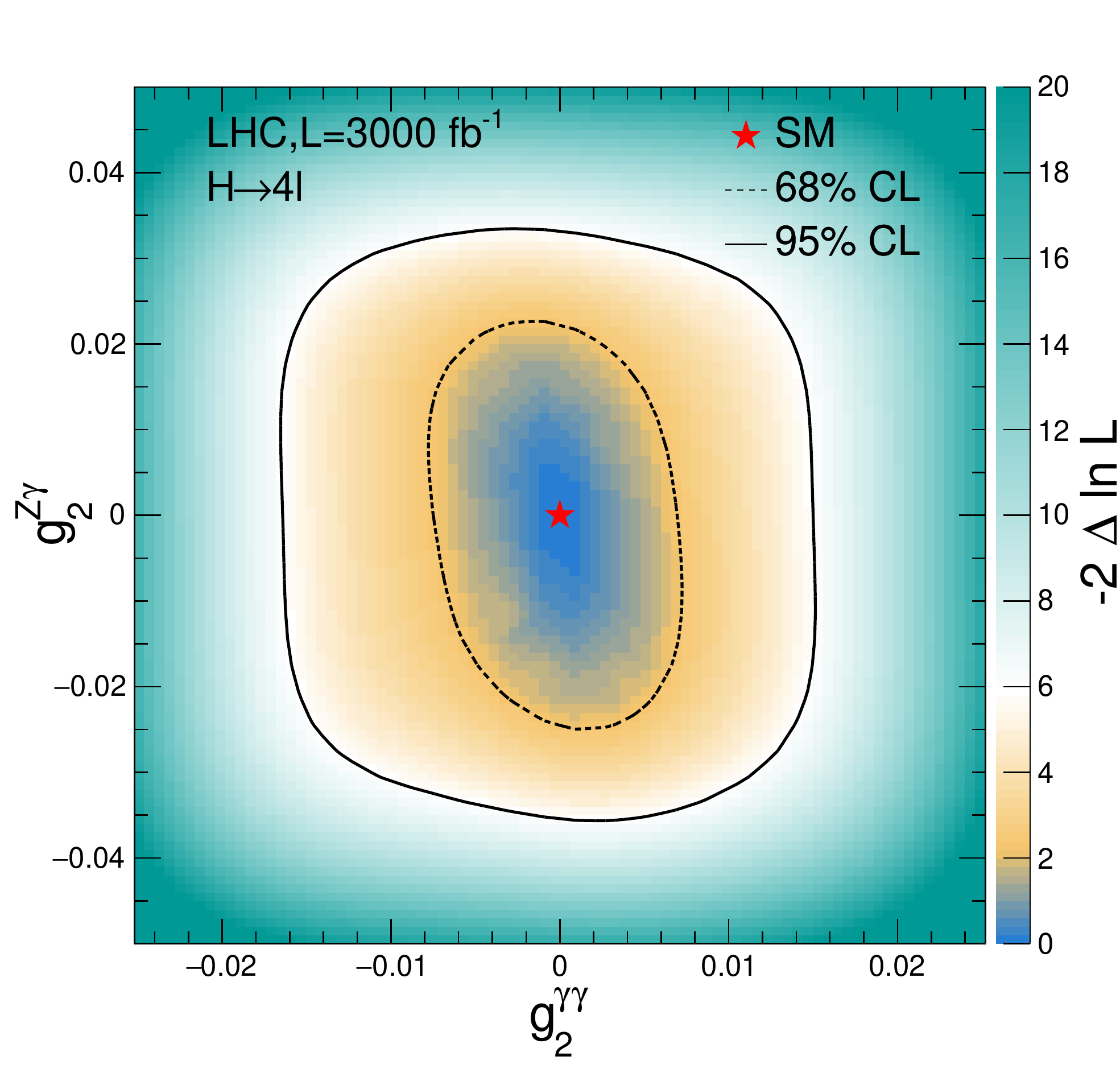}
\captionsetup{justification=centerlast}
\caption{
Expected two-dimensional constraints from a simultaneous fit of 
$g_2^{\gamma\gamma}$, $g_4^{\gamma\gamma}$, $g_2^{Z\gamma}$, and $g_4^{Z\gamma}$
using decay and production information as shown in Fig.~\ref{fig:fg_scan}.
}
\label{fig:g_scan_decay}
\end{figure}

\subsection{Expected constraints on photon couplings from VBF and $VH$ production}
\label{subsec:vbf}

While analysis of the $\PH\to4\ell$ decay can be performed inclusively, the VBF and $VH$ production channels
require analysis of associated particles. Therefore, we follow the approach adopted in our 
earlier studies~\cite{Gritsan:2020pib} to categorize events using the jet information to create the two-jet 
categories of events enhanced in either VBF or $VH$ events with hadronic decay of the $V$, respectively. 
In addition, the leptonic $VH$, one-jet VBF, and boosted 
categories are defined, where $p_T^{4\ell} >120$\,GeV is used to select the latter~\cite{Gritsan:2020pib}. 
The remaining events constitute the untagged category. 

We build the analysis following the MELA approach with the two types of optimal discriminants, as discussed
for decay above, using the full kinematic information in both the production and the four-lepton decay. 
In the VBF and $VH$ topologies with two associated jets, both production and decay information are used,
except for the interference discriminants, where production information is chosen.
In the untagged category, the $H\to4\ell$ information is used as in Section \ref{subsec:decay},
and in the other categories the transverse momentum $p_T^{4\ell}$ is used.

In the case of the VBF and $VH$ processes, the $HVV$ coupling appears on both the production and decay sides.
Therefore, the amplitude squared has a product of four couplings. 
Equation~(\ref{eq:psignal}) is obtained from Eq.~(\ref{eq:diff-cross-section2})
and has 70 terms:
\begin{eqnarray}
{\rm VBF}, VH: ~~~ &&
\mathcal{P} \left({\bf x}; \vec{f} \,\right) 
\propto\sum_{\substack{k,l,m,n=1\\k\le l\le m\le n}}^K
\mathcal{P}_{klmn}\left({\bf x}\right)
\sqrt{|f_{gk} \cdot f_{gl}\cdot f_{gm}\cdot f_{gn}|} ~\mathrm{sign}(f_{gk} \cdot f_{gl}\cdot f_{gm}\cdot f_{gn})\,,
\label{eq:psignal}
\end{eqnarray}
where the notation follows Eq.~(\ref{eq:psignalshorter}). 

In addition to the joint analysis of production and decay, we also perform an analysis with production information only.
In order to achieve this, no decay information is used in the construction of discriminants, and only the total yield of
events is used in the untagged category. It is not possible to completely 
decouple the analysis from decay information, as for example the relative fraction of $2e2\mu$ events 
is sensitive to couplings in the decay amplitude. 
For example, there is a strong destructive interference between the two diagrams with permutation of identical leptons
in the $H\to4e / 4\mu$ decay with the $H\gamma\gamma$ couplings, as illustrated in Fig.~\ref{fig:Lexicon-H4l} 
and Section~\ref{sect:eft-basis}, which leads to a modification of the $2e2\mu$ yield faction. 
However, such dependence on anomalous couplings is greatly reduced with consideration of yields only. 

In Fig.~\ref{fig:fg_scan} we show the expected constraints on the four parameters of interest
$f_{g2}^{\gamma\gamma}$, $f_{g4}^{\gamma\gamma}$, $f_{g2}^{Z\gamma}$, and $f_{g4}^{Z\gamma}$
with 300 and 3000\,fb$^{-1}$ data at 13 TeV, where for comparison constraints from production information 
alone are shown separately from the full constraints using both decay and production.
The information contained in decay can be deduced from the difference of the two constraints, 
with one exception. The analysis of production information is sensitive to the relative fraction 
of the $\PH\to2e2\mu$, $4e$, and $4\mu$ events. We observe that in the case of the
$\PH\to\gamma^*\gamma^*\to 4e$ and $4\mu$ processes, there is a sizable effect of interference
between diagrams with permutations of the leptons. This effect in decay competes with information from production
in constraining $g_{2}^{\gamma\gamma}$ and  $g_{4}^{\gamma\gamma}$. We do not find such an 
effect to be important in constraining $g_{2}^{Z\gamma}$ and  $g_{4}^{Z\gamma}$.

What we observe is that the constraints from decay are significantly more powerful than from production
when both are analyzed with the same channel $H\to4\ell$. This happens for two reasons. 
First of all, all reconstructed events contain decay information, while only a small fraction
carry production information. Second, the ratio $\alpha_{nn}^{(i)}/\alpha_{11}$ 
is reduced compared to the same ratio in decay $\alpha_{nn}^{(f)}/\alpha_{11}$
for the photon couplings. This effect is in contrast to the trend observed for the $ZZ$ couplings
(see, for example, Fig.\,10 of Ref.\,\cite{Gritsan:2020pib}), as indicated 
with the $g_4^{ZZ}$ example in Table~\ref{tab:ratios}. This trend explains the tighter constraints on the 
anomalous $ZZ$ couplings using production information, as opposed to decay. For the same reason, 
the constraints on the photon couplings ($H\gamma\gamma$ and $HZ\gamma$) are tighter in decay compared
to production.

In Fig.~\ref{fig:g_scan_decay}, these results for the joint analysis of production and decay are interpreted in terms 
of constraints on the $g_i$ couplings. As indicated above, these constraints are dominated by decay information, 
and the results would look similar if only decay information were employed. 
This interpretation uses the full expression in Eq.~(\ref{eq:diff-cross-section2}), including the total \Hboson 
width dependence on anomalous couplings appearing in the denominator, using expressions 
obtained in Section~\ref{sect:eft-xs}. It is assumed that there no decays of the \Hboson to unknown particles. 

We should point out that while decay information is limited to $H\to VV$ channels only, production 
information can be obtained by combining all possible decay channels of the \Hboson, such as 
$H\to 4\ell, \gamma\gamma, b\bar{b}, \tau^+\tau^-,$ and $W^+W^-$. In this study, we investigate 
only the $H\to 4\ell$ channel, but the relative importance of production information will increase as
other channels are analyzed. 
This observation is also valid for analysis of the $\gamma H$ production, discussed next. 
At the same time, the $H\to4\ell$ channel can also be further optimized for the measurements of 
the photon couplings by relaxing the invariant mass and transverse momentum constraints on the leptons.
For example, the requirement $|m_{\ell\ell}|>12$\,GeV can be relaxed to $|m_{\ell\ell}|>4$\,GeV,
or even further, with significant gain in sensitivity to the $H\gamma\gamma$ and $HZ\gamma$ 
couplings, as can be seen from the invariant mass $m_1$ and $m_2$ distributions in Fig.~\ref{fig:Lexicon-H4l}.

\subsection{Expected constraints on photon couplings from $\gamma H$ production}
\label{subsec:gammaH}

Setting constraints on or measuring the rate of the $\gamma H$ production is interesting on its own,
as this production mechanism of the \Hboson has not been tested on the LHC, in part because its SM
rate is not accessible yet. In addition, we would like to assess the feasibility of the 
$g_2^{\gamma\gamma}$, $g_4^{\gamma\gamma}$, $g_2^{Z\gamma}$, and $g_4^{Z\gamma}$
coupling measurement in this production process. 
In order to simplify these estimates and because NLO EW event simulation is not available for the $\gamma H$ process yet, 
we assume that the SM contributions are small compared to the accessible values and thus can be absorbed into these 
effective point-like couplings. The validity of these assumptions can be checked against the expected constraints.
Single loop calculations of the SM production cross section of $q\bar{q} \rightarrow \gamma H$ 
predict a cross section of less than 5\,fb at the LHC~\cite{Abbasabadi:1997zr}, which corresponds to less than 
two $H\to4\ell$ event at the HL-LHC before any selection requirements are applied.

Experimentally, the main distinguishing feature of the $\gamma H$ production mechanism is a high-momentum 
isolated photon, and we found that the requirement $p_T^\gamma > 400$\,GeV keeps about
12\% of signal events, which corresponds to about 0.9 signal events for $g_2^{\gamma\gamma}=0.1$
and 0.11 background events with $H\to4\ell$ and 3000\,fb$^{-1}$ data at 13 TeV. 
Therefore, for feasibility studies, we identify an additional category of events with a good
isolated photon candidate associated with the \Hboson candidate and the above  $p_T^\gamma$ requirement. 
 
\begin{figure}[t]
\centering
\includegraphics[width=0.32\textwidth]{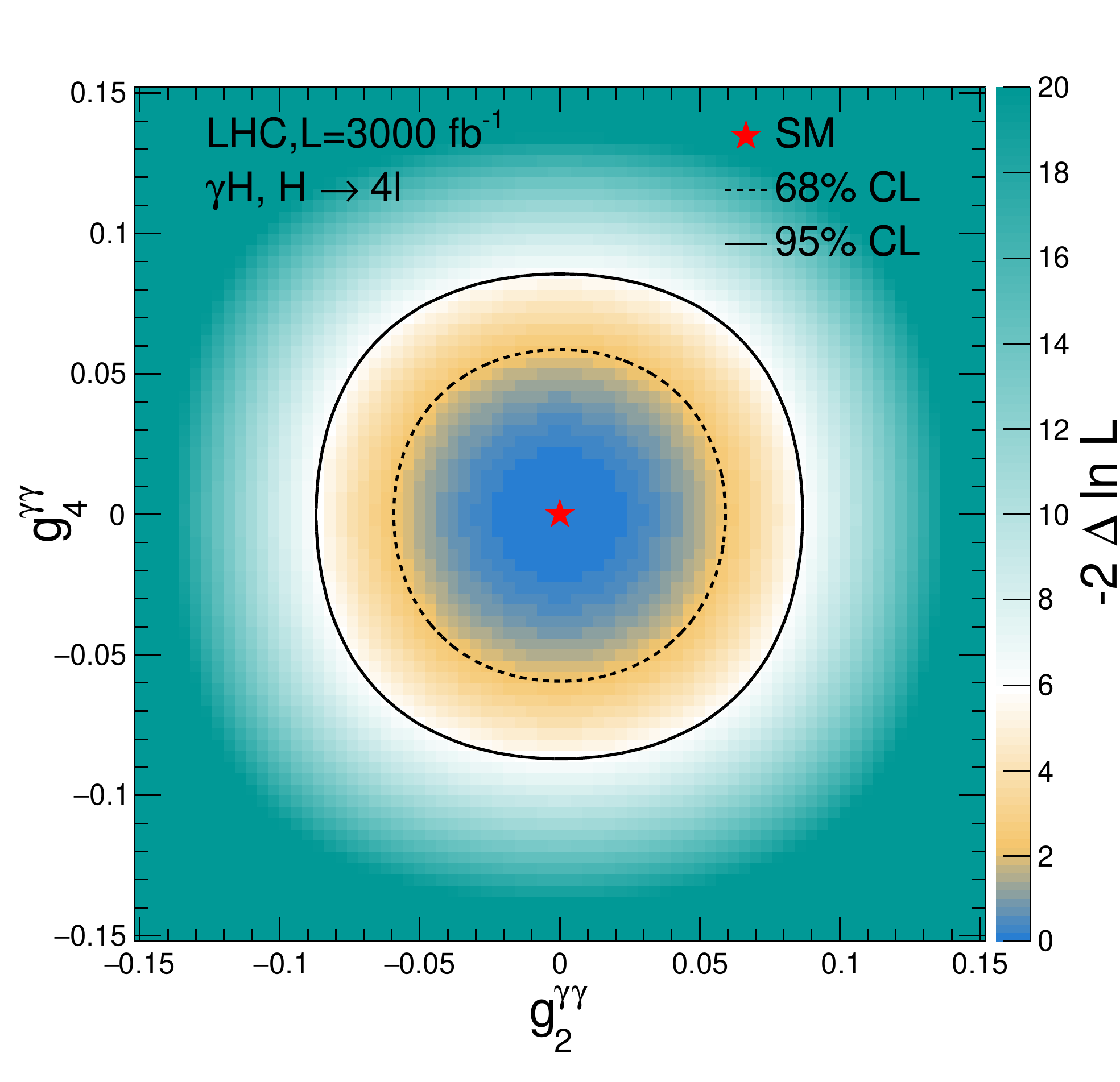}
\includegraphics[width=0.32\textwidth]{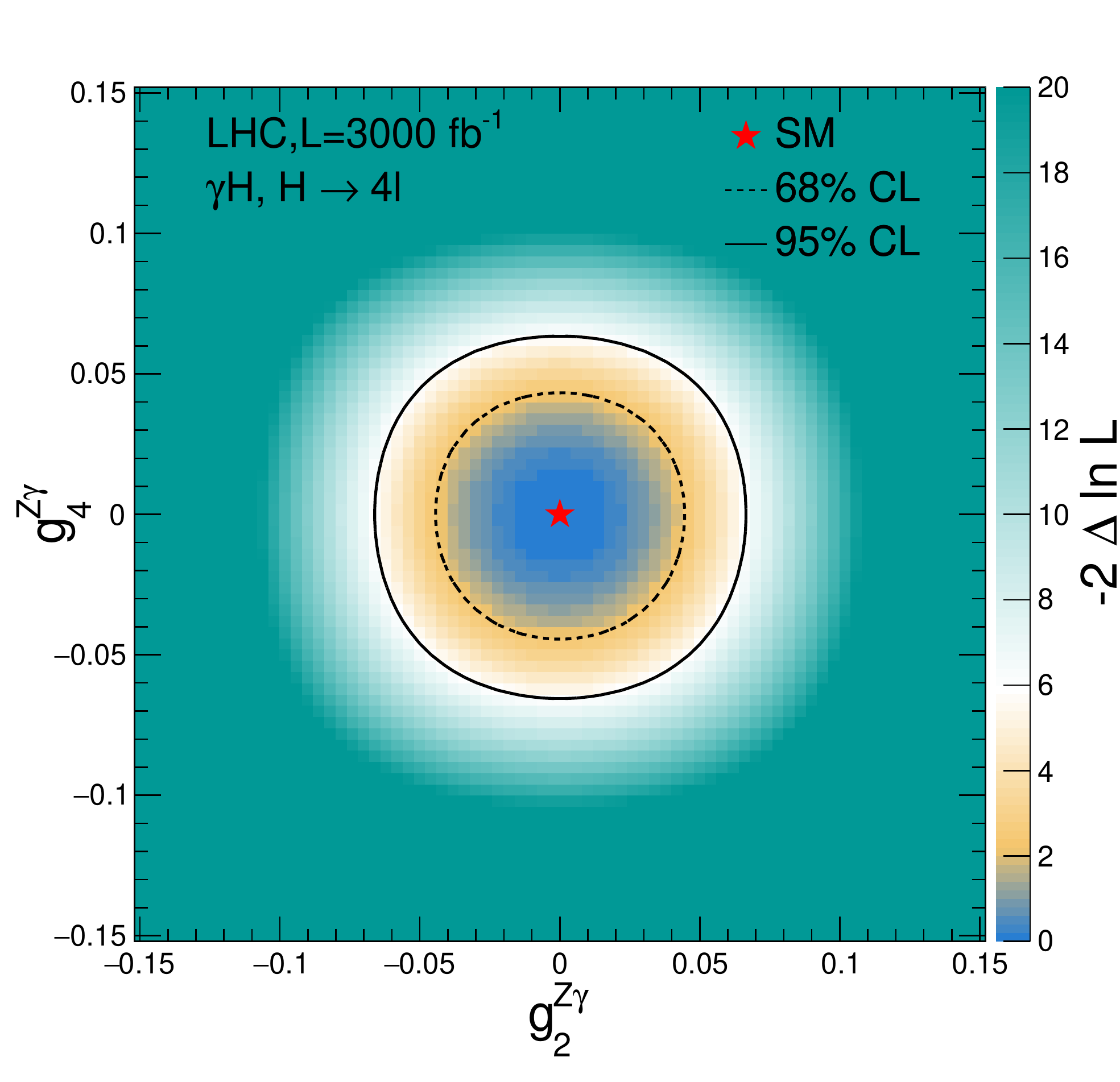}
\includegraphics[width=0.32\textwidth]{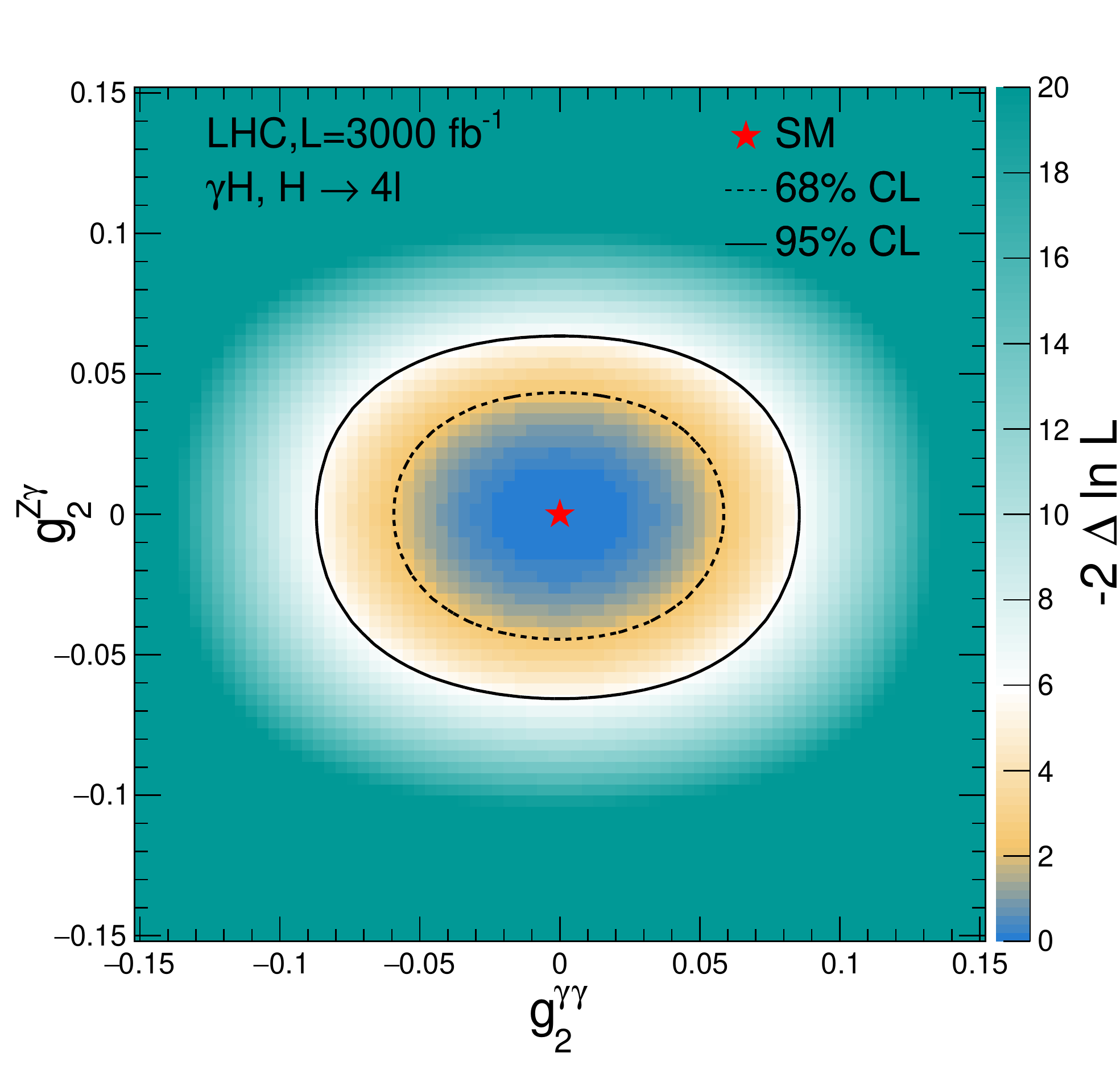}
\captionsetup{justification=centerlast}
\caption{
Expected two-dimensional constraints from a simultaneous fit of 
$g_2^{\gamma\gamma}$, $g_4^{\gamma\gamma}$, $g_2^{Z\gamma}$, and $g_4^{Z\gamma}$
using the $\gamma H$ production channel.
}
\label{fig:gammaH_scan}
\end{figure}
 
Since kinematic features of events do not differ between $CP$-odd and
$CP$-even couplings and the difference between the $HZ\gamma$ and $H\gamma\gamma$
couplings is weak, we perform a simple fit for excess of events over background 
without using kinematic distributions. The expected number of signal events is expressed 
through couplings as the product of coupling modifies on the decay and production sides. 
These expressions are similar to those in Eqs.~(\ref{eq:ratio-3}) and~(\ref{eq:cross-section-gammaH}),
but differ in two aspects. 
Equation~(\ref{eq:ratio-3}) has to be adjusted for the $H\to4\ell$ final state instead of the inclusive four-fermion 
final state and include the effect of acceptance requirements on the lepton quantities, as listed above. 
Equation~(\ref{eq:cross-section-gammaH}) has to take into account the effects of the photon
acceptance efficiency. 

With the above assumptions, the expected two-dimensional constraints on
$(g_2^{\gamma\gamma}, g_4^{\gamma\gamma})$, $(g_2^{Z\gamma}$, $g_4^{Z\gamma})$, 
and $(g_2^{Z\gamma}$, $g_2^{\gamma\gamma})$ are shown in Fig.~\ref{fig:gammaH_scan}. 
The expected $\gamma H$ constraints with the $H\to4\ell$ channel alone are not as powerful as 
those obtained from decay and shown in Fig.~\ref{fig:g_scan_decay}.
However, while this difference is sizable in the case of the $H\gamma\gamma$ couplings, 
the difference is not as large in the case of the $HZ\gamma$ couplings. 
Moreover, these constraints are comparable and even better than those obtained from production 
information in the VBF and $VH$ channels.  In both cases, significant gain will result from the analysis
of the other \Hboson decay channels, which are not considered in this feasibility study. 
Therefore, it is important to proceed with analysis of VBF, $VH$, and $\gamma H$ production
in all accessible \Hboson final states. 

Searches for heavy resonances decaying into a photon and a hadronically decaying \Hboson have been 
performed on the LHC~\cite{ATLAS:2018sxj,ATLAS:2020jeb}, where the topology of the final state
overlaps with the process of our interest, but the interpretation of the results is very different. 
A reinterpretation of these resonance-search results in terms of the EFT couplings of the \Hboson was
attempted in Ref.~\cite{Shi:2018lqf}. However, this study makes crude approximations in its
interpretation of LHC data, uses a different set of parameters in a different basis, applies additional
symmetry considerations, and is limited to $CP$-even couplings only, making a direct comparison 
with our results difficult.

\subsection{Expected constraints on photon couplings from decays with on-shell photons}
\label{subsec:RAA}

The above results can be compared to possible constraints from the measurements 
of $R_{\gamma\gamma}$ and $R_{\Z\gamma}$, defined in Section~\ref{sect:eft-xs}.
The projection of experimental measurements of the \Hboson branching fractions to 3000\,fb$^{-1}$ 
of LHC data has been performed in Ref.~\cite{Cepeda:2019klc}. 
In particular, Table 37 of this reference estimates
$R_{\gamma\gamma}\simeq1.00\pm0.05$ and $R_{\Z\gamma}\simeq1.00\pm0.24$ at 68\% CL. 
However, these measurements are estimated using the coupling modifier framework ($\kappa$-framework),
where the tensor structures of interactions of the \Hboson and the kinematic distributions are assumed
to be the same as in the SM. 
Though the kinematic distributions in decays $H\to{\gamma\gamma}$ and ${\Z\gamma}$ are not affected
by anomalous couplings, other aspects of the analyses, such as distributions of associated particles, 
would be affected. Nonetheless, one can use these estimates as optimistic expectations of constraints 
on $R_{\gamma\gamma}$ and $R_{\Z\gamma}$ with anomalous contributions. 

\begin{figure}[t]
\centering
\includegraphics[width=0.32\textwidth]{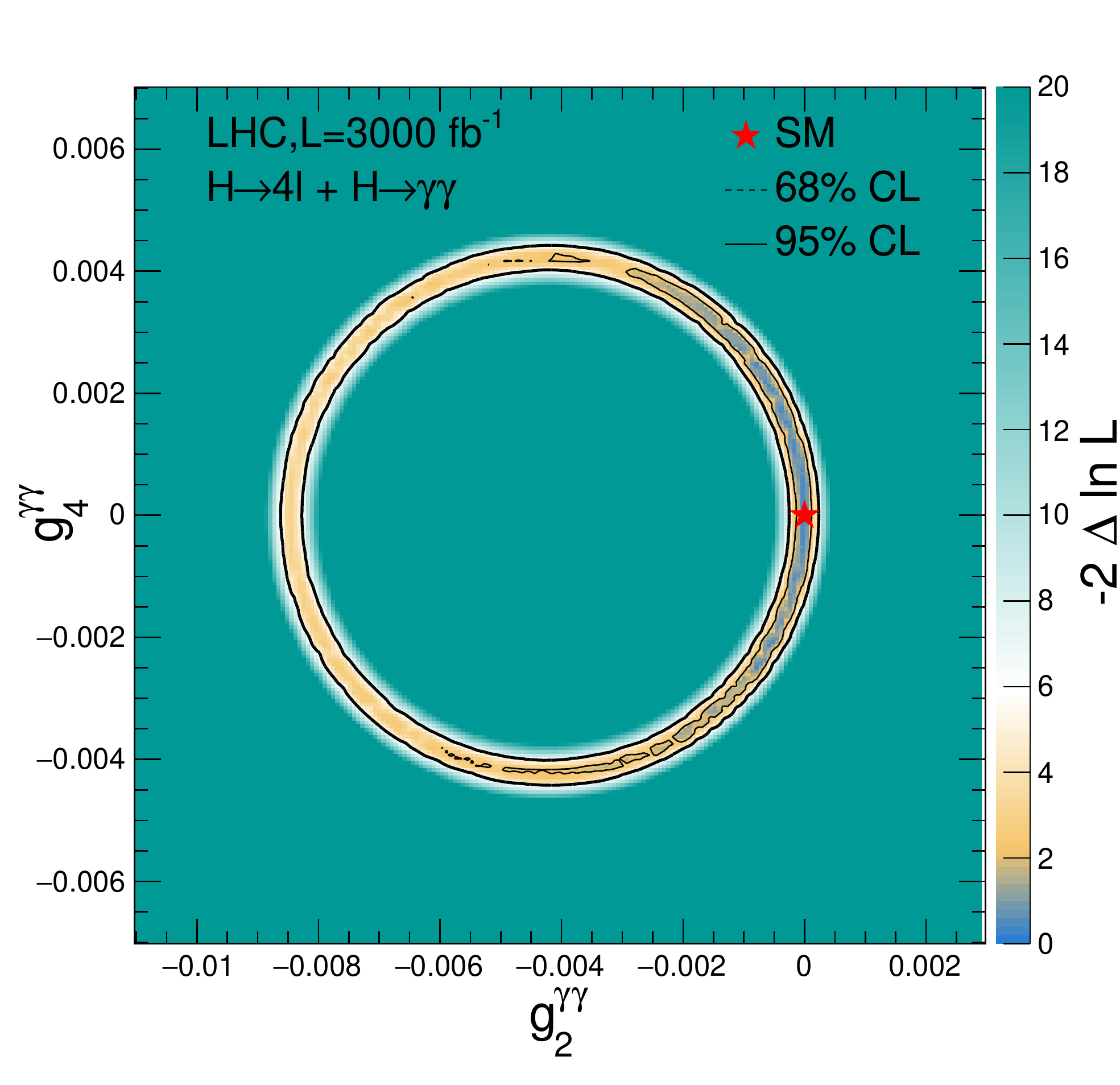}
\includegraphics[width=0.32\textwidth]{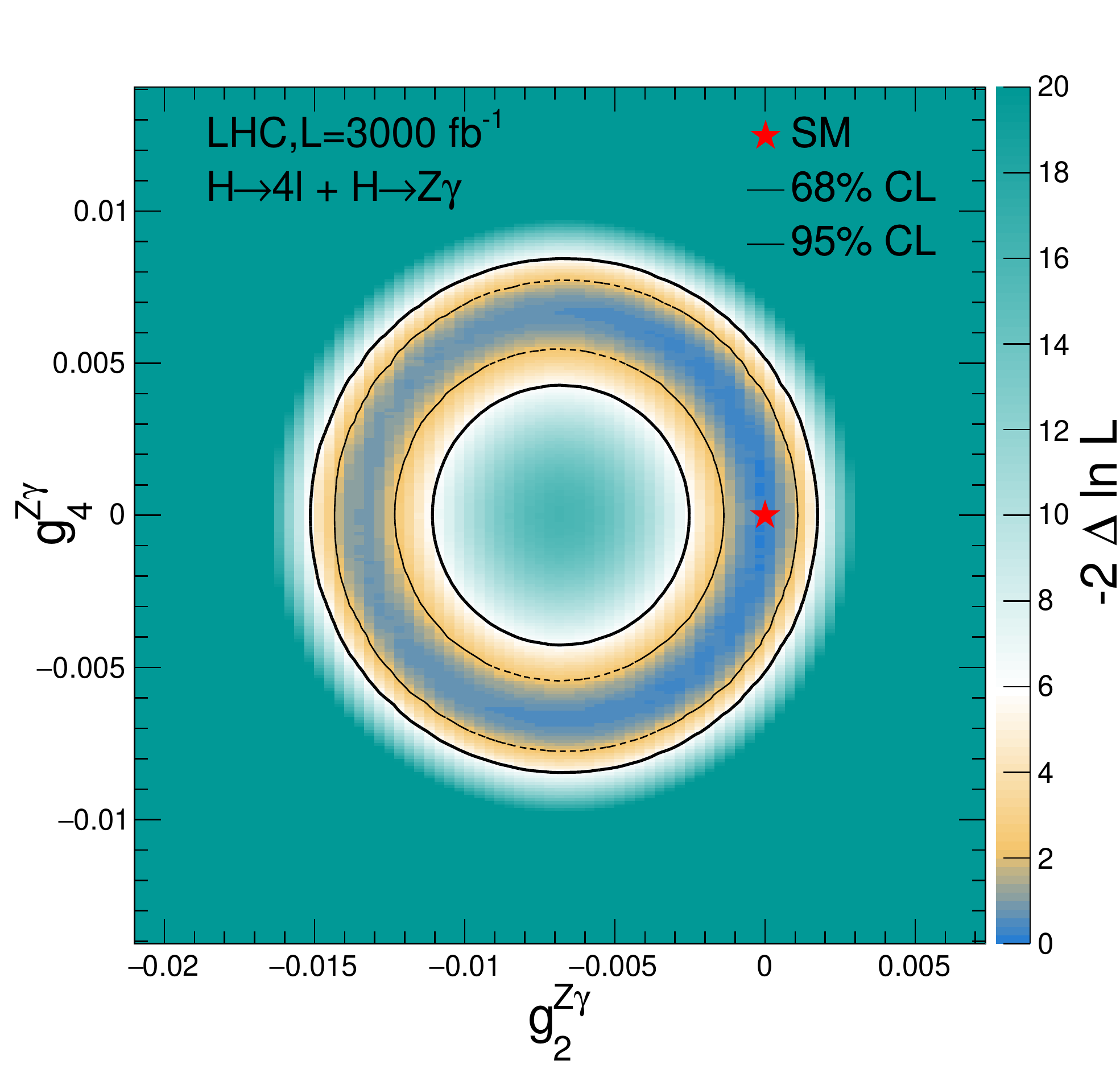}
\captionsetup{justification=centerlast}
\caption{
Expected two-dimensional constraints on
($g_2^{\gamma\gamma}$, $g_4^{\gamma\gamma})$ (left), and $(g_2^{Z\gamma}, g_4^{Z\gamma})$ (right)
using Eqs.~(\ref{eq:ratio-gammagamma-2}) and~(\ref{eq:ratio-zgamma-2}) and the HL-LHC projection
for experimental measurements of $R_{\gamma\gamma}$ and $R_{\Z\gamma}$ from Ref.~\cite{Cepeda:2019klc}.
Inclusion of the $H\to4\ell$ data with decay and production information
helps in resolving degenerate solutions on a ring of a given radius. 
}
\label{fig:RAA_scan}
\end{figure}

Relating the experimental constraints on $R_{\gamma\gamma}$ and $R_{\Z\gamma}$ to constraints
on $g_2^{\gamma\gamma}$, $g_4^{\gamma\gamma}$, $g_2^{Z\gamma}$, and $g_4^{Z\gamma}$
is not trivial, as various loop contributions are possible, as indicated in 
Eqs.~(\ref{eq:ratio-gammagamma}) and~(\ref{eq:ratio-zgamma}). 
In particular, modifications of the \Hboson couplings to fermions and $W$ boson, which are the
dominant SM contributions to the loop, cannot be disentangled from the effective point-like couplings
indicated with the $\kappa_{\cal Q}$ and $\tilde\kappa_{\cal Q}$ couplings. The latter are related to 
$g_{2,4}^{\gamma\gamma,Z\gamma}$ with $Q$ and $R_{\cal Q}$ in Eqs.~(\ref{eq:g2AA}) and (\ref{eq:g2ZA}).
However, assuming that the \Hboson couplings to fermions and $W$ boson can be constrained to 
high precision from other measurements, one can set those to SM values for the purpose of this comparison. 
In such a case, $R_{\gamma\gamma}$ and $R_{\Z\gamma}$ could be expressed directly through 
the $g_{2,4}^{\gamma\gamma}$ and $g_{2,4}^{Z\gamma}$ couplings, respectively, without 
further complication, as shown in Eqs.~(\ref{eq:ratio-gammagamma-2}) and~(\ref{eq:ratio-zgamma-2}). 

With the above assumptions, the expected two-dimensional constraints on
$(g_2^{\gamma\gamma}, g_4^{\gamma\gamma})$ and on $(g_2^{Z\gamma}, g_4^{Z\gamma})$ 
are circles in the 2D plane of the two couplings, as follows from Eqs.~(\ref{eq:ratio-gammagamma-2}) 
and~(\ref{eq:ratio-zgamma-2}), respectively. All points on the circle of a given radius are equally likely, 
because one cannot distinguish between the $CP$-odd and $CP$-even couplings from the rate
information alone. However, the addition of $H\to4\ell$ decay and production information, discussed above, 
will help to resolve degenerate solutions, as shown in Fig.~\ref{fig:RAA_scan}. 
The improvement from the inclusion of the $H\to4\ell$ data is more visible in the case
of the $H\gamma\gamma$ couplings. 
As we would expect, constraints from the $H\to{\gamma\gamma}$ and ${\Z\gamma}$ rates
are more restrictive than the constraints obtained from either decay or production information
shown in Figs.~\ref{fig:g_scan_decay} and~\ref{fig:gammaH_scan}. 

As we reach precision on anomalous $H\gamma\gamma$ and $HZ\gamma$ couplings
from the $H\to4\ell$ decay and from VBF, $VH$, and $\gamma H$ production 
comparable to that from the $H\to{\gamma\gamma}$ and ${\Z\gamma}$ decays, 
resolving the loop contributions in these production and decay processes will become important.
This is similar to the discussion of $H\to{\gamma\gamma}$ and ${\Z\gamma}$ in Section~\ref{sect:eft-xs}, 
but taking into account non-trivial $q^2$-dependence affecting kinematic distributions. 
This is equivalent to the NLO EW corrections discussed in Section~\ref{sect:eft-sm} in application to the 
SM processes, but would require consideration of anomalous couplings.


\section{Summary}
\label{sect:summary}

We have presented a study of electroweak production of the \Hboson in VBF and $VH$ and its decay to two vector bosons,
with a focus on the treatment of virtual loops of either SM particles or new states appearing in $HVV$ interactions. 
The treatment of virtual photons has been illustrated with the JHU generator framework, and comparisons have been made 
to several other frameworks, including \textsc{SMEFTsim} using \textsc{MadGraph5}, \textsc{HAWK}, and \textsc{Prophecy4f}.
A \textsc{JHUGenLexicon} program has been introduced for EFT parameterization and translation between frameworks. 
Overall, good agreement between the frameworks is found, including parameterization of EFT effects, 
once the sign conventions are matched, such as of the antisymmetric tensor 
$\varepsilon_{\mu\nu\rho\sigma}$ and the convariant derivative. The photon couplings appear to have a larger relative effect
in decay compared to production, due to the dynamics of these processes. Nonetheless, the $\gamma{H}$
production topology appears interesting for isolating the photon couplings. 

We have derived the scaling of the \Hboson production and decay rates with anomalous couplings, which are 
necessary for the total width calculations. From these, we also obtained the effective point-like couplings which 
reproduce the $H\to\mathrm{gg}$, $\gamma\gamma$, and $Z\gamma$ processes in the SM. We compared the 
effect of these  couplings to the NLO EW corrections in the other processes with \offshell\ vector bosons. 
While the effects may be reproduced to within an order of magnitude, the effective couplings are not adequate for 
a careful modeling of the EW correction. 
At the same time, however, the squared higher-order terms may become as important as linear terms at lower $q^2$ 
values of the virtual photons, something that is neglected in the the NLO EW corrections. 
Subsequently, we make a proposal on how to handle singularities involving \Hboson decays to light fermions 
via photons.  This procedure, based on matching amplitudes in the collinear limit, allows one to handle these 
singularities with efficient numerical evaluation. This approach can also incorporate the hadronic structure in
decays to quarks using the experimentally measured quantities in low-energy processes.  

We further make phenomenological observations on the special role of 
intermediate photons in analysis of LHC data in the EFT framework. Some of these features 
have been illustrated with projections for experimental measurements with the full LHC and HL-LHC datasets. 
The rates of the $H\to{\gamma\gamma}$ and ${\Z\gamma}$  processes appear the most
restrictive on the photon couplings, but cannot disentangle the $CP$-even and $CP$-odd couplings.
We observe that the decay information $H\to4\ell$ is more powerful in constraining the photon couplings
than information in the VBF and $VH$ production using the same \Hboson decay channel. 
However, these production channels, along with the $\gamma H$ production, can be analyzed using 
other \Hboson decays, with further gain in sensitivity to the photon couplings. 
The $H\to4\ell$ channel can also be further optimized for the measurements of the photon couplings 
by relaxing the invariant mass and transverse momentum constraints on the leptons in analysis of LHC data. 
This will also require careful consideration of NLO EW effects in these processes, while including effects 
of anomalous contributions. Calculation of these effects will be required for precise analysis of HL-LHC data. 

\bigskip


\noindent
{\bf Acknowledgments}:
This research is partially supported by the U.S. NSF under grant PHY-2012584. 
Calculations reported in this paper were performed on the Maryland Advanced Research Computing Center (MARCC). 
We thank our colleagues on the JHU event generator project for help and support, and in particular 
Renqi Pan, Ulascan Sarica, Yaofu Zhou, and Meng Xiao, and acknowledge the contributions of CMS collaboration 
colleagues to validation of the JHU event generator and the MELA project development. 
We thank Ritvik Gunturu and Valdis Slokenbergs for help with generating MC samples. 
We thank Marco Zaro, Fabio Maltoni, and Kentarou Mawatari for discussion of \textsc{MadGraph} sign conventions;
Ilaria Brivio and Michael Trott for discussion of \textsc{SMEFTsim};
Adam Falkowski and Ken Mimasu for discussion of \textsc{Rosetta};
Ansgar Denner, Stefan Dittmaier, and Alexander M\"uck for help with 
the \textsc{Prophecy4f} and \textsc{HAWK} generators;
Werner Bernreuther and Long Chen for discussion of loop calculations;
Ian Low and Roberto Vega-Morales for discussion of analytical calculations;
Andrew Gilbert and Jonathon Langford for discussion of EFT approaches on CMS.
We also thank other members of the LHC Higgs and EFT Working Groups for stimulating discussions. 
%


\appendix 

\section{Relative contributions of operators to the \Hboson processes on LHC}
\label{app:A}

In the following, we investigate the relative contributions of the operators listed in Table~\ref{tab:relate-couplings}
to the VBF, $VH$, and $H\to VV\to 4\ell$ processes. 
We obtain these contributions of individual terms in the mass-eigenstate basis relative to the overall cross section 
of a single operator in the Warsaw basis, excluding the SM coupling,
and relative to the SM cross section alone ($\sigma/\sigma_{\rm SM}$).
The results are shown in Tables~\ref{tab:warsaw-4l}--\ref{tab:warsaw-gammaH}, 
where the relative contributions are shown separately for the 
$H\to VV\to 4\ell$, VBF, $q\bar{q}$ or gg\,$\to V(\to\ell^+\ell^-)H$, and  $\gamma H$ processes. 
The fractions do not necessarily add up to 1 due to interference effects. 
The ratio to the SM expectation is shown for $C_{\PH X}=1$. 
As discussed in more detail in Sections~\ref{sect:eft-xs} and~\ref{sect:eft-lowq}, the presence of collinear singularities
requires a special treatment of the low-$q^2_V$ of the gauge bosons, and in the study in this Appendix
we choose to apply a requirement $q^2_V>1$\,GeV$^2$.  
The results of this study are discussed in Section~\ref{sect:eft-basis}. 

\begin{table}[h]
\centering
\captionsetup{justification=centerlast}
\caption{
Relative contributions of the individual terms in the mass-eigenstate amplitude to a single operator $C_{\PH X}$ in the Warsaw 
basis expressed as a fraction of the gg\,$\to H\to4\ell$ cross section. The SM contribution is excluded from the $HVV$ coupling,
and the cross section ratio to the SM expectation ($\sigma/\sigma_{\rm SM}$) is shown for $C_{\PH X}=1$ in the first column. 
The contributions to the $H\to4\ell$ process are shown with the requirement~$m_{\ell\ell}>1$\,GeV. 
}
\label{tab:warsaw-4l}
\begin{tabular}{lccccccccccccc}
\vspace{-0.3cm} \\
   \hline
            &  $\sigma/\sigma_{\rm SM}$  &  $\delta{g}_1^{ZZ}=\delta{g}_1^{WW}$  & $\kappa_1^{ZZ}$ 
            &  $g_2^{ZZ}$  &  $g_2^{Z\gamma}$ &  $g_2^{\gamma\gamma}$  
            &  $g_4^{ZZ}$  &  $g_4^{Z\gamma}$ &  $g_4^{\gamma\gamma}$ 
            & $\kappa_2^{Z\gamma}$ & $\kappa_1^{WW}$  & $g_2^{WW}$  &  $g_4^{WW}$  
             \\
   \hline
   $C_{\PH\Box}$    & $0.004$ & $1$ & $0$ & $0$ & $0$ & $0$ & $0$ & $0$ & $0$ & $0$ & $0$ & $0$ & $0$  \\
   $C_{\PH{D}}$    & $0.017$ & $1.078$ & $0.068$ & $0$ & $0$ & $0$ & $0$ & $0$ & $0$ & $0.486$ & $0$ & $0$ & $0$  \\
   $C_{\PH\PW}$    & $0.635$ & $0$ & $0$ & $0.00117$ & $0.685$ & $0.238$ & $0$ & $0$ & $0$ & $0$ & $0$ & $0$ & $0$  \\
   $C_{\PH\PW{B}}$    & $0.781$ & $0.007$ & $0.001$ & $0.00029$ & $0.268$ & $0.632$ & $0$ & $0$ & $0$ & $0.018$ & $0$ & $0$ & $0$  \\
   $C_{\PH{B}}$    & $2.215$ & $0$ & $0$ & $0.00003$ & $0.243$ & $0.759$ & $0$ & $0$ & $0$ & $0$ & $0$ & $0$ & $0$  \\
   $C_{\PH\widetilde{\PW}}$    & $0.579$ & $0$ & $0$ & $0$ & $0$ & $0$ & $0.00052$ & $0.713$ & $0.286$ & $0$ & $0$ & $0$ & $0$  \\
   $C_{\PH\widetilde{W}B} $    & $0.749$ & $0$ & $0$ & $0$ & $0$ & $0$ & $0.00012$ & $0.239$ & $0.683$ & $0$ & $0$ & $0$ & $0$  \\
   $C_{\PH\widetilde{B}}$    & $2.196$ & $0$ & $0$ & $0$ & $0$ & $0$ & $0.00001$ & $0.194$ & $0.720$ & $0$ & $0$ & $0$ & $0$  \\
   \hline
\end{tabular}
\end{table}

\begin{table}[h]
\centering
\captionsetup{justification=centerlast}
\caption{
Relative contributions, as in Table~\ref{tab:warsaw-4l}, to the cross section of the VBF process, with the requirement~$q^2_V>1$\,GeV$^2$. 
}
\label{tab:warsaw-VBF}
\begin{tabular}{lcccccccccccccc}
\vspace{-0.3cm} \\
   \hline
            &  $\sigma/\sigma_{\rm SM}$  & $\delta{g}_1^{ZZ}=\delta{g}_1^{WW}$  & $\kappa_1^{ZZ}$ 
            &  $g_2^{ZZ}$  &  $g_2^{Z\gamma}$ &  $g_2^{\gamma\gamma}$  
            &  $g_4^{ZZ}$  &  $g_4^{Z\gamma}$ &  $g_4^{\gamma\gamma}$ 
            & $\kappa_2^{Z\gamma}$ & $\kappa_1^{WW}$  & $g_2^{WW}$  &  $g_4^{WW}$  
             \\
   \hline
   $C_{\PH\Box}$    & $0.004$ & $1$ & $0$ & $0$ & $0$ & $0$ & $0$ & $0$ & $0$ & $0$ & $0$ & $0$ & $0$ \\
   $C_{\PH{D}}$    & $0.170$ & $0.105$ & $0.081$ & $0$ & $0$ & $0$ & $0$ & $0$ & $0$ & $0.154$ & $0.572$ & $0$ & $0$ \\
   $C_{\PH\PW}$    & $0.052$ & $0$ & $0$ & $0.159$ & $0.196$ & $0.059$ & $0$ & $0$ & $0$ & $0$ & $0$ & $0.839$ & $0$ \\
   $C_{\PH\PW{B}}$    & $0.086$ & $0.067$ & $0.086$ & $0.030$ & $0.046$ & $0.115$ & $0$ & $0$ & $0$ & $0.531$ & $0.059$ & $0$ & $0$ \\
   $C_{\PH{B}}$    & $0.063$ & $0$ & $0$ & $0.012$ & $0.159$ & $0.522$ & $0$ & $0$ & $0$ & $0$ & $0$ & $0$ & $0$ \\
   $C_{\PH\widetilde{\PW}}$    & $0.043$ & $0$ & $0$ & $0$ & $0$ & $0$ & $0.153$ & $0.207$ & $0.066$ & $0$ & $0$ & $0$ & $0.811$ \\
   $C_{\PH\widetilde{W}B} $    & $0.012$ & $0$ & $0$ & $0$ & $0$ & $0$ & $0.170$ & $0.304$ & $0.831$ & $0$ & $0$ & $0$ & $0$ \\
  $C_{\PH\widetilde{B}}$    & $0.059$ & $0$ & $0$ & $0$ & $0$ & $0$ & $0.010$ & $0.156$ & $0.520$ & $0$ & $0$ & $0$ & $0$ \\
   \hline
\end{tabular}
\end{table}

\begin{table}[h]
\centering
\captionsetup{justification=centerlast}
\caption{
Relative contributions, as in Table~\ref{tab:warsaw-4l}, to the cross section of the 
$\qqbar\to V(\to\ell^+\ell^-)H$ process, with the requirement~$m_{\ell\ell}>1$\,GeV. 
}
\label{tab:warsaw-VH}
\begin{tabular}{lcccccccccccccc}
\vspace{-0.3cm} \\
   \hline
            & $\sigma/\sigma_{\rm SM}$  & $\delta{g}_1^{ZZ}=\delta{g}_1^{WW}$  & $\kappa_1^{ZZ}$ 
            &  $g_2^{ZZ}$  &  $g_2^{Z\gamma}$ &  $g_2^{\gamma\gamma}$  
            &  $g_4^{ZZ}$  &  $g_4^{Z\gamma}$ &  $g_4^{\gamma\gamma}$ 
            & $\kappa_2^{Z\gamma}$ & $\kappa_1^{WW}$  & $g_2^{WW}$  &  $g_4^{WW}$  
             \\
   \hline
   $C_{\PH\Box}$    & $0.004$ & $1$ & $0$ & $0$ & $0$ & $0$ & $0$ & $0$ & $0$ & $0$ & $0$ & $0$ & $0$ \\
   $C_{\PH{D}}$    & $0.949$ & $0.019$ & $0.655$ & $0$ & $0$ & $0$ & $0$ & $0$ & $0$ & $1.026$ & $0$ & $0$ & $0$ \\
   $C_{\PH\PW}$    & $0.154$ & $0$ & $0$ & $1.087$ & $0.294$ & $0.017$ & $0$ & $0$ & $0$ & $0$ & $0$ & $0$ & $0$ \\
   $C_{\PH\PW{B}}$    & $2.265$ & $0.003$ & $0.151$ & $0.022$ & $0.008$ & $0.004$ & $0$ & $0$ & $0$ & $0.774$ & $0$ & $0$ & $0$ \\
   $C_{\PH{B}}$    & $0.125$ & $0$ & $0$ & $0.125$ & $0.366$ & $0.232$ & $0$ & $0$ & $0$ & $0$ & $0$ & $0$ & $0$ \\
   $C_{\PH\widetilde{\PW}}$    & $0.097$ & $0$ & $0$ & $0$ & $0$ & $0$ & $1.044$ & $0.330$ & $0.023$ & $0$ & $0$ & $0$ & $0$ \\
   $C_{\PH\widetilde{W}B} $    & $0.057$ & $0$ & $0$ & $0$ & $0$ & $0$ & $0.536$ & $0.218$ & $0.125$ & $0$ & $0$ & $0$ & $0$ \\
   $C_{\PH\widetilde{B}}$    & $0.090$ & $0$ & $0$ & $0$ & $0$ & $0$ & $0.106$ & $0.353$ & $0.263$ & $0$ & $0$ & $0$ & $0$ \\
   \hline
\end{tabular}
\end{table}

\begin{table}[h]
\centering
\captionsetup{justification=centerlast}
\caption{
Relative contributions, as in Table~\ref{tab:warsaw-4l}, to the cross section of the 
gg\,$\to Z(\to\ell^+\ell^-)H$ process, with the requirement~$m_{\ell\ell}>1$\,GeV. 
}
\label{tab:warsaw-ggVH}
\begin{tabular}{lcccccccccccccc}
\vspace{-0.3cm} \\
   \hline
            &  $\sigma/\sigma_{\rm SM}$ & $\delta{g}_1^{ZZ}=\delta{g}_1^{WW}$  & $\kappa_1^{ZZ}$ 
            &  $g_2^{ZZ}$  &  $g_2^{Z\gamma}$ &  $g_2^{\gamma\gamma}$  
            &  $g_4^{ZZ}$  &  $g_4^{Z\gamma}$ &  $g_4^{\gamma\gamma}$ 
            & $\kappa_2^{Z\gamma}$ & $\kappa_1^{WW}$  & $g_2^{WW}$  &  $g_4^{WW}$  
             \\
   \hline
     $C_{\PH\Box}$    & $0.009$ & $1$ & $0$ & $0$ & $0$ & $0$ & $0$ & $0$ & $0$ & $0$ & $0$ & $0$ & $0$\\
     $C_{\PH{D}}$    & $8.055$ & $0.006$ & $1.100$ & $0$ & $0$ & $0$ & $0$ & $0$ & $0$ & $0$ & $0$ & $0$ & $0$\\
     $C_{\PH\PW}$    & $0$ & $0$ & $0$ & $0$ & $0$ & $0$ & $0$ & $0$ & $0$ & $0$ & $0$ & $0$ & $0$\\
     $C_{\PH\PW{B}}$    & $4.495$ & $0.003$ & $1.066$ & $0$ & $0$ & $0$ & $0$ & $0$ & $0$ & $0$ & $0$ & $0$ & $0$\\
     $C_{\PH{B}}$    & $0$ & $0$ & $0$ & $0$ & $0$ & $0$ & $0$ & $0$ & $0$ & $0$ & $0$ & $0$ & $0$\\
     $C_{\PH\widetilde{\PW}}$    & $0$ & $0$ & $0$ & $0$ & $0$ & $0$ & $0$ & $0$ & $0$ & $0$ & $0$ & $0$ & $0$\\
     $C_{\PH\widetilde{W}B} $    & $0$ & $0$ & $0$ & $0$ & $0$ & $0$ & $0$ & $0$ & $0$ & $0$ & $0$ & $0$ & $0$\\
     $C_{\PH\widetilde{B}}$    & $0$ & $0$ & $0$ & $0$ & $0$ & $0$ & $0$ & $0$ & $0$ & $0$ & $0$ & $0$ & $0$\\
   \hline
\end{tabular}
\end{table}

\begin{table}[h]
\centering
\captionsetup{justification=centerlast}
\caption{
Relative contributions, as in Table~\ref{tab:warsaw-4l}, to the cross section of the $\qqbar\to\gamma H$ process.
}
\label{tab:warsaw-gammaH}
\begin{tabular}{lcccccccccccccc}
\vspace{-0.3cm} \\
   \hline
            & $\sigma/\sigma_{\rm SM}^{\gamma\PH}$  & $\delta{g}_1^{ZZ}=\delta{g}_1^{WW}$  & $\kappa_1^{ZZ}$
            &  $g_2^{ZZ}$  &  $g_2^{Z\gamma}$ &  $g_2^{\gamma\gamma}$
            &  $g_4^{ZZ}$  &  $g_4^{Z\gamma}$ &  $g_4^{\gamma\gamma}$
            & $\kappa_2^{Z\gamma}$ & $\kappa_1^{WW}$  & $g_2^{WW}$  &  $g_4^{WW}$
             \\
   \hline
     $C_{\PH\Box}$    & $0$ & $0$ & $0$ & $0$ & $0$ & $0$ & $0$ & $0$ & $0$ & $0$ & $0$ & $0$ & $0$ \\
     $C_{\PH{D}}$    & $0$ & $0$ & $0$ & $0$ & $0$ & $0$ & $0$ & $0$ & $0$ & $0$ & $0$ & $0$ & $0$ \\
     $C_{\PH\PW}$    & $60.3$ & $0$ & $0$ & $0$ & $1.190$ & $0.197$ & $0$ & $0$ & $0$ & $0$ & $0$ & $0$ & $0$ \\
     $C_{\PH\PW{B}}$    & $41.1$ & $0$ & $0$ & $0$ & $0.688$ & $0.956$ & $0$ & $0$ & $0$ & $0$ & $0$ & $0$ & $0$ \\
     $C_{\PH{B}}$    & $271.3$ & $0$ & $0$ & $0$ & $0.260$ & $0.472$ & $0$ & $0$ & $0$ & $0$ & $0$ & $0$ & $0$ \\
     $C_{\PH\widetilde{\PW}}$    & $60.1$ & $0$ & $0$ & $0$ & $0$ & $0$ & $0$ & $1.182$ & $0.198$ & $0$ & $0$ & $0$ & $0$ \\
     $C_{\PH\widetilde{W}B} $    & $41.7$ & $0$ & $0$ & $0$ & $0$ & $0$ & $0$ & $0.677$ & $0.930$ & $0$ & $0$ & $0$ & $0$ \\
     $C_{\PH\widetilde{B}}$    & $273.9$ & $0$ & $0$ & $0$ & $0$ & $0$ & $0$ & $0.263$ & $0.472$ & $0$ & $0$ & $0$ & $0$ \\     
   \hline
\end{tabular}
\end{table}

\section{Derivation of the collinear approximation to \Hboson width with intermediate photons }
\label{app:eft-lowq}

\subsection{Phase space}

The general \Hboson decay phase space reads  
\begin{eqnarray} \label{app-eq:phasespace}
	\mathrm{dPS}^{(N)}(E^2,m_1^2,...,m_N^2) = (2\pi)^{4-3N} \left[ \prod_{i=1}^N \mathrm{d}^4p_i \, \delta(p_i^2-m_i^2) \, \theta(p_i^0) \right] \delta^{(4)}\left(p_H-\sum_{i=1}^N p_i\right).
\end{eqnarray}

We rewrite the 4-particle phase space in Eq.~(\ref{app-eq:phasespace})
\begin{eqnarray}
	 \int \! \mathrm{dPS}^{(4)}(m_H^2,m_1^2,m_2^2,m_3^2,m_4^2) = (2\pi)^{-8} \int \! \left[ \prod_{i=1}^{4} \mathrm{d}^4 p_i \, \delta(p_i^2-m_i^2) \, \theta(p_i^0) \right] \delta^{(4)}(p_H-p_1-p_2-p_3-p_4)
\end{eqnarray}
by inserting the additional integrals 
\begin{eqnarray}
	1 &=& \int\!\mathrm{d}^4 p_{ij} \; \delta^{(4)}(p_{ij}-p_i-p_j),
	\\
	1 &=& \int_{q_{ij,\mathrm{min}}^2}^{q_{ij,\mathrm{max}}^2}\! \mathrm{d} q_{ij}^2 \; \delta(q_{ij}^2-p_{ij}^2)
\end{eqnarray}
for $(ij)=(12)$ and $(ij)=(34)$.
This yields 
\begin{eqnarray}
	 \int \! \mathrm{dPS}^{(4)}(...) &=& (2\pi)^{-6}
	 \int_{q_{12,\mathrm{min}}^2}^{q_{12,\mathrm{max}}^2}\! \frac{\mathrm{d}q_{12}^2}{2\pi} \, 
	 \int_{q_{34,\mathrm{min}}^2}^{q_{34,\mathrm{max}}^2}\! \frac{\mathrm{d}q_{34}^2}{2\pi} \, 
	 \mathrm{d}^4p_{12} \, \delta(p_{12}^2-q_{12}^2) \,
	 \mathrm{d}^4p_{34} \, \delta(p_{34}^2-q_{34}^2) \,
	 \delta^{(4)}(p_H-p_{12}-p_{34})
	 \nonumber \\ &&
	 \int \! \prod_{(ij)=(12),(34)} \mathrm{d}^4 p_i \, \mathrm{d}^4 p_j \, \delta(p_i^2-m_i^2) \, \delta(p_j^2-m_j^2) \, \theta(p_i^0) \, \theta(p_j^0) \, \delta^{(4)}(p_{ij}-p_i-p_j)
	 \nonumber \\
	 &=&
	 \int_{q_{12,\mathrm{min}}^2}^{q_{12,\mathrm{max}}^2}\! \frac{\mathrm{d}q_{12}^2}{2\pi} \, 
	 \int_{q_{34,\mathrm{min}}^2}^{q_{34,\mathrm{max}}^2}\! \frac{\mathrm{d}q_{34}^2}{2\pi} \, 
	 \int\! \mathrm{dPS}^{(2)}\!\!\left(m_H^2,q_{12}^2,q_{34}^2\right) \, \mathrm{dPS}^{(2)}\!\!\left(q_{12}^2,m_1^2,m_2^2\right) \, \mathrm{dPS}^{(2)}\!\!\left(q_{34}^2,m_3^2,m_4^2\right),
	 \nonumber\\ \label{app-eq:finalps4}
\end{eqnarray}
where $q_{ij,\mathrm{min}}^2=4 m_i^2$, $q_{12,\mathrm{max}}^2=m_H^2$ and $q_{34,\mathrm{max}}^2=(m_H-\sqrt{q_{12}^2})^2$.
In this form, the two integrals over $q_{ij}^2$ directly correspond to the $V$ and $V'$ squared invariant masses of the matrix element $\mathcal{M}_{H\to 4f}$. 
This will be useful later. 
\\

In case of identical fermions, it is useful to symmetrize the phase space in the last line of Eq.~(\ref{app-eq:finalps4}) because the 
matrix element has resonances not only in $q_{12}^2$ and $q_{34}^2$ but also in $q_{14}^2$ and $q_{32}^2$, i.e. 
$\mathcal{M} = A(1234) + A(1432)$. Therefore, in practice we run the numerical simulation using 
\begin{eqnarray}
	 \int \! \mathrm{dPS}^{(4)}(...) =
	  \frac12 \left[ \bigg( \text{Eq.~(\ref{app-eq:finalps4})}  \bigg) + \bigg( \text{Eq.~(\ref{app-eq:finalps4})}  \bigg)\bigg|_{2\leftrightarrow 4} \; \right].
\end{eqnarray}	 
In this write-up, however, we do not symmetrize the phase space for clarity of presentation. 
The analytic result is of course equivalent because 
\begin{eqnarray}
	  \frac12 \left[ \bigg( \text{Eq.~(\ref{app-eq:finalps4})}  \bigg) + \bigg( \text{Eq.~(\ref{app-eq:finalps4})}  \bigg)\bigg|_{2\leftrightarrow 4} \; \right]
	  \times |A(1234) + A(1432)|^2
	  = \bigg( \text{Eq.~(\ref{app-eq:finalps4})}  \bigg) \times |\mathcal{M}|^2.
\end{eqnarray}	 
\\

In a similar fashion, we rewrite the 3-particle phase space in Eq.~(\ref{app-eq:phasespace}) as
\begin{eqnarray}
	 \int \! \mathrm{dPS}^{(3)}(...) &=& 
	 \int_{q_{12,\mathrm{min}}^2}^{q_{12,\mathrm{max}}^2}\! \frac{\mathrm{d}q_{12}^2}{2\pi} \, 
	 \int\! \mathrm{dPS}^{(2)}\!\!\left(m_H^2,q_{12}^2,0\right) \, \mathrm{dPS}^{(2)}\!\!\left(q_{12}^2,m_1^2,m_2^2\right),
	 \label{app-eq:finalps3}
\end{eqnarray}
where the integration boundaries are $q_{12,\mathrm{min}}^2=4 m_1^2$ and $q_{12,\mathrm{max}}^2=m_H^2$.

\subsection{Collinear approximation}

Collinear factorization properties of a general process with leptons ($f=e,\mu,\tau$) yields  
\begin{eqnarray} \label{app-eq:collapprox}
	\int\! \mathrm{dPS}^{(N)} \, |\mathcal{M}_{X+\ell\bar\ell}|^2
	\; \xrightarrow[]{\,(p_\ell+p_{\bar{\ell}})^2 \ll \mu^2 \,} \;
	\frac\alpha{2\pi} \, S_\ell(\mu^2) \, \int\! \mathrm{dPS}^{(N-1)} \, |\mathcal{M}_{X+\gamma}|^2
	+ \mathcal{O}(\mu^2\big/\hat{s}),
\end{eqnarray}
where $\hat{s} \gg \mu^2$ is a typical momentum scale of the process and~\cite{Denner:2019zfp}
\begin{eqnarray}
	S_\ell(\mu^2) &=& \int_0^1 \! \mathrm{d}x \left[ \left( x^2 + (1-x)^2 \right) \log\left(\frac{\mu^2}{m^2_\ell} x (1-x) \right) + 2x(1-x) \right]
	\nonumber \\
	&=& \frac23 \log\left(\frac{\mu^2}{m^2_\ell}\right) - \frac{10}{9}.
\end{eqnarray}
\\
In case of hadronic final states $\gamma^* \to jj$ Eq.~(\ref{app-eq:collapprox}) becomes  
\begin{eqnarray}
	\int\! \mathrm{dPS}^{(N)} \, |\mathcal{M}_{X+jj}|^2
	\; \xrightarrow[]{\,(p_q+p_{\bar{q}})^2 \ll \mu^2 \,} \;
	\frac\alpha{2\pi} \, S_\mathrm{had}(\mu^2)\,  \int\! \mathrm{dPS}^{(N-1)} \, |\mathcal{M}_{X+\gamma}|^2
	+ \mathcal{O}(\mu^2\big/\hat{s}),
\end{eqnarray}
where $S_\mathrm{had}(\mu^2)$ contains non-perturbative contributions and is closely related to $F_\mathrm{had}(\mu^2)$ in Ref.~\cite{Denner:2019zfp}. 
It can be written as
\begin{eqnarray}
	S_\mathrm{had}(\mu^2) =  \frac{22}{9} \log\left(\frac{\mu^2}{M_Z^2}\right) + 2\pi \, \frac{\Delta \alpha_\mathrm{had}^{(5)}(M_Z^2)}{\alpha}.
\end{eqnarray}
$\Delta \alpha_\mathrm{had}^{(5)}(s)$ is the hadronic contribution of the shift in the running electromagnetic coupling (with five active quark flavors) in 
$\alpha(s) = \alpha(0) \big/ (1-\Delta \alpha(s))$ 
and is measured to be $\Delta \alpha_\mathrm{had}^{(5)}(M_Z^2)=(276.11 \pm 1.11) \times 10^{-4}$~\cite{Eidelman:1995ny,Keshavarzi:2018mgv}.

\subsection{Collinear approximation applied to $H\to 2\ell\gamma$}    \label{app-subsec:Hga2l}

We split the invariant mass integration in Eq.~(\ref{app-eq:finalps3})
\begin{eqnarray}\label{app-eq:pssplit1}
	\left( \int_{q_{12,\mathrm{min}}^2}^{\mu^2}\! \mathrm{d} q_{12}^2 + \int_{\mu^2}^{q_{12,\mathrm{max}}^2}\! \mathrm{d}q_{12}^2 \right)
\end{eqnarray}
and define two regions I and II, respectively. This yields region I where the collinear approximation for intermediate photons can be applied 
\begin{eqnarray}
	\Gamma_{H\to \gamma^*\gamma\to 2\ell\gamma}^\mathrm{I}(\mu^2) &=& 
	\frac1{2m_H}
	\int_{q_{12,\mathrm{min}}^2}^{\mu^2}\! \frac{q_{12}^2}{2\pi} \, 
	\int\! \mathrm{dPS}^{(2)}\!\!\left(m_H^2,q_{12}^2,0\right) \, \mathrm{dPS}^{(2)}\!\!\left(q_{12}^2,m_1^2,m_2^2\right)
	\; |\mathcal{M}_{H\to \gamma^*\gamma\to 2\ell\gamma}|^2 
	\nonumber \\
	&=&
	\frac1{2m_H} \frac{\alpha}{2\pi} S_\ell(\mu^2) \int\! \mathrm{dPS}^{(2)}\!\!\left(m_H^2,0,0\right) 
	\; |\mathcal{M}_{H\to \gamma \gamma}|^2
	+ \mathcal{O}(\mu^2\big/\hat{s})
	\nonumber \\
	&=& 
	\frac{\alpha}{2\pi}  S_\ell(\mu^2) \; (2_{\gamma\gamma}) \, \Gamma_{H \to \gamma \gamma}
	+ \mathcal{O}(\mu^2\big/\hat{s}).
\end{eqnarray}
For intermediate $Z$ bosons in region I, we just keep it as it is
\begin{eqnarray}
	\Gamma_{H\to \Z^*\gamma\to 2\ell\gamma}^I(\mu^2) = \Gamma_{H\to \Z^*\gamma\to 2\ell\gamma} \Big|_{q_{12}^2 < \mu^2}.
\end{eqnarray}
The interference between $\gamma^*$ and $Z^*$ states is zero in the collinear approximation in region I. 
In region II, the collinear approximation for intermediate photons is not applied and includes intermediate photons and $Z$s and their interference 
\begin{eqnarray}
	\Gamma_{H \to 2\ell\gamma}^\mathrm{II}(\mu^2) &=&
	\frac1{2m_H}
	\int_{\mu^2}^{q_{12,\mathrm{max}}^2}\! \frac{\mathrm{d} q_{12}^2}{2\pi} 
	\int\! \mathrm{dPS}^{(2)}\!\!\left(m_H^2,q_{12}^2,0\right) \, \mathrm{dPS}^{(2)}\!\!\left(q_{12}^2,m_1^2,m_2^2\right) 
	\; |\mathcal{M}_{H\to 2\ell\gamma}|^2 
	\nonumber \\
	&=& 
	\Gamma_{H\to 2\ell\gamma} \Big|_{q_{12}^2 \ge \mu^2}.
\end{eqnarray}
Hence, we end up with the sum of the above contributions
\begin{eqnarray} \label{app-eq:sumpwidths2lga}
	\Gamma_{H\to 2\ell\gamma} = \Gamma_{H\to \gamma^*\gamma\to 2\ell\gamma}^\mathrm{I}(\mu^2) + \Gamma_{H\to \gamma^*\gamma \to 2\ell\gamma}^\mathrm{II}(\mu^2)+
	\Gamma_{H\to Z^*\gamma\to 2\ell\gamma}^\mathrm{I}(\mu^2).
\end{eqnarray}

\subsection{Collinear approximation applied to $H\to 4\ell$}  \label{app-subsec:H4l}

Now, we use the phase space parameterization in Eq.~(\ref{app-eq:finalps4}) and split the two $q^2_{ij}$ integrations into a {\it low} mass and {\it high} mass region, which are
separated by $\mu^2$. Hence, we write it as
\begin{eqnarray}\label{app-eq:pssplit2}
	\left( \int_{q_{12,\mathrm{min}}^2}^{\mu^2}\! \mathrm{d} q_{12}^2 + \int_{\mu^2}^{q_{12,\mathrm{max}}^2}\! \mathrm{d}q_{12}^2 \right)
	\times
	\left( \int_{q_{34,\mathrm{min}}^2}^{\mu^2}\! \mathrm{d} q_{34}^2 + \int_{\mu^2}^{q_{34,\mathrm{max}}^2}\! \mathrm{d}q_{34}^2 \right),
\end{eqnarray}
resulting in four regions I--IV. 
If we choose $\mu^2$ such that $q^2_{ij,\mathrm{min}} \ll \mu^2 \ll \hat{s}$ then 
we can apply the collinear approximation of Eq.~(\ref{app-eq:collapprox}) in the respective 
sectors\footnote{We assume that in the collinear regions the matrix element is dominated by $\gamma^*$ exchange such that we can neglect $Z^*$ exchange.}.
Let's start with region I, where $q_{12}^2 \le \mu^2$ and $q_{34}^2 > \mu^2$
\begin{eqnarray}
	\Gamma_{H\to 2\ell2\ell'}^\mathrm{I}(\mu^2) &=& 
	\frac{1}{2m_H} \frac1{(4_{\ell\ell'})}
	\int_{q_{12,\mathrm{min}}^2}^{\mu^2}\! \frac{\mathrm{d} q_{12}^2}{2\pi} \int_{\mu^2}^{q_{34,\mathrm{max}}^2}\! \frac{\mathrm{d}q_{34}^2}{2\pi}
	\int\! \mathrm{dPS}^{(2)}\!\!\left(m_H^2,q_{12}^2,q_{34}^2\right) \, \mathrm{dPS}^{(2)}\!\!\left(q_{12}^2,m_1^2,m_2^2\right) 
	\nonumber \\
	&& \times \, \mathrm{dPS}^{(2)}\!\!\left(q_{34}^2,m_3^2,m_4^2\right) 
	\; |\mathcal{M}_{H\to 2\ell 2\ell'}|^2 
	\nonumber \\
	&=&
	\frac{1}{2m_H} \frac1{(4_{\ell\ell'})} \frac{\alpha}{2\pi}  S_\ell(\mu^2) \int_{\mu^2}^{q_{34,\mathrm{max}}^2}\! \frac{\mathrm{d}q_{34}^2}{2\pi} 
	\int\! \mathrm{dPS}^{(2)}\!\!\left(m_H^2,0,q_{34}^2\right) \; \mathrm{dPS}^{(2)}\!\!\left(q_{34}^2,m_3^2,m_4^2\right)
	\; |\mathcal{M}_{H\to \gamma 2\ell'}|^2
	+ \mathcal{O}(\mu^2\big/\hat{s})
	\nonumber \\
	&&+\frac{1}{2m_H} \frac1{(4_{\ell\ell'})}
	\int_{q_{12,\mathrm{min}}^2}^{\mu^2}\! \frac{\mathrm{d} q_{12}^2}{2\pi} \int_{\mu^2}^{q_{34,\mathrm{max}}^2}\! \frac{\mathrm{d}q_{34}^2}{2\pi}
	\int\! \mathrm{dPS}^{(2)}\!\!\left(m_H^2,q_{12}^2,q_{34}^2\right) \, \mathrm{dPS}^{(2)}\!\!\left(q_{12}^2,m_1^2,m_2^2\right) 
	\nonumber \\
	&& \times \, \mathrm{dPS}^{(2)}\!\!\left(q_{34}^2,m_3^2,m_4^2\right) 
	\; |\mathcal{M}_{H\to Z^*(\to 2\ell) + Z^*\!\!/\gamma^*(\to 2\ell')}|^2 
	\nonumber \\
	&=& 
	\frac{\alpha}{2\pi} \, \frac1{(2_{\ell\ell'})} \, S_\ell(\mu^2) \; \Gamma_{H \to \gamma 2\ell'} \Big|_{q_{34}^2 \ge \mu^2}
	+ \mathcal{O}(\mu^2\big/\hat{s}) 
	\nonumber \\
	&&+ \Gamma_{H\to Z^*(\to 2\ell) + Z^*\!\!/\gamma^*(\to 2\ell')}\Big|_{q_{12}^2 < \mu^2, \; q_{34}^2 \ge \mu^2}.
\end{eqnarray}
In a completely analog way, we find region II with  $q_{12}^2 > \mu^2$ and $q_{34}^2 \le \mu^2$
\begin{eqnarray}
	\Gamma_{H\to 2\ell 2\ell'}^\mathrm{II}(\mu^2) &=&
	\frac{1}{2m_H} \frac1{(4_{\ell\ell'})}
	\int_{\mu^2}^{q_{12,\mathrm{max}}^2}\! \frac{\mathrm{d}q_{12}^2}{2\pi}  \int_{q_{34,\mathrm{min}}^2}^{\mu^2}\! \frac{\mathrm{d}q_{34}^2}{2\pi}
	\int\! \mathrm{dPS}^{(2)}\!\!\left(m_H^2,q_{12}^2,q_{34}^2\right) \, \mathrm{dPS}^{(2)}\!\!\left(q_{12}^2,m_1^2,m_2^2\right) 
	\nonumber \\
	&& \times \, \mathrm{dPS}^{(2)}\!\!\left(q_{34}^2,m_3^2,m_4^2\right) 
	\; |\mathcal{M}_{H\to 2\ell 2\ell'}|^2 
	\nonumber \\
	&=& 
	\frac{\alpha}{2\pi} \frac1{(2_{\ell\ell'})} S_{\ell'}(\mu^2) \; \Gamma_{H \to 2\ell\gamma} \Big|_{q_{12}^2 \ge \mu^2}
	+ \mathcal{O}(\mu^2\big/\hat{s})
	 \nonumber \\
	&&+ \Gamma_{H\to Z^*\!\!/\gamma^*(\to 2\ell) + Z^*(\to 2\ell') }\Big|_{q_{12}^2 \ge \mu^2, \; q_{34}^2 < \mu^2}.
\end{eqnarray}
The region III contains the case where both, $q^2_{12}$ and $q^2_{34}$, are smaller than $\mu^2$
\begin{eqnarray}
	\Gamma_{H\to 2\ell 2\ell'}^\mathrm{III}(\mu^2) &=&
	\frac{1}{2m_H} \frac1{(4_{\ell\ell'})}
	\int_{q_{12,\mathrm{min}}^2}^{\mu^2}\! \frac{\mathrm{d} q_{12}^2}{2\pi}  \int_{q_{34,\mathrm{min}}^2}^{\mu^2}\! \frac{\mathrm{d}q_{34}^2}{2\pi}
	\int\! \mathrm{dPS}^{(2)}\!\!\left(m_H^2,q_{12}^2,q_{34}^2\right) \, \mathrm{dPS}^{(2)}\!\!\left(q_{12}^2,m_1^2,m_2^2\right) 
	\nonumber \\
	&& \times \, \mathrm{dPS}^{(2)}\!\!\left(q_{34}^2,m_3^2,m_4^2\right) 
	\; |\mathcal{M}_{H\to 2\ell 2\ell'}|^2 
	\nonumber \\
	&=& 
	\frac{1}{2m_H} \frac1{(2_{\ell\ell'})} \left(\frac{\alpha}{2\pi}\right)^2 S_\ell(\mu^2) S_{\ell'}(\mu^2) 
	\int\! \mathrm{dPS}^{(2)}\!\!\left(m_H^2,0,0\right) 
	\; |\mathcal{M}_{H\to \gamma\gamma}|^2
	+ \mathcal{O}(\mu^2\big/\hat{s})
	\nonumber \\	
	&&+\frac{1}{2m_H} \frac1{(4_{\ell\ell'})}
	\int_{q_{12,\mathrm{min}}^2}^{\mu^2}\! \frac{\mathrm{d} q_{12}^2}{2\pi}  \int_{q_{34,\mathrm{min}}^2}^{\mu^2}\! \frac{\mathrm{d}q_{34}^2}{2\pi}
	\int\! \mathrm{dPS}^{(2)}\!\!\left(m_H^2,q_{12}^2,q_{34}^2\right) \, \mathrm{dPS}^{(2)}\!\!\left(q_{12}^2,m_1^2,m_2^2\right) 
	\nonumber \\
	&& \times \, \mathrm{dPS}^{(2)}\!\!\left(q_{34}^2,m_3^2,m_4^2\right) 
	\; |\mathcal{M}_{H\to Z^* Z^* \to 2\ell 2\ell'}|^2 
	\nonumber \\
	&=& 
	\left(\frac{\alpha}{2\pi}\right)^2 \frac{(2_{\gamma\gamma})}{(2_{\ell\ell'})}  S_\ell(\mu^2) S_{\ell'}(\mu^2) 
	\; \Gamma_{H \to \gamma\gamma} 
	+ \mathcal{O}(\mu^2\big/\hat{s})
	\nonumber \\
	&& + \Gamma_{H\to Z^* Z^* \to 2\ell 2\ell'}\Big|_{q_{12}^2 < \mu^2, \; q_{34}^2 < \mu^2}.
\end{eqnarray}	
Region IV contains the case where both, $q^2_{12}$ and $q^2_{34}$, are larger than $\mu^2$. Therefore, no collinear approximation is applied
\begin{eqnarray}
	\Gamma_{H\to 2\ell2\ell'}^\mathrm{IV}(\mu^2) &=&
	\frac{1}{2m_H} \frac1{(4_{\ell\ell'})}
	\int_{\mu^2}^{q_{12,\mathrm{max}}^2}\! \frac{\mathrm{d} q_{12}^2}{2\pi} \int_{\mu^2}^{q_{34,\mathrm{max}}^2}\! \frac{\mathrm{d}q_{34}^2}{2\pi}
	\int\! \mathrm{dPS}^{(2)}\!\!\left(m_H^2,q_{12}^2,q_{34}^2\right) \, \mathrm{dPS}^{(2)}\!\!\left(q_{12}^2,m_1^2,m_2^2\right) 
	\nonumber \\
	&& \times \, \mathrm{dPS}^{(2)}\!\!\left(q_{34}^2,m_3^2,m_4^2\right) 
	\; |\mathcal{M}_{H\to 2\ell 2\ell'}|^2 
	\nonumber \\
	&=& 
	\Gamma_{H\to 2\ell 2\ell'} \Big|_{q_{12}^2 \ge \mu^2,q_{34}^2 \ge \mu^2}
\end{eqnarray}		
\\

Finally, we can write Eq.~(\ref{eq:hwidth4f}) as 
\begin{eqnarray} \label{app-eq:sumpwidths4l}
	\Gamma_{H\to 2\ell 2\ell'} = \Gamma_{H\to 2\ell 2\ell'}^\mathrm{I}(\mu^2) \,+\, \Gamma_{H\to 2\ell 2\ell'}^\mathrm{II}(\mu^2) \,+\, \Gamma_{H\to 2\ell 2\ell'}^\mathrm{III}(\mu^2) \,+\, \Gamma_{H\to 2\ell 2\ell'}^\mathrm{IV}(\mu^2).
\end{eqnarray}
Note that the left-hand side is independent of the arbitrary value $\mu$, as long as $m_f^2 \ll \mu^2 \ll M_H^2$ is fulfilled. 
It is a nice feature that the anomalous couplings don't have to be made explicit in the above derivation. They are always contained in the partial widths $\Gamma_{H\to 4\ell}$, $\Gamma_{H\to 2\ell+\gamma}$, and $\Gamma_{H\to \gamma\gamma}$.

\subsection{Decays to quarks}

The derivations in the previous Sections~\ref{app-subsec:H4l} and~\ref{app-subsec:Hga2l} remain valid,  if we 
sum over all five light quarks and replace $S_\ell(\mu^2) \to S_\mathrm{had}(\mu^2)$.

%
%
%
\providecommand{\href}[2]{#2}\begingroup\raggedright\endgroup

\end{document}